\documentclass[preprint2]{proto}
\pdfoutput=1
\usepackage{times}
\usepackage{xcolor}
\usepackage{graphicx}
\usepackage{wasysym}
\usepackage{float}
\usepackage{textcomp}
\usepackage{gensymb}
\usepackage{tabularx}
\usepackage{multirow}
\usepackage[draft]{todonotes}   % notes showed
\usepackage[normalem]{ulem} % for /sout strike out

\graphicspath{{./}{Figures/}}

\newcommand{\kms}{\mbox{km~s$^{-1}$}}

\newcommand{\add}[1]{\textcolor{cyan}{\textbf{#1}}}

\newcommand{\remove}[1]{\textcolor{red}{\sout{#1}}}
\begin{document}

\title{\textbf{\LARGE CHEMISTRY OF COMET ATMOSPHERES}}

\author {\textbf{\large N. Biver}}
\affil{\small\em LESIA, Observatoire de Paris, PSL Research University, CNRS, 
  Sorbonne Universit\'e, Universit\'e de Paris, Meudon, France}

\author {\textbf{\large N. Dello Russo}}
\affil{\small\em Johns Hopkins University Applied Physics Laboratory, Laurel, MD 20723, USA}

\author {\textbf{\large C. Opitom}}
\affil{\small\em Institute for Astronomy, University of Edinburgh, Royal Observatory, Edinburgh, EH9 3HJ, UK}

\author {\textbf{\large M. Rubin}}
\affil{\small\em Physics Institute, Space Research and Planetary Sciences, University of Bern, Bern, Switzerland}

\begin{abstract}

\begin{list}{ } {\rightmargin 1in}
\baselineskip = 11pt
\parindent=1pc
{\small 
The composition of cometary ices provides key information on the thermal and chemical properties of the outer parts of the protoplanetary disk where they formed 4.6 Gy ago. This chapter reviews our knowledge of composition of cometary comae based on remote spectroscopy and in-situ investigations techniques. Cometary comae can be dominated by water vapour, CO or CO$_2$. The abundances of several dozen of molecules, with a growing number of complex organics, have been measured in comets. Many species that are not directly sublimating from the nucleus ices have also been observed and traced out into the coma in order to determine their production mechanisms. Chemical diversity in the comet population and compositional heterogeneity of the coma are discussed. With the completion of the Rosetta mission, isotopic ratios, which hold additional clues on the origin of cometary material, have been measured in several species. Finally, important pending questions (e.g., the nitrogen deficiency in comets) and the need for further work in certain critical areas are discussed in order to answer questions and resolve discrepancies between techniques. 
\\~\\~\\~}%leave this in to get the correct vertical space after the abstract
\end{list}
\end{abstract}  

%-----------------------------------------------------------------------------------------------------------
% For end of June, everybody should have started writing their assigned section
%-----------------------------------------------------------------------------------------------------------

%-----------------------------------------------------------------------------------------------------------
%--------------------------------- Cyrielle
\section{\textbf{INTRODUCTION}}
%---------- since I have written a few things on my side here, I put them here (Nicolas)
\label{sec:intro}

We discuss the current knowledge of the composition of comets as derived from observations of molecular species in their comae. These species either sublimate directly into the coma from the nucleus ices (parent molecules) or they are secondary products. 
These secondary products can arrive in the coma from either the dissociation of a parent molecule or as a distributed source contained in released ice and dust. We review the investigative techniques which have made considerable progress in determining the composition of cometary comae over the last two decades including remote spectroscopic observation (from UV to Radio wavelengths) and in-situ mass spectrometry (\textit{Giotto} and \textit{Rosetta} missions).
%The \textit{Rosetta} mission from European Space Agency was equipped with several instruments to probe the coma of the Jupiter Family Comet (JFC) 67P. It escorted the comet for two years around its perihelion on 13 August 2015. The remote instruments used to probe the gas coma composition were Alice, the Ultraviolet Imaging Spectrograph \citep{Parker2007}, OSIRIS the visible scientific cameras with narrow band filters for the Wide Angle Camera \citep{Keller2007}, the Visual IR Thermal Imaging Spectrometer \citep[VIRTIS, $\lambda=0.5-5\mu$m][]{Coradini2007} and the Microwave Instrument for Rosetta Orbiter \citep[MIRO, $\lambda\sim$0.6~mm,][]{Gulkis2007}.
%For in-situ composition analysis, the Rosetta Orbiter Spectrometer for Ion and Neutral Analysis \citep[ROSINA,][]{Balsiger2007}, could probe molecular gas composition with a high sensitivity and mass resolution m/$\Delta$m up to 3000. 
The measured column density (remote observation) or local density (mass spectrometry) of the given molecule characterizes the local state of the cometary coma at a given time. Observational techniques with a range of spatial coverage and resolution provide information on how molecules are distributed in the coma. The basic assumption of steady state continuous production of gas or dust from the nucleus is generally used to retrieve the molecular production rates $Q_{molec}$. 
We discuss the molecules that dominate the gas flow of the nucleus and drive cometary activity. We then present our current knowledge of all molecules that have been detected in the coma and their abundances relative to other volatiles, most often water ($Q_{molec}$/$Q_{H_2O}$), and the source of possible discrepancies between various measurements. Composition can vary significantly from one comet to another, and we review this variability and link it to possible comet origins. Spatial, short, and long term monitoring of comets, especially as undertaken by the \textit{Rosetta} mission to comet 67P/Churyumov-Gerasimenko (hereafter 67P), have also shown that the composition of the coma can vary both locally and with time or heliocentric distance. Isotopic abundances also provide important insights into the origin of cometary matter. Several key isotopic measurements, such as the D-to-H and $^{14}$N/$^{15}$N ratios, which are fundamental to understanding the delivery of cometary material to planets are presented here. We conclude with needs and perspectives that should improve our current understanding of comet composition.

% this version does not accept ps files so I had to convert them to pdf, but need to crop them afterwards... any better solution? Nicolas - figure 1 move to section 4

%--------------------------------- Nicolas
  \subsection{Radio spectroscopy of rotational lines}
  \label{sec:radio-spectro}
\begin{figure}[ht!]
 %\begin{center}
 \includegraphics[angle=0,width=\columnwidth]{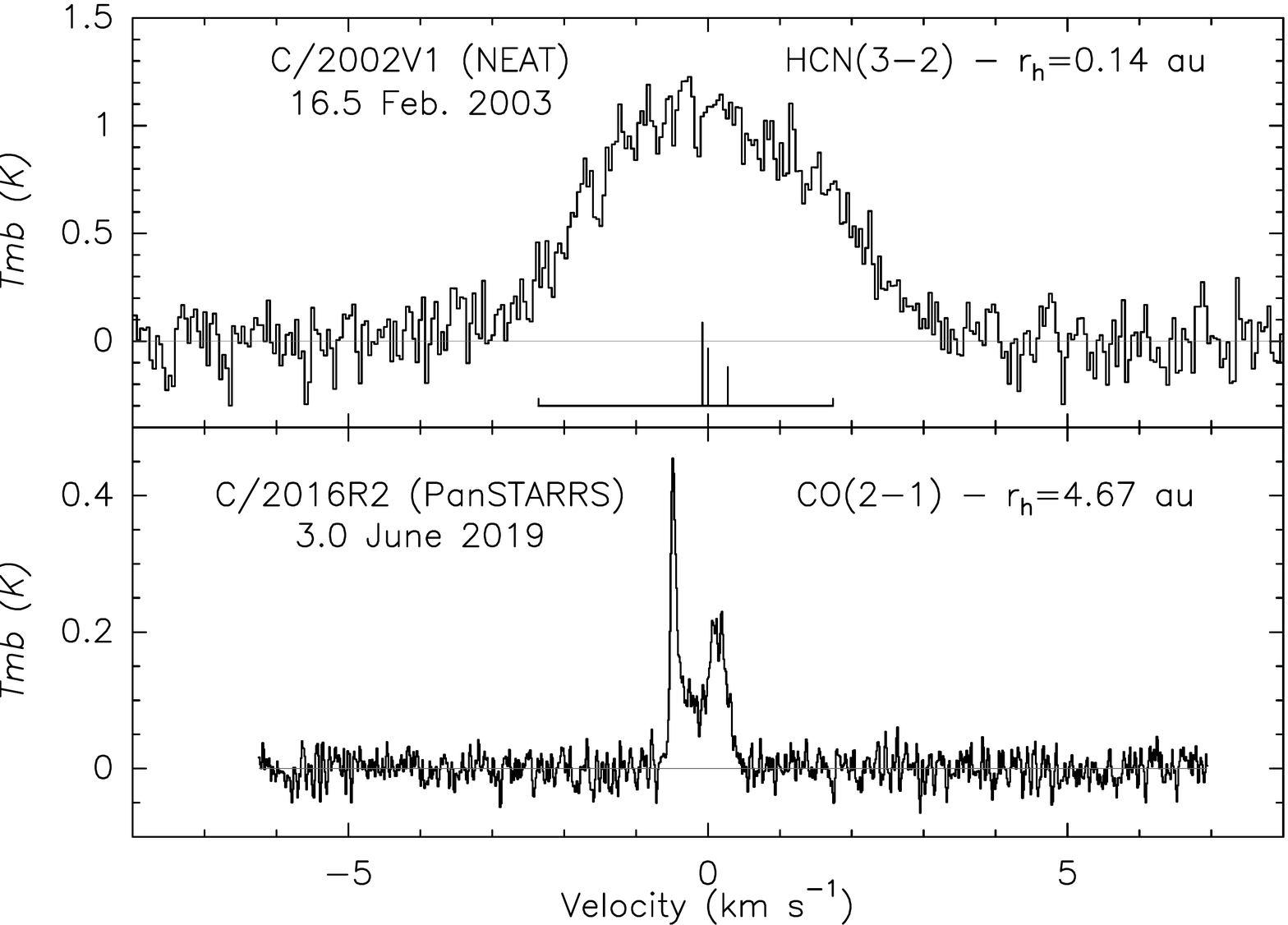}%\vspace{-0.5cm}
 \caption{Cometary lines observed at 1~mm wavelength at high spectral resolution (10-40 kHz). The horizontal axis is the Doppler velocity relative to the comet reference frame. The vertical axis is the line intensity in unit of K in the main beam antenna temperature scale. 
 The top spectrum is the HCN(3--2) line at 265886.434~MHz showing a broad line width due to the large expansion velocity (2.0 \kms) at a very small heliocentric distance ($r_h$), \citep{Biver2011}. The positions and intensities of the hyperfine components are indicated below the line.
 The bottom spectrum is the CO(2--1) line at 230538.000 MHz showing a narrow line in comet C/2016~R2 far from the Sun. It also exhibits a very narrow blue-shifted component (Full width at half maximum = 0.08 \kms) indicative of a very cold ($\sim$3 K) sunward jet.}
 \label{fig-linewidth}
 %\end{center}
 \end{figure}

   Radio spectroscopy is a powerful technique to probe the composition of cometary atmospheres. All asymmetric molecules have a dipole moment and can be observed in the millimeter to the submillimeter  domain via their rotational lines. At the temperature of coma gas (typically 5 - 200 K), most molecules have their peak rotational emission in the 80--800~GHz frequency domain, probed by ground- and space-based radio telescopes.
   In addition, the excitation process is mostly spontaneous emission due to radiative decay of the low energy rotational levels and can be efficient up to large heliocentric distances, such as in the case of CO detected in at least three comets (C/1995 O1 (Hale-Bopp), 29P, and C/2016 R2 (PanSTARRS)) beyond 5~au from the Sun.
   Of the $\sim$30 molecular species detected by remote spectroscopy in comets, 80\% have been detected in the radio \citep{Bockelee2004,Biver2015}. Radio spectroscopy of comets has the additional unique capability of resolving the line Doppler velocity profile due to its very high spectral resolution ($\nu/\delta\nu = 10^6$ to $10^8$). Cometary lines are typically 1--4~\kms~wide due to Doppler broadening (see example in Fig.~\ref{fig-linewidth}) and resolving the line enables measurement of the gas expansion velocity, asymmetry in outgassing, and characterization of opacity effects, such as in comet 67P with \textit{Rosetta}/MIRO \citep{Biver2019b}.

    \subsubsection{Wide band high resolution spectroscopy}
\begin{figure*}[ht!]
 \begin{center}
 \includegraphics[angle=0,width=0.88\textwidth]{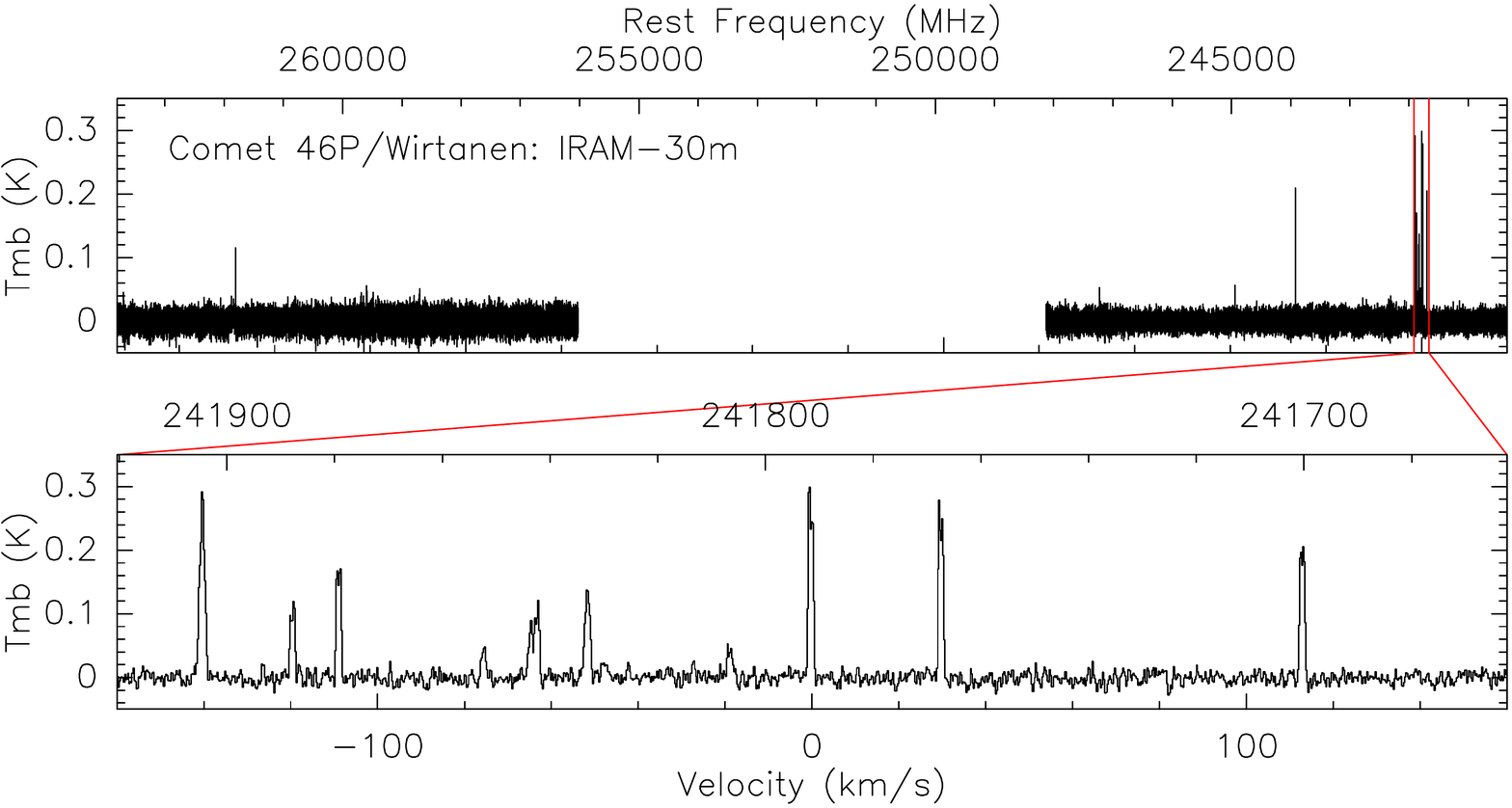}\vspace{-3.0cm}
 \caption{Spectrum of comet 46P/Wirtanen observed in December 2018 with the IRAM EMIR230 receiver coupled to the FTS spectrometer. With such a setup 2$\times$8~GHz, on each vertical and horizontal polarization (averaged here), is instantaneously covered, with a 200~kHz resolution (zoom on one series of methanol lines at 242~GHz which are resolved).}
 \label{fig-specwideband}
 \end{center}
 \end{figure*}

    Since the advent of millimeter and submillimeter astronomy in the late 1980's there has been steady progress in receiver and spectrometer performance. For the last ten years, new wide band receivers (typically 4 to 16~GHz instantaneous bandwidth) have equipped many facilities such as IRAM \citep[i.e., EMIR,][]{Carter2012}, NOEMA or ALMA. High spectral resolution ($\nu/\delta\nu > 10^6$) is also necessary to resolve cometary lines in spectra, and benefit from the full sensitivity of the receiver. This has become possible with the most recent correlators or Fourier Transform Spectrometers (FTS) \citep{Klein2012} which can offer a spectral resolution better than 200~kHz over several GHz of bandwidth. An example of the combined (instantaneous) frequency coverage ($2\times8$~GHz) and high resolution (200~kHz) spectrum of comet 46P is shown in Fig.~\ref{fig-specwideband} \citep{Biver2021a}.
  
    \subsubsection{Using a large number of lines to detect molecules}
    Another advantage of the wide frequency coverage of radio receivers and spectrometers is the ability to detect more complex molecules, which have line emission spectra that encompass a large number of lines with similar intensities. The number of lines increases rapidly with molecular complexity; for example, in the 210--270~GHz domain, CO has one line, H$_2$CO three lines, and CH$_3$CHO 64 lines, with intensities comparable within a factor three. Averaging many lines of similar expected intensity (weighted according to their S/N) for a given species increases its detectability; for example, the ability to average 100 lines of similar strength (and S/N) improves detection sensitivity by a factor of ten over a single line. This method has been applied recently in the detection of complex organics in comets even when individual lines are too weak by themselves to be seen in spectra \citep[e.g.,][]{Biver2015}. This technique is also applied to infrared echelle spectra that sample many ro-vibrational transitions of a given molecule \citep{Paganini2017}. The high spectral resolution of radio spectroscopy and the relatively narrow width of cometary lines also limits confusion between lines of different molecules.

%--------------------------------- Neil
  \subsection{Infrared spectroscopy of vibrational bands}
  \label{sec:ir-spectro}
\begin{figure*}[ht!]
      \begin{center}
      \includegraphics[angle=0,width=1.03\textwidth]{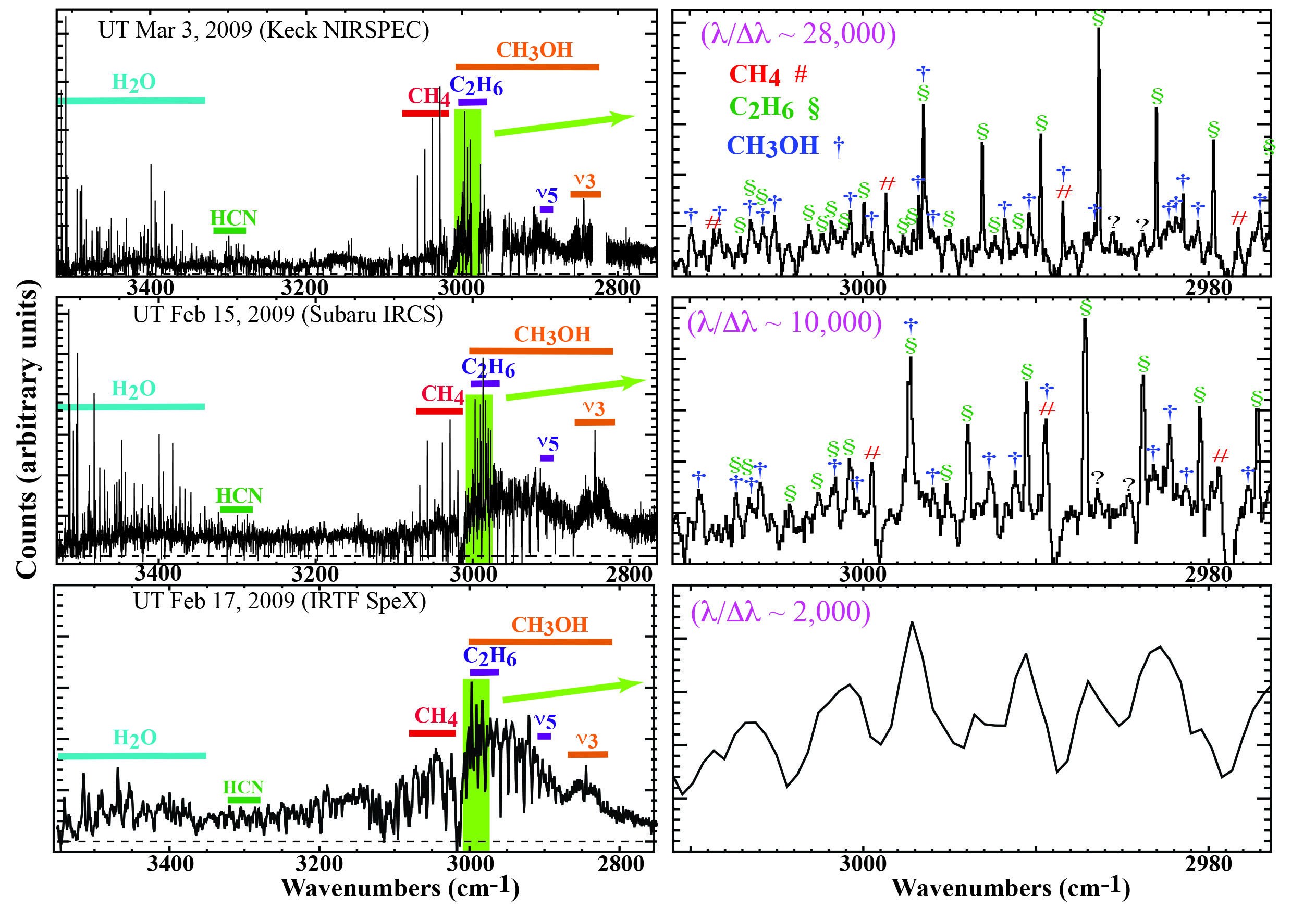}
      \caption{Near-IR spectra of comet C/2007 N3 Lulin obtained at different spectral resolving powers. This illustrates the density of ro-vibrational emissions from multiple species with overlapping vibrational bands in some spectral regions. High spectral resolving power is essential for disentangling and interpreting molecular contributions at IR wavelengths.}
      \label{fig:IR resolution}
      \end{center}
  \end{figure*}

  Many parent molecules have strong vibrational bands in the near-infrared (IR) in the 2 – 5 $\mu$m spectral region. Because there are several windows of high atmospheric transmittance in the near-IR, ground-based observations have been the primary driver of molecular identifications in this spectral region in comets. Since many species have IR vibrational bands in overlapping spectral regions, high spectral resolving power ($\lambda$/$\delta\lambda$ $>$ 10$^4$) is needed to detect ro-vibrational lines within vibrational bands (Fig.~\ref{fig:IR resolution}). Even relatively simple molecules can have complex ro-vibrational spectra creating a “forest” of lines with multiple contributing species, making high-resolution necessary in order to disentangle these highly diagnostic emissions in IR spectra (Fig.~\ref{fig:IR resolution}).
  
  The IR spectral region provides the only means to detect certain species. For example, symmetric hydrocarbons (e.g., CH$_4$, C$_2$H$_6$, C$_2$H$_4$, and C$_2$H$_2$) can only be investigated in comets with remote sensing techniques at IR wavelengths. This is also the case for CO$_2$, one of the main cometary volatiles that under many circumstances drives activity, but can only be directly detected from space-based platforms. Additionally, IR wavelengths provide the most efficient and straightforward method of studying H$_2$O, the most abundant volatile in comets, through multiple non-resonance fluorescence emissions from ground-based observatories or through its strong fundamental vibrational bands from space. Other molecules typically sampled at IR wavelengths (HCN, CH$_3$OH, H$_2$CO, NH$_3$, CO, and OCS) can also be detected at radio wavelengths (and in the case of CO both radio and UV wavelengths), providing additional independent methods for studying these volatiles in comets.
  
    \subsubsection{Factors that affect detectability of molecules}
    Derived volatile production rates are determined by temperature-dependent fluorescence efficiencies (g-factors) for individual ro-vibrational emissions for the given species. Over the years IR fluorescence models have been developed and improved for all molecules that have been detected in comets as well as additional molecules with potentially detectable IR bands \citep[e.g.,][]{Villanueva2018}.  The temperature-dependent strengths and positions of ro-vibrational emissions for H$_2$O, OH, CO, OCS, CH$_4$, C$_2$H$_6$, C$_2$H$_2$, CH$_3$OH, H$_2$CO, HCN, NH$_3$, and NH$_2$ are well enough understood to enable spectral blends to be interpreted and corrected for and production rates to be accurately determined with uniform methodology. 

    Many additional more complex molecules have strong IR bands that are theoretically detectable even at their lower abundances with the sensitivities of modern ground-based high-resolution IR spectrometers, but are in practice largely inaccessible. The reasons for this are related to spectral complexity, which increases fast for larger molecules. First, many parameters needed for developing accurate fluorescence models are unavailable for these complex molecules. Second, spectral confusion from many lines of simpler but more abundant molecules makes definitive identification of more complex but less abundant molecules very difficult. For example, spectral complexity in the 3.35 $\mu$m region of comets even at high spectral resolution is dominated by multiple dense emissions of primarily C$_2$H$_6$ and CH$_3$OH (Fig.~\ref{fig:IR resolution}), which interferes with the detection of more complex hydrocarbons (e.g., C$_3$H$_8$, C$_4$H$_{10}$).  The spectral complexity near 3.0 $\mu$m is dominated by the presence of many simple molecules (e.g., H$_2$O, OH, NH$_3$, NH$_2$, HCN, C$_2$H$_2$), which makes detection of more complex long-chain molecules such as HC$_3$N and C$_4$H$_2$ problematic.

    Detection of species from ground-based IR observations can also be hindered by extinction from the terrestrial atmosphere. An example of this is CO$_2$, an abundant species in comets that has a very strong band near 4.26 $\mu$m, but is completely obscured in ground-based observations by terrestrial CO$_2$. Detection of CO and CH$_4$ in comets depends on a sufficient large relative geocentric velocity to shift cometary lines from their terrestrial counterparts. Strong lines from other species (e.g., C$_2$H$_2$, C$_2$H$_4$, and NH$_3$) can coincidentally be in spectral regions of poor transmittance making their detection difficult unless atmospheric water vapor is low.

    Ground-based observations of comets at mid- to far-IR wavelengths ($\sim$ 5 - 100 $\mu$m) are similarly hindered by terrestrial atmospheric extinction in addition to increasing and rapidly changing thermal emission from telescope and atmosphere. Thus, studies in this wavelength region are mostly done from airborne observatories (e.g., SOFIA) or space-based telescopes (e.g., Spitzer). While most volatile coma molecules are best detected at near-IR and radio wavelengths, the mid- to far-IR can provide information on the composition of the less-volatile dusty coma components \citep[e.g.,][]{Lisse2020}.

%\com{Cyrielle}{Are we missing anything for X-rays far-UV regions?}
%--------------------------------- Cyrielle
  \subsection{UV-Visible spectroscopy of electronic transitions}
  \label{sec:vis-spectro}
  
  \begin{figure}[ht!]
%      \begin{center}
      \includegraphics[angle=0,width=1.0\columnwidth,trim=56 10 55 50,clip]{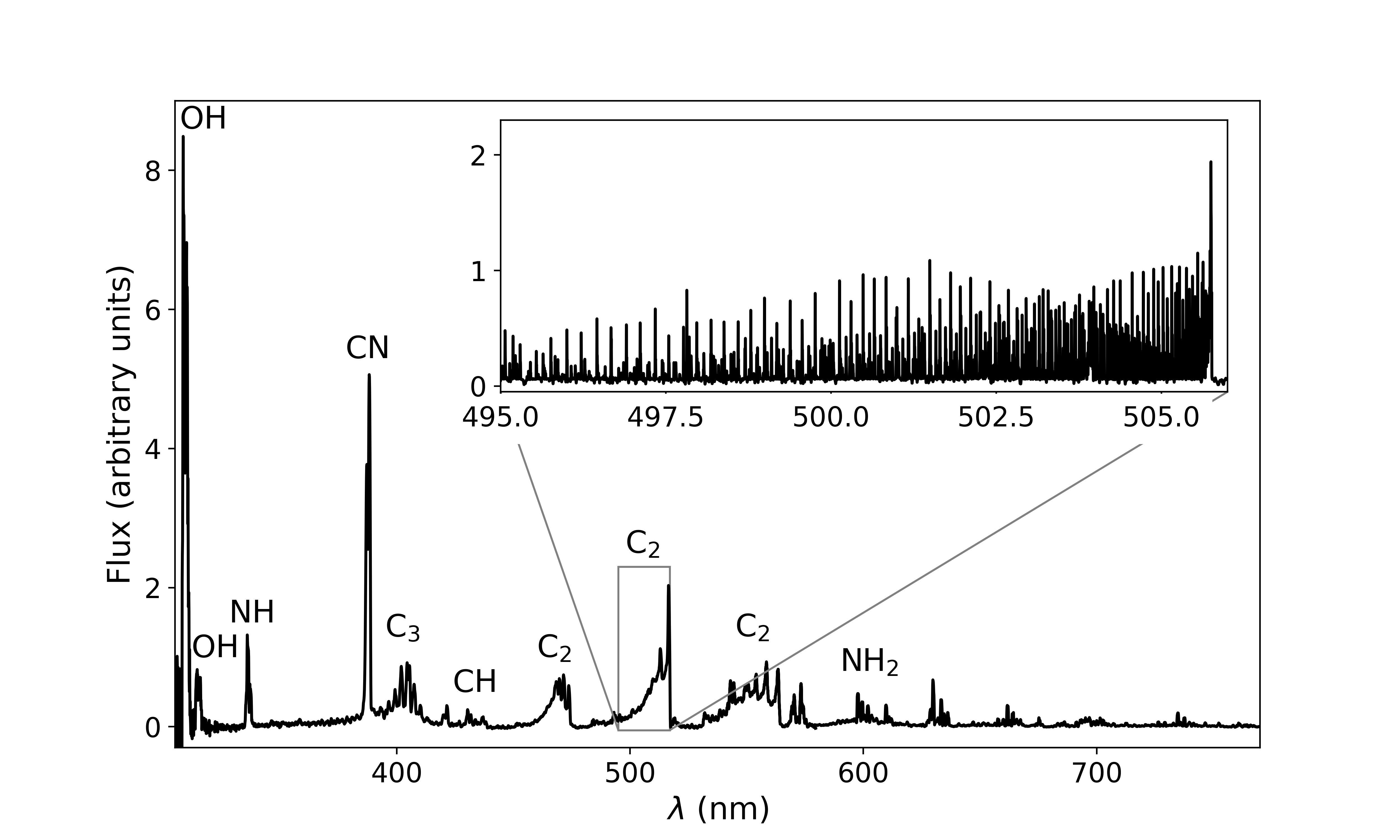}
      \caption{Optical spectrum of comet C/2015 ER61 (PANSTARRS) obtained with the ISIS spectrograph at the William Herschel Telescope (A. Fitzsimmons and M. Hyland, priv. comm.). The inset shows the region of the C$_2$ ($\Delta\nu=0$) band observed with a much higher spectral resolution using the UVES spectrograph at the VLT \citep{Yang2018}.}
      \label{fig:Vis_ER61}
%      \end{center}
  \end{figure}  
  
    Spectroscopic observations of cometary atmospheres at UV and visible wavelengths sample mostly electronic transitions from secondary products such as OH, NH, CH, CN, C$_2$, C$_3$, or NH$_2$, as illustrated in Fig. \ref{fig:Vis_ER61}. Some parent species such as CO or S$_2$ can also be probed at UV wavelengths. In addition to purely electronic transitions, some prompt emission lines \citep[see, e.g.,][]{Bodewits2022} from O, C, or OH are also detected at visible wavelengths, which can give information on their parent molecules like H$_2$O and CO$_2$. X-ray emissions have also been seen in comets, but this reveals the composition of the solar wind rather than the cometary coma \citep{Bodewits2022}.

%More details about emission processes in the coma of comets are given in Chapter XXX. In this section we review some of the major advances in UV-visible spectroscopy of comets over the past 20 years.
  	
    \subsubsection{Low-res spectroscopy and narrow-band imaging}
    Comets have been observed with low resolution spectroscopy at visible wavelengths for over a century, focusing on secondary products as illustrated in Fig.\ref{fig:Vis_ER61}. Even though spectra often have a high density of lines due to the number of molecular emission bands in the visible spectral region, narrow-band imaging and low resolution spectroscopy were among the first tools available to constrain the composition of cometary atmospheres. Narrow-band filters isolate the light emitted by several gas species as well as the dust-reflected sunlight. When combined with a camera, narrow-band filters enable compositional and morphological studies of the coma. Narrow-band imaging and low-resolution spectroscopy of comets at visible wavelengths are among the most sensitive and routine observations available to study the composition of comets. Visible databases contain several hundred comets observed over decades, providing a means to study comet composition and its diversity statistically \citep{AHearn1995,Schleicher2008,Fink2009,Langland2011,Cochran2012}. Those two techniques are covered in detail by \cite{Schleicher2004} and \cite{Feldman2004} and we refer the reader to these publications for an extended discussion. 
     
    Because of terrestrial atmospheric extinction, observations blueward of 300 nm can only be performed from space. However, they are instrumental in constraining coma abundances of the main constituents of cometary ices. The main features observable in the UV spectra of comets from parent and fragment species are discussed by \cite{Bockelee2004} and \cite{Feldman2004}. An important example is Hydrogen, one of the main photodissociation products of water, which is detectable through its Lyman-$\alpha$ emission at 121.6~nm.
	
    The last decade has seen a significant increase in the use of Integral Field Unit spectrographs (IFUs), particularly on large telescopes. This technique is especially advantageous for the observation of extended objects because it provides simultaneous spatial and spectral information. Observing comets with an IFU combines the advantages of spectroscopy and narrow-band imaging by enabling mapping of the spatial distributions of several gas species and the dust continuum at different wavelengths simultaneously. Given the rotational variability of comets on timescales as brief as a few hours, IFU observations have the advantage of simultaneous measurements of multiple species in contrast to narrow-band observations where different filters need to be cycled to sample each species. The first observations of comets with IFUs at visible wavelengths are relatively recent \citep{Dorman2013,Vaughan2017,Opitom2019} but they have demonstrated the potential of this type of observation to study the spatial distributions of secondary products and their production mechanisms in the coma.

     \subsubsection{High-resolution spectroscopy}
    The last 30 years have seen the advent of high resolution spectroscopic observations of comets. Fabry-Perot interferometers, providing high spectral resolutions but covering very limited wavelengths ranges were among the first instruments used to perform high spectral resolution observations of comets \citep{Roesler1985}. Crossed-dispersed spectrographs with spectral resolutions from a few tens of thousands to as high as 200,000 and covering larger wavelength ranges are now available on large telescopes (UVES at the Very Large Telescope, HET at McDonald Observatory, HIRES at Keck, or HDS at the Subaru Telescope). This new generation of optical instrumentation allows emission lines composing molecular bands to be resolved and has opened a new era of visible observations of comets, as illustrated in Fig. \ref{fig:Vis_ER61}.

    A resolving power of $\sim$ 40,000 is sufficient to separate emission lines from isotopologes of the same species. 
%The first measurements of isotopic ratios using high resolution optical spectroscopy were performed in the late 1990s-early 2000s at McDonald observatory \citep{Kleine1995,Wyckoff2000,Arpigny2003}, using $^{12}$C$^{13}$C, $^{13}$C$^{14}$N, and $^{12}$C$^{15}$N emission lines. More recently, the D/H ratio was measured from the OH (0-0) band around 309 nm \cite{Hutsemekers2008} and Hubble STIS observations of D Lyman-$\alpha$ \citep{Weaver2008}. 
    UV and visible high resolution spectroscopic observations provide the opportunity to measure the isotopic abundance ratios of several key elements (H, C, N, and O).  High resolution spectroscopic observations of comets at optical wavelengths can also resolve ortho and para spin isomers, enabling estimates of the ortho-to-para abundance ratios (OPR) of water and NH$_2$ \citep[][and  section~\ref{sec:opr}]{Kawakita2001,Shinnaka2012,Aikawa2022}. %Isotopic ratios and OPRs are discussed in more detail later in this chapter. 

    High resolution spectroscopic observations are also necessary to resolve telluric and cometary emission lines. For example, forbidden oxygen lines [OI] around 557, 630, and 636 nm and the sodium D doublet at 589.0 and 589.6 nm are blended with telluric lines at lower spectral resolution. High resolution spectroscopy is currently only feasible for relatively active comets (with high gas production rates) that are close to the Sun (typically within 3 au); however, high resolution spectrographs mounted on the next generation of extremely large telescopes, such as HIRES on the E-ELT, will more routinely allow these studies in a larger subset of comets. 

%--------------------------------- Martin
  \subsection{In-situ mass spectrometry}
  Another way to investigate the neutral gas environment around comets is in-situ mass spectrometry. This method provides neutral gas abundances along the trajectory of the spacecraft and is therefore limited to dedicated missions. Neutral gas mass spectrometry has been carried out at two comets, the profiles of the \textit{Giotto} and \textit{Rosetta} comet missions are discussed below.

  \subsubsection{Comet mission profiles} \label{sec:mass-spectro}
\begin{figure}[ht!]
       \includegraphics[angle=0,width=1.0\columnwidth,trim=10 5 10 5,clip]{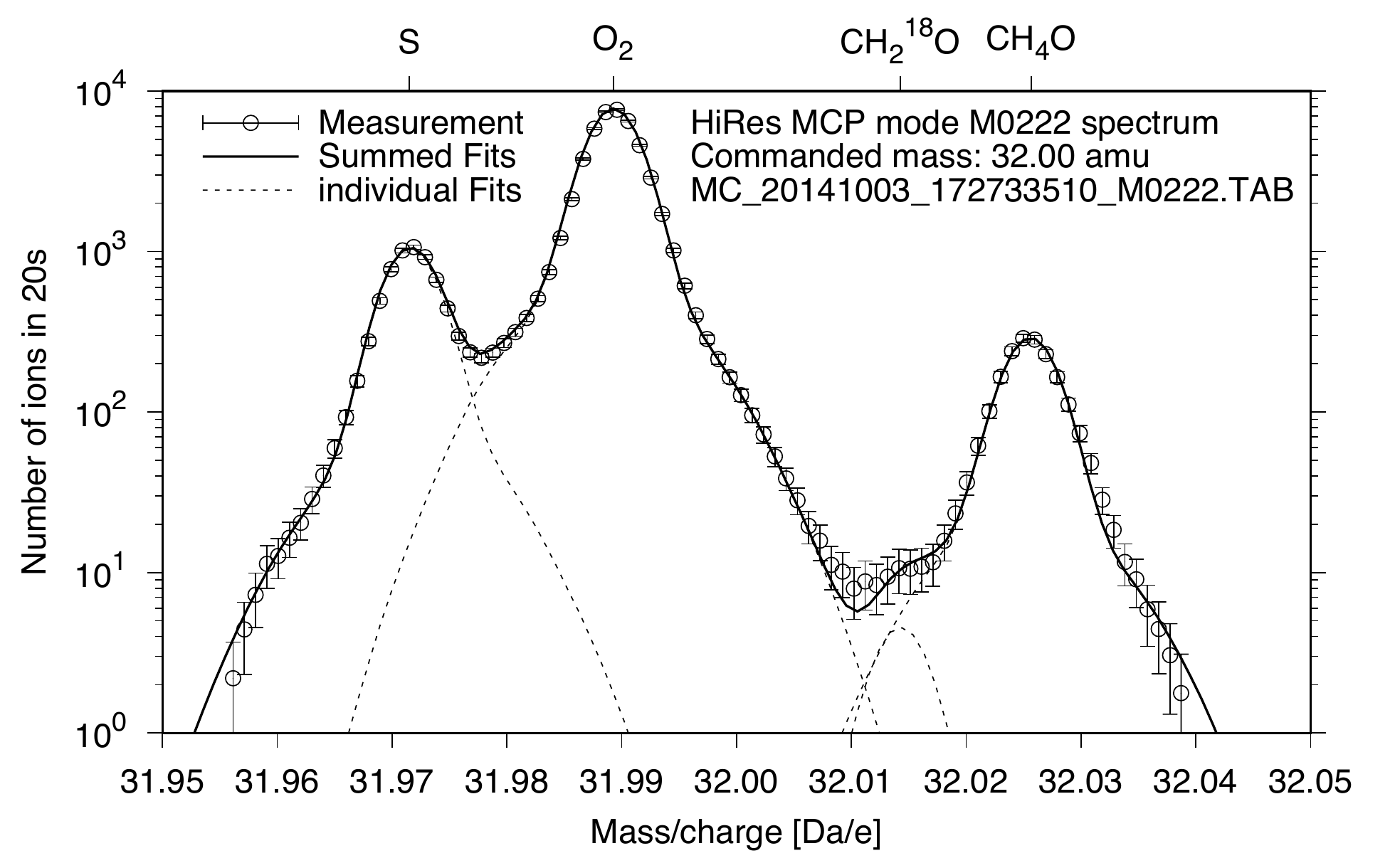}
       \caption{ROSINA DFMS mass spectra \citep{Balsiger2007} of mass/charge 32~u/e with sulfur S (in parts a fragment of S-bearing species), molecular oxygen O$_2$, and methanol CH$_3$OH. Also, H$_2$C$^{18}$O due to formaldehyde and fragmentation of methanol, both with the heavy oxygen isotope, can be found \citep[cf.][]{Altwegg2020b}.}
       \label{fig:MassSpectra}
     \end{figure}

     In 1986 the \textit{Giotto} spacecraft flew by comet 1P/Halley \citep{Reinhard1986}. During the inbound portion of the fast, 68.4~\kms~flyby the Neutral Mass Spectrometer \citep[NMS; ][]{Krankowsky1986} carried out continuous measurements of the neutral gas abundances \citep{Eberhardt1999} until, near closest approach, impacting dust grains led to the failure of the sensor.\newline
     The second comet visited by neutral gas mass spectrometers was 67P \citep{Glassmeier2007}. \textit{Rosetta} carried the Rosetta Orbiter Spectrometer for Ion and Neutral Analysis \citep[ROSINA; ][]{Balsiger2007} Double Focusing Mass Spectrometer DFMS and Reflectron-type Time-Of-Flight (RTOF) on the orbiter. Furthermore, the two lander mass spectrometers Ptolemy \citep{Wright2007} and the COmetary SAmpling and Composition experiment \citep[COSAC;][]{Goesmann2007} were carried by Philae. \textit{Rosetta} carried out an extended 2-years investigation following 67P from beyond 3~au, through perihelion at 1.24~au, and out again past 3~au. It was at times gravitationally bound to the comet and hence typical relative velocities between \textit{Rosetta} and 67P were on the order of 1~m/s, much smaller compared to typical neutral gas velocities on the order of 1~\kms. Fig.~\ref{fig:MassSpectra} shows an example mass spectrum containing parent and fragment species as well as minor isotopologues.
  
     \subsubsection{Instrument capabilities, mass resolution}
     %In neutral gas mass spectrometry atoms and molecules are first ionized, for instance through electron impact. Afterwards the corresponding ions are accelerated, separated by mass, e.g., in electric and/or magnetic fields or by their time-of-flight, and then registered on a detector with suitable amplification through, e.g., micro channel plates (MCP). The signal of the detected ions can then be converted to the density of the corresponding parent species. During the ionization process cometary parent molecules can be shattered leading to a pattern of fragment ions that is characteristic for the parent molecule, e.g., H$_2$O$\rightarrow$OH$^+$, H$_2$O$\rightarrow$O$^+$, etc. Hence, in order to obtain for example the amount of cometary CO, first the fragmentation of CO$_2\rightarrow$CO$^+$ from inside the instrument has to be subtracted.  

     Neutral gas mass spectrometry has some advantages and disadvantages. For instance, the observed atoms and molecules do not require a strong electric dipole moment and hence volatile species such as O$_2$ and N$_2$ can be investigated. Furthermore, being in-situ, Earth's atmosphere will not interfere and there are no optical depth issues requiring reversion to a minor isotopologue with only limited information on the specific isotope ratio (cf. section~\ref{sec:SOC-isotopes}). On the down side, the ability to distinguish isomers is limited and is only possible when the differences associated to the fragmentation occurring during ionization (e.g., H$_2$O$\rightarrow$OH$^+$, H$_2$O$\rightarrow$O$^+$, etc.) are larger than the measurement uncertainties.

     Table \ref{tab:MassSpectrometers} lists the mass spectrometers that have carried out neutral gas measurements at comets. The instruments either employed electric and magnetic fields or time-of-flight to separate ions by their mass-to-charge. In order to resolve ionized atoms and molecules on the same integer mass line, e.g., O$^+$ and CH$_4^+$ at 15.9944 and 16.0308~u/e, respectively, a mass resolution on the order of $m/\Delta m > 1000$ (FWHM) is required. If the resolution does not allow separation of the different contributors to a given mass line , e.g., O$^+$, NH$_2^+$, and CH$_4^+$ on mass/charge 16~u/e, it may still be possible to disentangle their relative proportions: based on the calibrated fragmentation patterns, the contributions to a given mass line, e.g. H$_2$O$\rightarrow$O$^+$ and NH$_3\rightarrow$NH$_2^+$ can be subtracted from the total signal on mass/charge 16~u/e leaving only CH$_4^+$. The signal of CH$_4^+$ can then be related to the abundance of the cometary parent species methane, CH$_4$, in the local coma. This approach quickly becomes very complex in gas mixtures due to the large number of involved fragments and isomers.

     \subsubsection{\label{sec:IMS}Ion mass spectrometry}
     The composition of the neutral gas coma can also be constrained from in-situ plasma measurements \citep[see chapter][]{Beth2022}. However, this approach requires additional modeling of the involved ionization, dissociation, and chemical reactions between neutrals, ions, and electrons. A suite of processes, such as photoionization, photodissociation, ion-neutral reactions and  charge exchange, as well as ion-electron recombination, have to be included.
     
     For instance, \cite{Allen1987} found evidence for the presence of ammonia and methane in comet 1P/Halley based on Giotto ion mass spectrometer {\citep[IMS; ][]{Balsiger1987}} measurements. Furthermore, the D/H ratio in the water of the same comet was derived from  IMS \citep{Balsiger1995} and ion mode measurements of NMS \citep{Eberhardt1987}. At 67P, the total outgassing activity was estimated from the increasing He$^+$/He$^{2+}$ ratio of solar wind helium crossing the gas coma and charge-exchanging with neutrals \citep{SimonWedlund2016,Hansen2016}. The same process, when high charge state solar wind minor ions undergo charge exchange excitation with neutrals, led to the X-ray emission observed at comet C/1996 B2 (Hyakutake) {\citep[][ cf. section~\ref{sec:vis-spectro}]{Haeberli1997}}.

     \begin{table*}
     \caption[]{Neutral gas mass spectrometers operated at comets.}\label{tab:MassSpectrometers} 
     \begin{center} 
%     \begin{tabular}{l|l|c|c|c|l}
     \begin{tabular}{llcccl}
     \hline
        Instrument & Craft & Mass separation & Mass range & Resolution & Reference \\
         &  & technique & [u/e] & $m/\Delta m$ &  \\
        \hline
        NMS  & Giotto & EM-fields & 1-56$^a$ & unit$^b$ & \cite{Krankowsky1986} \\
        ROSINA DFMS & Rosetta & EM-fields & 12 to 180 & 3000 @ 1\% & \cite{Balsiger2007} \\
        ROSINA RTOF & Rosetta & Time of Flight & 1 to $>$300 & $>$500  @ 50\% & \cite{Balsiger2007} \\
        COSAC & Philae & Time of Flight & 1 to 1500 & 300 @ 50\% & \cite{Goesmann2007} \\
        Ptolemy & Philae & Time of Flight & 12 to 150 & unit & \cite{Wright2007} \\
        \hline
     \end{tabular}
     \end{center}
     $^a$1-38~u/e in nominal double focusing mode, $^b$ unit mass resolution in the range 12-56~u/e. \\
     \end{table*}

%--------------------------------- Uncomment the lines below for a figure.
%\begin{center}
%\includegraphics[width=7.5cm]{sample.jpg}
%\caption{Caption of the figure.}
%\label{fig:niceart}
%\end{center}
%\end{figure}
%---------------------------------

%-----------------------------------------------------------------------------------------------------------
%--------------------------------- Nicolas
\section{\textbf{FROM OBSERVATIONS TO ABUNDANCES}}
\label{sec:obs-to-abund}
 For estimating relative abundances in ices that sublimate near the surface of the nucleus, either for the parent molecules directly observed or the assumed parent of a secondary product species, we convert the measured quantities, a line intensity converted into a column density or a local density (in-situ measurements) into production rate $Q$. The first step for spectroscopic observations requires modelling the excitation of considered energy levels of the molecule and radiative transfer. In all cases a knowledge of the spatial distribution of the molecule is needed. Most studies use the Haser model \citep{Haser1957} to describe the density of a sublimating parent molecule. Remote spectroscopic observations are nowadays often 2-D or 3-D data-cubes that combine spectral information at high resolution and 1-D (long slit spectroscopy) to 2-D (data-cubes, interferometric maps) spatial information. This is used to characterize the local density of the molecules and reveal evidence for departure from the basic Haser model: outgassing asymmetry and radial distribution for non-nucleus sources.
%---------------------------------  Nicolas
  \subsection{Deriving production rates}
  \label{sec:prod-rates}
  The processes that lead to excitation of molecular energy levels are multiple: collisions with neutral gas that tend to establish a local thermal equilibrium (LTE) at the local gas temperature, collisions with electrons created in the coma and from the solar wind, and radiative processes. The radiative environment for the inner coma is dominated by the solar radiation in the UV to IR range, the cosmic black body background at 2.73 K in the radio range and dust and nucleus thermal emission in the IR. Self absorption by optically thick lines in the IR to submillimeter range can also play an important role. Generally UV to IR emission will result from fluorescence excitation (absorption followed by rapid emission of photons due to the short lifetime of the electronic and vibrational states excited in this process) while rotational transitions in the radio domain result from spontaneous emission due to slow decay of the rotational levels (lifetimes of $10^2$ to $10^6$ s) populated by collisions or radiative processes, including pumping via vibrational bands. Dissociation products of molecules can also radiate via prompt emission when they are created in an excited electronic state, such as CO Cameron bands in the UV or OH in the IR. All processes have been reviewed in detail in \citet{Bockelee2004} or \citet{Bodewits2022}.
  State-of-the-art models adapted to comets need to take into account non-steady-state calculation specific for cometary comae in which the gas is flowing away from the nucleus and the density seen by a given molecule decreases with time \citep{Marschall2022}.

  Several parameters, such as the gas temperature and the expansion velocity can be inferred directly from observations: rotational temperatures measured in the IR or the radio from series of ro-vibrational or rotational lines and analysis of the line profiles resolved in the radio where width is directly related to the gas expansion velocity.

%--------------------------------- Neil
%  \subsection{Investigating release mechanisms through molecular spatial distributions in the coma}
  \subsection{Spatial distribution of molecules in the coma}
  \label{sec:spatial-distri}
  
  While imaging cometary dust is often the easiest way to study coma morphology it provides an incomplete picture as gas and dust spatial distributions are generally distinct. Determining and comparing the spatial distributions of volatile gases in the coma of comets can provide information both about their sources (direct sublimation from ice and/or extended coma sources) and about how ices are associated or separated in the nucleus. Both spacecraft and ground-based studies have provided abundant evidence that some molecules are principally released directly from their sublimating ices while others are released from the dissociation of other molecules in the coma; however, the contribution from theses sources for a given molecule often differs from comet to comet and the progenitor species of secondary products are often difficult to identify. When no spatial information is available, it is difficult to assess whether some molecules are primarily parents or products. This can affect the accuracy of derived production rates if the parentage is wrongly assumed.
 %This suggests that abundances of these unknown progenitor species of extended sources are variable within the comet population just like their measured products. 
 Spatial studies of comets span a large range in scale and resolution, from the highest resolution studies of the inner coma by in-situ spacecraft to ground-based studies with various aperture sizes and different comet geocentric distances. The addition of spacecraft data and increasing ground-based capabilities has allowed more detailed studies of molecular spatial distributions in the coma.
  
  \subsubsection{Narrowband imaging and interferometric maps}
  Studies of cometary coma features have been conducted for many years at optical wavelengths and extending from the near-UV to the near-IR. Modern telescopes and imagers allow extensive spatial coverage in the coma while also providing sub-arcsecond spatial resolution over large format CCD arrays. Broadband continuum filters have provided detailed information on the dust structure in the comae of comets. Using targeted narrowband filters and continuum removal techniques or IFUs has allowed the isolation of coma gas structure, which is needed to provide complete information on coma morphology because coma gas and dust distributions are generally distinct. Gas phase species sampled at these wavelengths are product species or ions (e.g., OH, NH, CN, C$_2$, C$_3$, CO$^+$, H$_2$O$^+$) \citep[e.g.,][]{Schleicher2004}. For certain bright comets the consistency of assumed parent scalelengths for these species can be tested, which is important as many of these values are poorly constrained. With increasing capabilities at radio and IR wavelengths, coordinated observations of the spatial distributions of parents and secondary products can be performed providing a more direct comparison of parent-product associations in individual comets.
     
\begin{figure}[ht!]
  \begin{center}
  \includegraphics[angle=0,width=0.78\columnwidth]{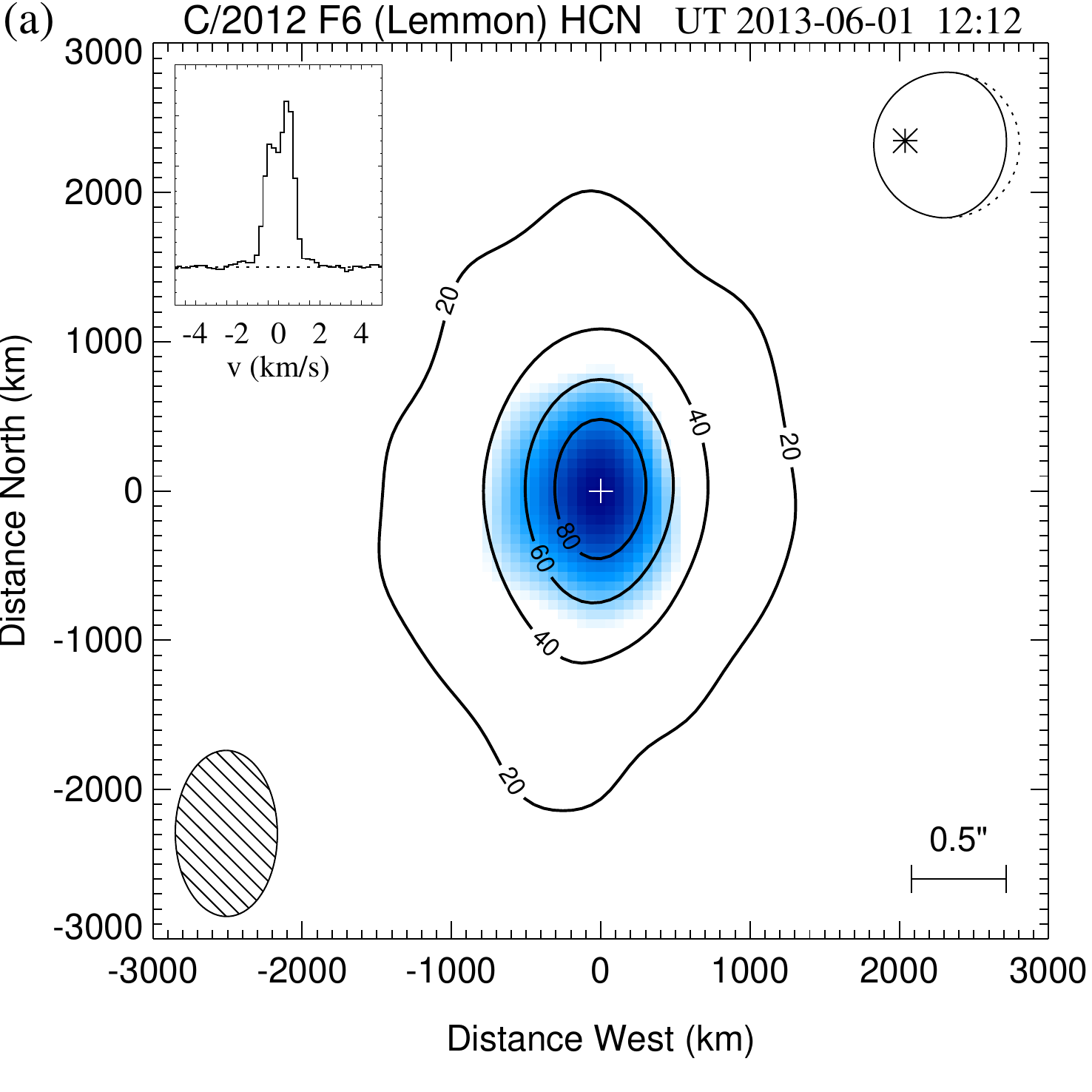}
  \includegraphics[angle=0,width=0.78\columnwidth]{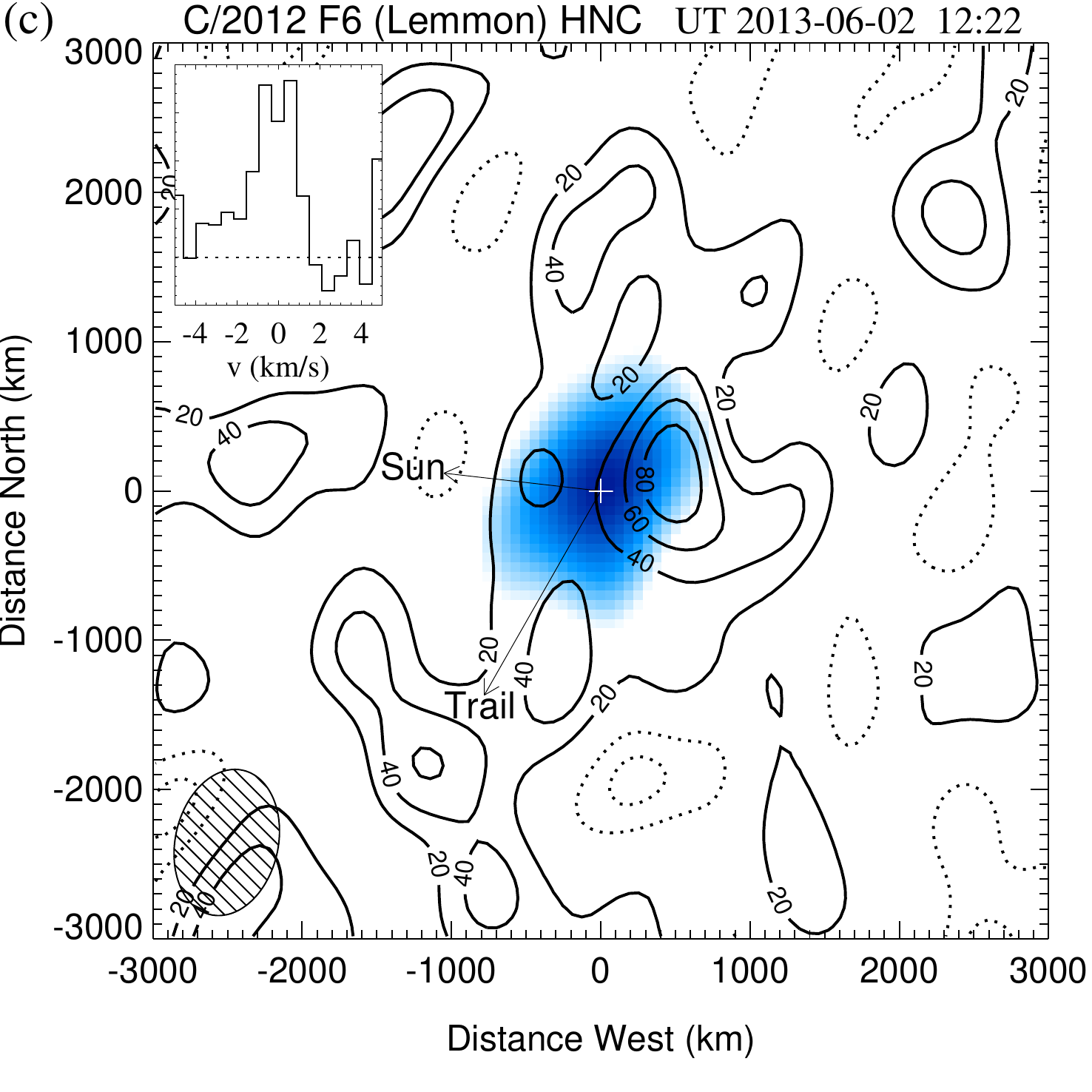}
  \includegraphics[angle=0,width=0.78\columnwidth]{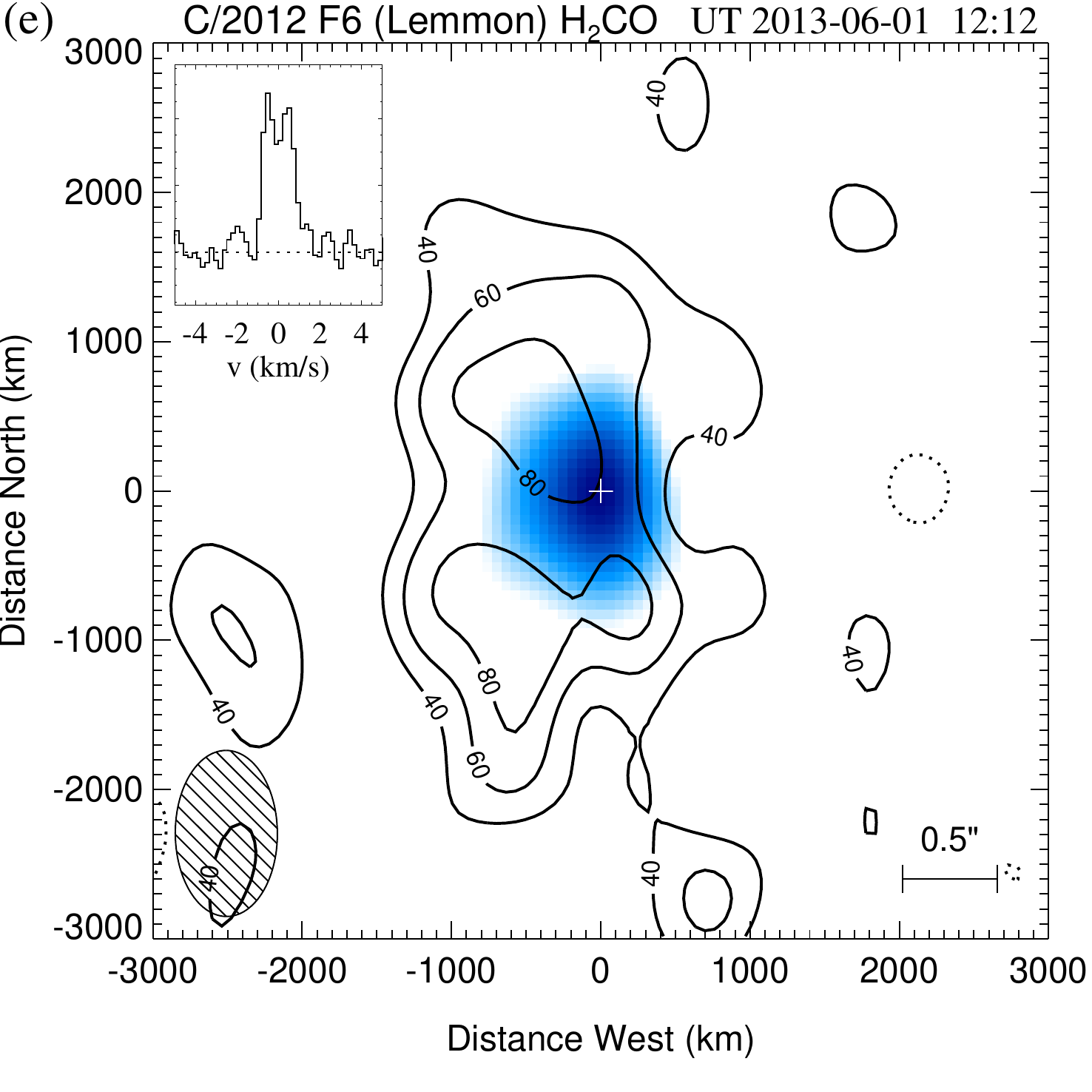}\vspace{-0.4cm}
  \caption{ALMA interferometric contour maps of (a) HCN, (c) HNC, and (e) H$_2$CO in C/2012 F6 (Lemmon). The signal distributions of HNC and H$_2$CO are much less peaked due to their production in the coma in contrast to HCN coming mostly from the nucleus \citep{Cordiner2014}. In the lower left corners the hatched ellipse shows the shape and size of the synthetic beam. The spectrum at he central point is shown in the upper left and the continuum flux is displayed in blue.}
   \label{fig:interfero-map}
  \end{center}
\end{figure}

  Interferometric maps at radio wavelengths provide information on the spatial and velocity distributions of gas molecules and dust. Modern heterodyne receivers allow the determination of high-resolution line-of-sight velocities which enable three dimensional spatial structures of measured molecules to be obtained \citep[e.g.,][]{Qi2015}. The importance of radio interferometry in obtaining spatial maps for molecules released from different sources has been demonstrated \citep[e.g.,][]{Boissier2007, Cordiner2014, Cordiner2017, Roth2021}. Maps for molecular species show dramatic differences in their spatial distributions suggesting heterogeneity in the way ices are stored and associated within the nucleus. Under favorable circumstances radio interferometery with facilities such as NOEMA, ALMA or the SMA have determined the spatial distribution of parent and product species such as H$_2$CO, HCN, HNC (e.g. Fig~\ref{fig:interfero-map}), CH$_3$OH, CO, or CS. Future capabilities with facilities such as the ngVLA will allow higher sensitivities and cover longer wavelengths allowing NH$_3$ and OH to be mapped in the comae of comets at high spatial resolution \citep[e.g.,][]{Cordiner2018}.

\clearpage

  \subsubsection{Infrared spatial profiles}
\begin{figure*}[ht!]
     \begin{center}
         \includegraphics[angle=0,width=0.6\textwidth]{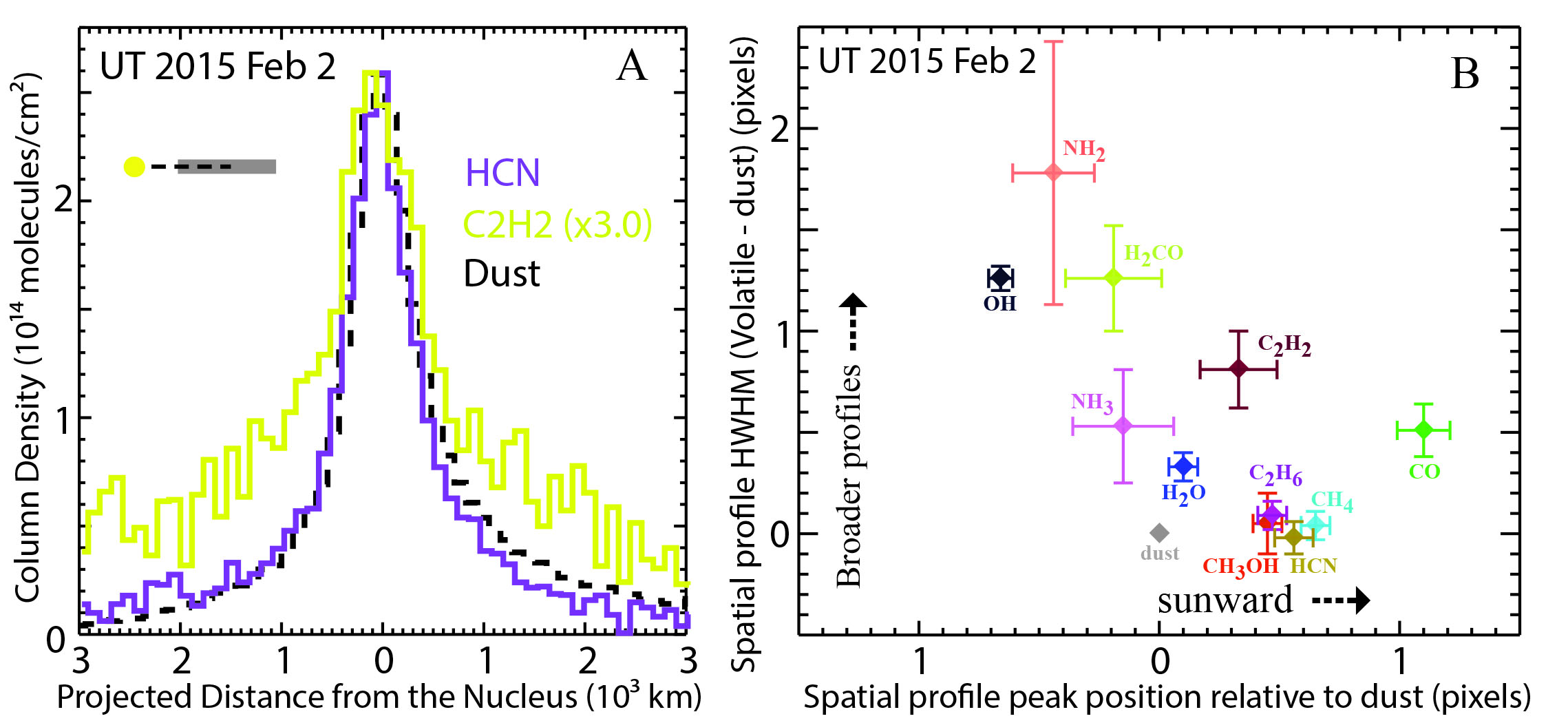}
         \caption{(A) Spatial profiles of C$_2$H$_2$, HCN, and dust in C/2014 Q2 (Lovejoy) showing evidence for a C$_2$H$_2$ extended source due to excess flux on the profile wings compared to HCN. (B) Spatial profile properties for volatiles and dust in C/2014 Q2 (Lovejoy) showing the distinct characteristics of spatial profiles based on profile peak position and width \citep{DelloRusso2021}.} %This suggests the following nucleus ice associations and extended sources: (1) CH$_3$OH, C$_2$H$_6$, CH$_4$, and HCN are from direct release of associated ices in the nucleus. (2) H$_2$O ice may also be associated with the above species in the nucleus, but its anti-sunward spatial profile position relative to these species is likely due to an additional contribution from icy grains in the coma. (3) H$_2$CO, NH$_3$, and C$_2$H$_2$ spatial profiles suggest significant contributions from extended sources in the coma. (4) CO ice in the nucleus is not associated with other measured volatiles.}
         \label{fig:IR-spatial-profiles}
         \end{center}
     \end{figure*}

     The long-slit capabilities of IR ground-based observing enables the determination and comparison of spatial distributions of molecules in the inner coma. These comparisons address whether species observed simultaneously relate to a common or different outgassing sources and suggest potential nucleus associations. IR spatial analysis of comets have routinely determined the spatial distributions of H$_2$O, C$_2$H$_6$, CH$_3$OH, and HCN in many comets observed to date. Spatial distributions for CO and CH$_4$ are also readily obtained when the comet’s geocentric velocity is sufficient to Doppler-shift cometary emissions from their atmospheric counterparts. For some of the brightest comets, spatial distributions have also been obtained for H$_2$CO, C$_2$H$_2$, NH$_3$, and OCS. Previous ground-based IR studies have suggested separate polar and apolar ice phases in the nucleus of some comets \citep[e.g., ][]{Mumma2011}; however, this association by ice polarity is not seen in some cases \citep[e.g.,][]{DelloRusso2021}. IR spatial analysis has revealed surprising evidence for extended sources for some species (Fig.~\ref{fig:IR-spatial-profiles}A), as well as showing possible volatile associations of ices in the nucleus (Fig.~\ref{fig:IR-spatial-profiles}B) \citep[e.g., ][]{DelloRusso2021}. Technological advances in ground-based IR instrumentation has allowed spatial analysis to be done in more detail and on a rapidly increasing number of comets. This work is beginning to address which spatial properties are specific to individual comets and which are more general properties of all comets.
     
    \subsubsection{In-situ radial profiles}
    Spacecraft missions have performed the highest resolution studies of how volatiles are released into the inner comae of comets. In contrast to remote sensing observations which derive spatial properties over the global coma, in-situ studies can reveal volatile and dust release from specific areas on the nucleus. The \textit{EPOXI} flyby revealed a two-lobed nucleus for 103P/Hartley 2 with the bulk of the activity emanating from the smaller lobe \citep{AHearn2011}. Release from the small lobe was dominated by the spectral signature of CO$_2$ gas with entrained H$_2$O ice chunks, whereas H$_2$O vapor primarily sublimated from a different region along the waist between the lobes \citep{AHearn2011}. \textit{EPOXI} spatial studies of volatiles were based on low-resolution IR spectroscopic measurements during a flyby so were a brief snapshot of the volatile coma structure of 103P and mainly sensitive to H$_2$O and CO$_2$, the strongest IR molecular bands. \textit{Rosetta} studies of 67P enabled tracking of long-term molecular associations in the coma of many species with multiple instruments. In contrast to 103P, \textit{Rosetta} observed dominant H$_2$O outgassing directly from ices in the nucleus of 67P and not from coma icy grains \citep[e.g.,][]{Luspay2015}, suggesting a more direct association between coma gas and nucleus ice distributions.
   
    Processes like sublimating icy grains and photochemistry are imprinted in the gas densities along radial profiles from the nucleus. Marked deviations from $1/r^2$ %(cf. Eq.~\ref{eq:parentdensity}) 
    (assuming constant expansion velocity) indicate contribution from a distributed source, either through sublimation from grains in the coma or dissociation of larger molecules. At 67P, \cite{Altwegg2016} reported that the total gas density dropped with cometocentric distance as expected but glycine was most likely released from water ice on dust particles \citep{Hadraoui2019}. 
    Based on \textit{Giotto} NMS results at 1P/Halley, substantial distributed sources were reported for H$_2$CO, CO \citep{Meier1993,Eberhardt1999}, and possibly H$_2$S, while CH$_3$OH behaved differently \citep{Eberhardt1994}. During the flyby, the \textit{Giotto} spacecraft covered a wide range in phase angle and cometocentric distances \citep{Reinhard1986} and hence NMS measured gas released during different times from distinct regions on the nucleus with varying illumination. Therefore, temporally and spatially inhomogeneous outgassing could result in similar spatial profiles \citep[cf. Fig. 3 of][]{Rubin2009}.

   % Martin: Neil, I incorporated this paragraph 6.2.3: Volatile production rates measured over a large timescale suggested two distinct ice phases in 67P, associated with either H$_2$O or CO$_2$ release \citep{Hassig2015, Fink2016, Gasc2017}. CO$_2$ appeared correlated with C$_2$H$_6$, CO, H$_2$S, and CH$_4$ \citep{Luspay2015, Hassig2015, Gasc2017}, whereas H$_2$O appeared correlated with CH$_3$OH, NH$_3$21P Giacobini-Zinner, and O$_2$ (\cite{Luspay2015, Gasc2017}). Some relationships varied with time; for example, HCN was sometimes correlated with CO$_2$ (\cite{Gasc2017}) and at other times with H$_2$O (\cite{Luspay2015}). As the volatile coma structure in comets is explored by spacecraft it becomes important to connect these high-resolution spatial results with the global coma results obtained for a large number of comets from remote sensing observations.
%-----------------------------------------------------------------------------------------------------------
\section{\textbf{DRIVERS OF COMET ACTIVITY}}
\label{sec:hypervolatiles}

Water is the most abundant species in cometary ices and the main driver of activity for comets within 2-3 au from the Sun. At larger heliocentric distances, the water sublimation rate decreases and water alone is not sufficient to sustain cometary activity. More volatile ices, like CO and CO$_2$, are believed to drive the distant activity of comets.
%Review of the three main molecules than can drive the activity of a comet: when/where they dominate the coma and how it is measured.

%--------------------------------- Nicolas
  \subsection{CO and distant activity}\label{sec:CO}
  CO was first detected in comets close to the Sun in the UV via resonant fluorescence in the Fourth Positive Group ($A^1\Pi-X^1\Sigma^+$) with sounding rockets and satellites \citep[][and reference therein]{Bockelee2004}.
  Since its first direct cometary detection in 29P/Schwassmann-Wachmann 1 in 1993 by \citet{Senay1994} and \citet{Crovisier1995}, CO has been detected in the radio and in the IR in several comets at large heliocentric distances, and up to 14~au in comet C/1995~O1 (Hale-Bopp) \citep{Biver2002}. CO is one of the most volatile molecules, along with N$_2$, detected in comets. Due to its volatility, it can sublimate as far out as the Kuiper Belt \citep[$T_{sub}\sim$25~K,][]{Fray2009}. 
  Hence its signature has been searched for in many icy objects including Trans-Neptunian Objects (TNOs) and Centaurs \citep{Bockelee2001,Jewitt2008,Womack2017}. CO sublimation might have been responsible for transient or outburst activity in Centaurs like 95P/Chiron \citep{Womack2017} or 174P/Echeclus \citep{Wierzchos2017}, but this has not been confirmed due to the difficulty to detect CO in these distant objects.
  Nevertheless, in several comets observed beyond 2--3~au (Table~\ref{tab-cometco}), a substantial outgassing of CO has been detected and often dominates the activity of the comet ($Q_{CO}/Q_{H_2O}> 1$). In those comets, CO sublimation can lift dust into the coma even at distances more than 20~au from the Sun. Radio lines also often show a strong blueshift (e.g., the bottom spectrum of Fig.~\ref{fig-linewidth}), suggesting that CO sublimation is enhanced at the warmer subsolar point.
  This does not necessarily mean that the ices of these comets are dominated by CO: comets like C/1995~O1 (Hale-Bopp) or C/2009~P1 (Garradd), when they were closer to the Sun (1.5~au or less) had comae dominated by water vapour with CO abundances around 5-25\%. In fact the abundance of CO relative to water in comets varies by at least two orders of magnitude (0.3 to 35\%, Table~\ref{tab:abundances}, Figs~\ref{fig-histomolec} and \ref{fig-histo1}) in comets observed within 1.5~au from the Sun. %(Generally the IR and UV techniques are more sensitive close to the Sun than radio).
  JFCs \citep[e.g., 67P,][]{Lauter2020} tend to be depleted in CO with abundances generally not exceeding 3\% relative to water. Comets with low CO abundances may have most of their CO trapped in water ice, so even beyond the sublimation of water\add{,} CO generally doesn't control activity in JFCs. %On the other hand, CO$_2$, which is less volatile than CO, is abundant enough to drive activity in some JFCs.
  Finally\add{,} some comets like C/2016~R2 \citep[][ and Sect.~\ref{Sec:outliers}]{Biver2018,McKay2019} seem to be dominated by CO outgassing even at a distance where H$_2$O outgassing becomes significant in most comets. 
  %One might wonder if these "blue" comets (their visible spectrum is dominated by the blue emission of CO$^+$ lines) do not belong to a category of comet with a radically different composition, due to, e.g., a different location of formation in the early solar system.

\begin{table*}
\caption[]{Comets dominated by CO outgassing}\label{tab-cometco}
%\begin{center}
\begin{tabular}{lccrcl}
\hline
Comet    & Date & $r_h$   & $Q_{\rm CO}$         & $Q_{\rm CO}/Q_{\rm H_2O}$ & Ref.\\  
         &      & (au)    & (molec.s$^{-1}$) &                 &     \\
\hline
C/1995~O1 (Hale-Bopp)$^a$ & May  1996 & 4.6 &  $6\times10^{28}$ &  4.5    & \citep{Biver2002} \\
     & Sep. 1996 & 3.3 & $15\times10^{28}$ & $\sim1$ & \citep{Biver2002} \\
     & Aug. 1997 & 2.3 & $24\times10^{28}$ & $\sim1$ & \citep{Biver2002} \\
     & Dec. 1997 & 3.9 & $10\times10^{28}$ &  3.8    & \citep{Biver2002} \\
C/1997~J2 (Meunier-Dupouy) & Mar. 1998 & 3.1 &  $0.39\times10^{28}$ &  ($^b$) & \citep{Biver2018} \\
29P/Schwassmman-Wachmann 1 & May  2010 & 6.2 &  $4\times10^{28}$ & $\sim10$ & \citep{Bockelee2010b} \\
C/2006~W3 (Christensen)   & Nov. 2009 & 3.3 &  $3\times10^{28}$ & $>2.2$  & \citep{Bockelee2010a} \\
                          & Aug. 2010 & 4.9 &  $1.2\times10^{28}$ & $\sim5$  & \citep{Bonev2017} \\
                          &&&&& +\citep{deValborro2014} \\
C/2009~P1 (Garradd)       & Apr. 2012 & 2.1 &  $3\times10^{28}$ & $0.6$   & \citep{Feaga2014} \\
C/2016~R2 (PanSTARRS)     & Feb. 2018 & 2.8 &  $8\times10^{28}$ & $\sim300$ & \citep{McKay2019} \\
C/2017~K2 (PanSTARRS)     & Feb. 2021 & 6.7 &  $0.16\times10^{28}$ & ($^b$) & \citep{Yang2021} \\
174P/Echeclus             & Jun. 2016 & 6.1 & $0.08\times10^{28}$ & ($^b$) & \citep{Wierzchos2017} \\
\hline
\end{tabular}
$^a$: for comet Hale-Bopp, which activity was monitored from 7~au pre-perihelion to 14~au post-perihelion we provide the CO/H$_2$O ratio for the most distant detections of H$_2$O and the time when it crossed the $Q_{\rm CO} = Q_{\rm H_2O}$. Water outgassing dominated between September 1996 and August 1997. $^b$: $Q_{\rm H_2O}$ not measured, likely low.

%\end{center}
\end{table*}

%--------------------------------- Martin
%  \subsection{\label{sec:CO2-driver} CO$_2$ as driver of activity in 67P}
\subsection{CO$_2$ as driver of activity}\label{sec:CO2-driver}
The CO$_2$ abundance is very difficult to estimate from ground-based observations and most measurements of the CO$_2$ abundance in comets come from infrared space observatories (ISO, Spitzer, AKARI) or spacecraft (Deep Impact, EPOXI, Rosetta). Indirect measurements can be made from observation of the CO Cameron band in the UV and forbidden oxygen lines in the optical.
The AKARI survey \citep{Ootsubo2012} has revealed that in most cases CO$_2$ is a major constituent of cometary atmospheres, with abundances typically ranging from 5 to 30\% with no clear differences based on dynamical origin. Beyond 2.5 au from the Sun the CO$_2$ abundance relative to water increases and CO$_2$ can become the major constituent of the coma, driving the activity of the comet.
Even closer to the Sun, the outgassing of CO$_2$ can dominate the activity of the nucleus such as in the case of (3552) Don Quixote \citep[Spitzer, ][]{Mommert2020} or 103P/Hartley \citep[EPOXI,][]{AHearn2011}. 
%However, a Spitzer survey of comets \citet{Reach2013} find a group of comets depleted in CO$_2$ (and CO as they cannot distinguish contribution from each species) with CO$_2$/H$_2$O $<$1\%.  
%% I removed this comment as it seems that Reach2013 values are biased by wrongly extrapolated H2O references and results not compatible with Akari for the same comet 

  {\sl CO$_2$ behavior in 67P:}
  The intricate pattern of activity in 67P is a result of the peculiar shape of its nucleus and associated illumination together with the obliquity of its rotation axis of 52$^o$ \citep{Sierks2015} and the different volatility of the involved gases \citep[][]{Fray2009}. The gas activity induced a decrease in the rotation period (from 12.4 to 12.0 hr) as the comet passed perihelion \citep{Kramer2019,Keller2015b}. Furthermore, the transport and re-deposition of (icy) grains around the nucleus \citep{Keller2017} leads to very different surface morphologies \citep{ElMarry2015} leaving their imprint on the structures of the gas coma.
  
\begin{figure*}[ht!]
    \centering
    \includegraphics[width=0.6\textwidth]{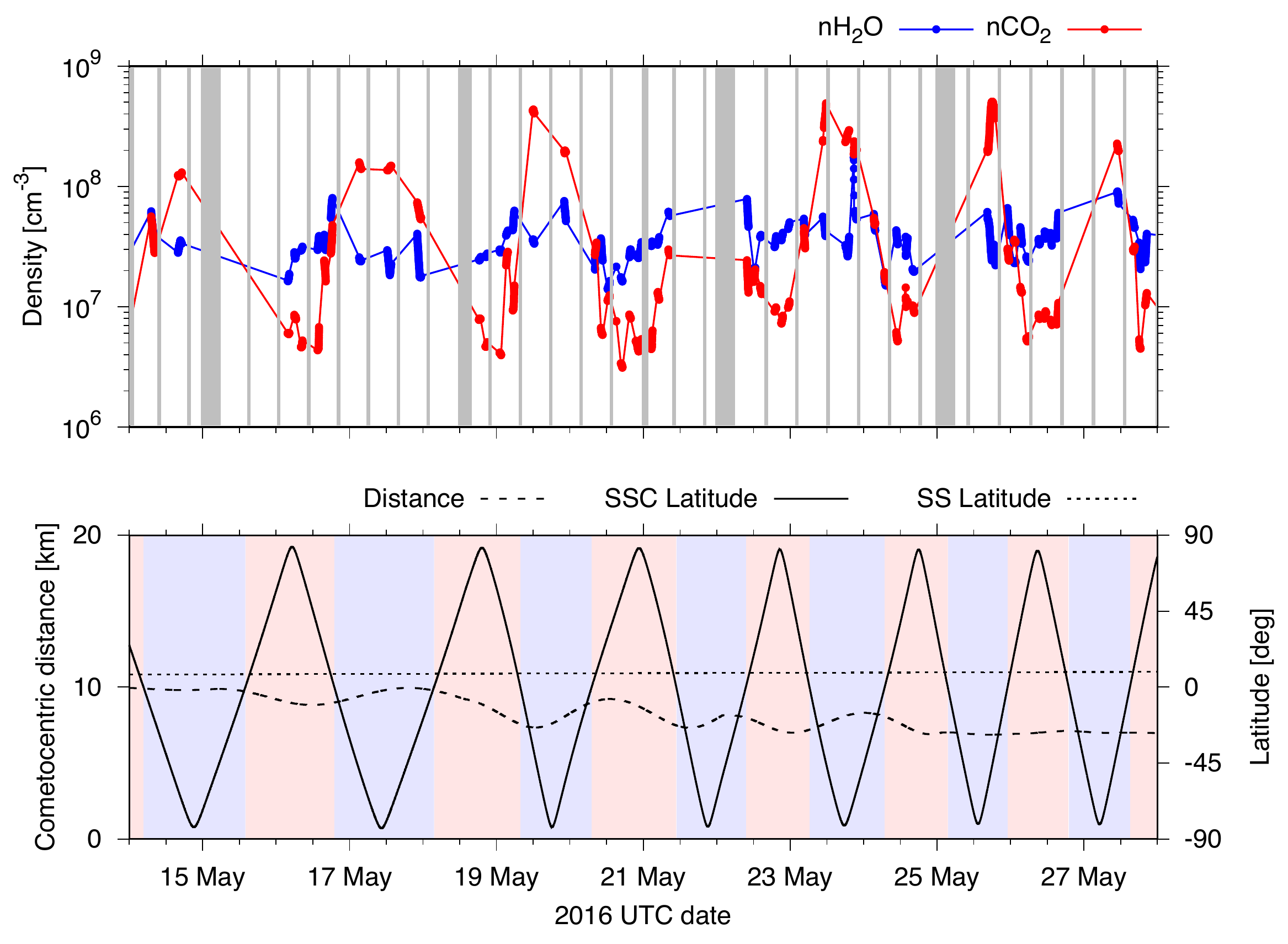}
    \caption{Top panel: H$_2$O and CO$_2$ densities at 67P measured in May 2016 (post outbound equinox) at $\sim$3~au by ROSINA DFMS \citep[cf. ][]{Gasc2017} during close terminator orbits. Grey areas mark times of thurster operations when the instrument was off, other gaps due to different measurement mode. Bottom panel: cometocentric distance ($<$10~km), sub-spacecraft latitude (red: north, summer; blue: south, winter) and sub-solar latitude.}
    \label{fig:Heterogeneity}
  \end{figure*}

  Fig.~\ref{fig:Heterogeneity} shows a time series of H$_2$O and CO$_2$ measurements late in the \textit{Rosetta} mission at a heliocentric distance of $\sim$3~au, just after the outbound equinox. While the spacecraft orbited the comet at close distances in the terminator plane, it passed over regions in the south where CO$_2$ was the main driver of activity while H$_2$O dominated above the northern hemisphere. At that time the sub-solar latitude climbed from southern to northern latitudes, however, the south did remain the more active hemisphere for the remaining part of the mission out to almost 4~au despite the sub-solar latitude increasing again above 20~degrees northern. CO$_2$ remained the dominant molecule and a similar picture was obtained on a global scale from studies of the H$_2$O and CO$_2$ gas production rates of 67P at this distance \citep{Combi2020,Lauter2020}. The same behavior was also observed very early on in the mission, beyond 3~au on the inbound path of the comet. \cite{Hassig2015} reported an elevated CO$_2$/H$_2$O ratio above the the southern hemisphere from which pristine ices are sublimating due to the substantial erosion rates \citep{Keller2015a} observed during the intense but short summer months \citep{Capria2017}. Local enhancements in CO$_2$/H$_2$O ratio have also been identified by remote sensing observations of the [O~I] emission  \citep{Bodewits2016}.
  
% 67P may be the best studied comet in this respect, however, similar behavior was revealed at comet 103P/Hartley 2 \citep{AHearn2011} and 9P/Tempel 1 \citep{Feaga2007} as distinct regions of H$_2$O and CO$_2$ activity were observed and in the latter case a distributed source of icy grains further modified the gas coma \citep{Fougere2013}.
 
%--------------------------------- Neil
  \subsection{Monitoring of H$_2$O production}\label{sec:H2O}
  
  H$_2$O is the dominant volatile ice in comets and as such determining its production rate provides a measure of the overall comet volatile productivity when the comet is within about 2 - 3 au from the Sun. H$_2$O is also the molecule to which the abundances of all other volatile species are generally compared. Thus, determining H$_2$O production rates and coma spatial distribution is an important goal in overall investigations of comet composition. H$_2$O production in comets can be directly measured or inferred through its secondary product species by several techniques over a range of wavelengths from both ground-based and space-based observatories.
  
     \subsubsection{Direct observations of H$_2$O}
     H$_2$O has strong IR vibrational bands near 2.7 and 6.3-$\mu$m associated with the O-H stretching and bending modes respectively that can be easily detected from remote or in-situ space-based observations \citep[e.g., ][]{Weaver1987, Crovisier1997, AHearn2011, Bockelee2015b}. H$_2$O can also be detected from space through rotational line transitions at radio wavelengths \citep[e.g.,][]{Neufeld2000, Lecacheux2003, Hartogh2011}. Because of extinction from atmospheric H$_2$O, IR fundamental bands and rotational lines are generally inaccessible in comets from ground-based observing platforms. However, ground-based IR observations have routinely detected H$_2$O in comets through the years via non-resonance fluorescence emissions that are not opaque in the Earth’s atmosphere \citep[e.g., ][]{Mumma1996, DelloRusso2000, Villanueva2012, Faggi2016}. The spectral coverage of modern ground-based IR instruments allows the simultaneous sampling of H$_2$O with many other volatile species providing a direct measure of relative volatile abundances and comparison of coma spatial distributions that eliminate uncertainties due to temporal variability in volatile release and coma dynamics.
     
     \subsubsection{Tracers of H$_2$O: OH, H, OI}
     Because H$_2$O is by far the dominant source of OH, H and O in the coma of comets close to the Sun, H$_2$O production rates can be inferred from measurements of these species, as branching ratios of H$_2$O to these secondary products are generally well known. Observations of OH at radio wavelengths allows a determination of H$_2$O production rates, gas outflow velocity and asymmetries \citep[e.g., ][]{Crovisier2013}. OH can also be detected through prompt emission at IR wavelengths and has been used as a tracer for H$_2$O production in comets \citep[e.g., ][]{Bonev2006}. Optical and near-UV wavelengths enable the detection of H, O, and OH from both ground- and space-based observatories  \citep[e.g.,][]{Combi2019, McKay2013, Opitom2015}. Because of the sensitivity of radio, optical, and near-UV techniques for observing H$_2$O product species, these have been the main methods over the years for long-term monitoring of volatile production in comets within an apparition. IR observatories and instrumentation now also have the sensitivity for long-term monitoring of H$_2$O production in comets, but are generally limited by telescope availability for such studies.
     
     \subsubsection{Water from icy grains}
     H$_2$O is characterized by its low volatility compared to other measured coma species, which has implications for its outgassing behavior. First, H$_2$O sublimation is generally not fully activated in a comet until it is between about 2 and 3~au from the Sun. Second, while H$_2$O sublimates into the coma directly from nucleus ices, H$_2$O icy grains can also be dragged into the coma by more volatile gases and survive as ice in the coma longer than other measured volatiles. The prevalence of H$_2$O icy grains in the coma may depend on the abundances of more volatile species such as CO$_2$ or CO that can also drive activity.  For example, the correlation of CO$_2$ gas and H$_2$O ice in the inner coma of 103P/Hartley 2 suggests that areas of gas sublimation rich in CO$_2$ may have dragged H$_2$O ice into the coma from below the nucleus surface \citep{AHearn2011}. From ground-based observatories, the presence of icy grains in the coma can often be inferred by the spatial distribution of H$_2$O molecules in the global coma compared to other volatiles. Because H$_2$O icy grains can survive longer in the coma than other volatiles, care must be taken when (1) determining abundances of volatiles relative to H$_2$O especially when the aperture size is small when projected on the comet, and (2) comparing H$_2$O production rates determined by techniques with significantly different aperture sizes.
     
    \subsubsection{Discrepancies in derived water production rates} 
    Measurements of water abundances from different tracers are not always consistent, even for measurements close in time. Water production rates measured at 2~au pre-perihelion in comet C/2009 P1 (Garradd) ranged from about $8\times 10^{28}$ molecules/s from high-resolution IR observation of H$_2$O with VLT/CRIRES \citep{Paganini2012} to $2-2.9\times 10^{29}$ molecules/s from  H$_2$O observations with the Herschel HIFI instrument, OH observations with UVOT/Swift in the near-UV \citep{Bockelee2012,Bodewits2014}, and from H observation with SOHO/SWAN \citep{Combi2013} on the largest spatial scale. Several factors could play a role in the discrepancies between production rates measured with different techniques. First, instruments used for measurements at different wavelengths and from ground- and space-based observatories have different fields of view (FoV). For example, the SWAN instrument onboard SOHO, used to derived water production rates from HI emission, has a FoV of  $1\degree \times 1 \degree$. On the other end, near-IR spectrographs (such as VLT/CRIRES or Keck/NIRSPEC) tend to have small slits of the order of a few arcseconds. %In some comets, a significant percentage of icy grain sources could sublimate outside the FoV of instruments with limited spatial coverage, explaining why for some comets variable H$_2$O production rates are measured using different techniques. 
    Second, the use of different coma density models \citep[e.g. Haser model,][]{Haser1957} versus the vectorial model \citep{Festou1981b} can also lead to differences in the derived production rates \citep{Cochran1993}. Finally, model parameter assumptions, such as scalelengths or gas velocities can also impact derived production rates \citep{Cochran1993,Fink2009}.

%-----------------------------------------------------------------------------------------------------------
%--------------------------------- Martin
\section{\textbf{VOLATILE COMPOSITION OF THE COMA}}
\label{sec:composition}
\begin{figure}[ht!]
 %\begin{center}
 \includegraphics[angle=270,width=\columnwidth]{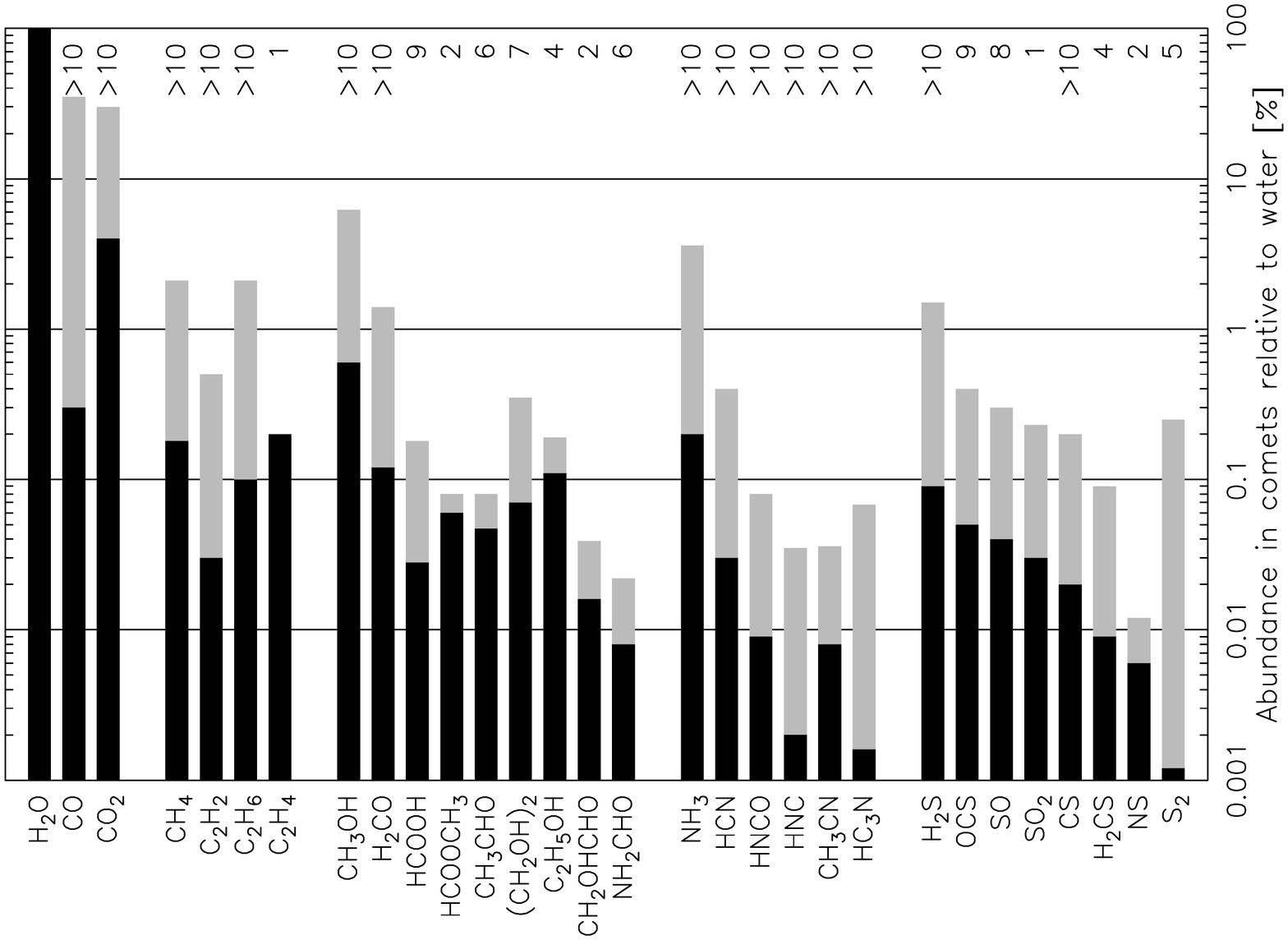}%\vspace{-2cm}
 \caption{The range of abundances relative to water for the molecules observed remotely in cometary comae. The number of comets in which the molecule has been detected is indicated to the right \citep[updated from ][]{Bockelee2017}.}
 \label{fig-histomolec}
 %\end{center}
 \end{figure}

Around 30 molecules have been identified remotely in cometary atmospheres with an abundance relative to water often varying by one order of magnitude. Fig.~\ref{fig-histomolec} shows
the range of abundances measured and the number of comets in which they have been detected. The \textit{Rosetta} mission has nearly doubled this number of molecules. The present status of abundances measured in comets and in 67P is summarized in Table~\ref{tab:abundances} and detailed hereafter.
%I have added this introduction and moved figure 1 to here, Nicolas.

\begin{table*}
\caption[]{Molecular abundances in cometary atmospheres}\label{tab:abundances}\renewcommand{\arraystretch}{0.82}
\vspace{-0.2cm}
\begin{center}
\begin{tabular}{llllll}
\hline
Molecule & Name &\multicolumn{3}{c}{Abundance relative to water in \%} \\  
         &                      & from radio  & from IR  & in-situ in 67P$^a$ \\
\hline
CO           & carbon monoxide & $<1.23$--35   & 0.3--26     &  0.3--3$^b$ \\
CO$_2$       & carbon dioxide  &     -         & 4--30       &  7.0$^b$   \\  
\hline
CH$_4$      & methane          &   -           & 0.15--2.7 &  0.4$^b$    \\ 
C$_2$H$_6$  & ethane           &   -           & 0.1--2.7 &  0.8$^b$    \\ 
C$_2$H$_2$  & acetylene        &   -           & 0.03--0.37 &   -     \\ 
C$_2$H$_4$  & ethylene          &   -           & 0.2 &        \\ 
C$_3$H$_8$  & propane          &   -           &  -         & $0.018\pm0.004$     \\ 
C$_6$H$_6$  & benzene          &   -           &  -         & $0.00069\pm0.00014$     \\ 
C$_7$H$_8$  & toluene          &   -           &  -         & $0.0062\pm0.0012$     \\ 
\hline
CH$_3$OH     & methanol        & 0.7--6.1      & $<0.13$--4.3  & 0.5--1.5$^b$   \\
H$_2$CO      & formaldehyde    & 0.13--1.4$^d$ & $<0.02$--1.1 & 0.5 \\
HCOOH        & formic acid     & 0.03--0.18    &  -          & 0.013 \\ %m=46
CH$_3$CHO    & acetaldehyde    & 0.05--0.08    &  -          & \vline~0.047 \\ \smallskip %m=44
c-C$_2$H$_4$O  & ethylene oxide & $<0.006$     &  -          & \vline       \\ 

(CH$_2$OH)$_2$& ethylene glycol& 0.07--0.35    &  -          & \vline~ 0.011 \\ \smallskip %m=62
CH$_3$OCH$_2$OH & methoxymethanol & $<9$       &  -          & \vline  \\ 

HCOOCH$_3$   & methyl formate  & 0.06--0.08    &  -          & \vline~0.0034\\ %m=60
CH$_2$OHCHO  & glycolaldehyde  & 0.016--0.039  &  -          & \vline~ \\  \smallskip
CH$_3$COOH   & acetic acid     & $<0.026$     &  -          & \vline~ \\

C$_2$H$_5$OH & ethanol         & 0.11--0.19    &  -          & \vline~0.10$^b$ \\\smallskip %m=46
CH$_3$OCH$_3$ & dimethyl ether & $<0.025$      &  -          & \vline~  \\ 

CH$_3$COCH$_3$ & acetone       & $\leq0.011$  &  -          & \vline~0.0047 \\ %m=58
C$_2$H$_5$CHO & propanal       &  -            &  -          & \vline  \\ %\smallskip
CH$_2$CO     & ketene          & $\leq0.0078$  &  -          &      \\ %m=44
\hline
N$_2$       & molecular nitrogen & \multicolumn{2}{c}{ ($<0.002-1000$ from N$_2^+$)$^c$}     &  $0.089\pm0.024$ \\
NH$_3$      & ammonia          & 0.18--0.60    & 0.1--3.6  & 0.4$^b$        \\ %Lauter
HCN         & hydrogen cyanide & 0.05--0.25    & 0.03--0.5 & \vline~0.20$^b$ \\ \smallskip %Lauter
HNC      & hydrogen isocyanide & 0.0015--0.035 &  -        & \vline \\ 
CH$_3$CN     & methyl cyanide  & 0.008--0.054  &  -        & 0.0059 \\
HC$_3$N      & cyanoacetylene  & 0.002--0.068  &  -        & 0.0004 \\
HNCO         & isocyanic acid  & 0.009--0.080  &  -        & 0.027 \\
NH$_2$CHO    & formamide       & 0.015--0.022  &  -        & 0.004 \\
C$_2$H$_3$CN & vinyl cyanide   & $<0.0027$     &  \\
C$_2$H$_5$CN & ethyl cyanide   & $<0.0036$     &  \\
C$_2$N$_2$   & cyanogen        &     -         &  -          & 0.0004$\pm$0.0002 \\ %Hanni+Lauter
\hline
H$_2$S      & hydrogen sulphide  & 0.09--1.5   &  -          & 2.0$^b$ \\
SO          & sulphur monoxide   & 0.04--0.30   &  -         & 0.071 \\
SO$_2$      & sulphur dioxide    & 0.03--0.23   & -          & 0.127  \\
CS          & carbon monosulphide& 0.03--0.20  &  -          & - \\
CS$_2$      & carbon disulphide  &   -         &   -         & 0.02$^b$ \\
OCS         & carbonyl sulphide  & 0.05--0.40  &  0.04-0.40  & 0.07$^b$ \\
H$_2$CS     & thioformaldehyde   & 0.009--0.090 &  -         & 0.0027 \\
S$_2$       & sulphur dimer      & \multicolumn{2}{c}{(0.001-0.25)$^c$} & 0.002 \\
NS          & nitrogen sulphide  & 0.006--0.012 &  -         &      \\
CH$_3$SH    & methyl mercaptan   & $<0.023$     &  -         & 0.038 \\
\hline
HF          & hydrogen fluoride  & 0.018        &  -         & 0.003--0.048 \\
HCl         & hydrogen chloride  & $<0.011$     &  -         & 0.002--0.059 \\
HBr         & hydrogen bromide   &  -           &  -         & 0.00012--0.00083 \\
PH$_3$      & phosphine          & $<0.07$      & -          & $<0.003$ \\
PN          & phosphorus nitride & $<0.003$     & -          & $<0.001$ \\
PO          & phosphorus oxide   & $<0.013$     & -          & 0.011  \\
CH$_3$NH$_2$ & methylamine     & $<0.055$      &  \\
NH$_2$CH$_2$COOH & glycine I     & $<0.18$      & -          & 0.000017  \\
\hline
Ar          & argon              &             &           & $0.00058\pm0.00022$ \\
Kr          & krypton            &             &           & $0.000049\pm0.000022$ \\
Xe          & xenon              &             &           & $0.000024\pm0.000011$ \\
\hline
O$_2$        & molecular oxygen &    -         & -           &  2.0$^b$   \\  
\hline
\end{tabular}
\end{center}
Table updated from information in \citet{Bockelee2017,Biver2021a,Rubin2019a,DelloRusso2016a,Lippi2021}. %\\
  $^a$: Based on \citet{Biver2019b,Rubin2019a,Lauter2020,Hanni2021}.
  $^b$: value based on the integrated gas loss over the whole Rosetta mission (2~years).
  $^c$: From observations in the visible-UV.
  $^d$: Assuming a daughter distribution.\\
\end{table*}
%\com{Martin}{N$_2$, Ar, Kr, and Xe added, maybe add remote observations of N$_2$? Add Ar from \cite{Stern2000}?}

%--------------------------------- Neil
  \subsection{Hydrocarbons}
  \label{sec:hydrocarbons}
  Symmetric hydrocarbons such as CH$_4$, C$_2$H$_6$, C$_2$H$_2$, and C$_2$H$_4$ are investigated remotely in the IR. Based on the growing database of comet observations there is evidence that CH$_4$, C$_2$H$_6$, and C$_2$H$_2$ are depleted in Jupiter family comets compared to long-period comets from the Oort cloud. These and more complex hydrocarbons were detected in the coma of 67P using the ROSINA mass spectrometer during the \textit{Rosetta} mission.
  
     \subsubsection{CH$_4$ (methane)}
     CH$_4$ is the simplest and most volatile saturated hydrocarbon (C$_n$H$_{2~n+2}$), so its relative abundance in comets may provide clues about the region of formation and processing history of cometary ices \citep{Gibb2003}. CH$_4$ has a strong IR fundamental band ($\nu_3$) near 3.3 $\mu$m; however, its detection in comets from Earth-based observatories requires a sufficient cometary geocentric velocity ($|$d$\delta$/dt$|$ $>\sim$ 15 \kms) to shift comet lines away from corresponding terrestrial counterparts.  CH$_4$ is typically one of the most abundant hydrocarbons in comets with mixing ratios CH$_4$/H$_2$O ranging from  $\sim$ 0.15 – 3\%, with typical abundances  $\sim$ 0.8\% \citep[][and references therein]{DelloRusso2016a}.
 
     \subsubsection{C$_2$H$_6$ (ethane)}
     C$_2$H$_6$ $\nu_7$ band near 3.35 $\mu$m is favorable for detection in comets owing to the large intrinsic strength and a pile-up of ro-vibrational lines within Q-branches. Lines from the $\nu_5$ band of C$_2$H$_6$ near 3.45 $\mu$m are also generally detected although the strongest lines are several times weaker than the $\nu_7$ Q-branches. The terrestrial C$_2$H$_6$ component is very weak so no specific geocentric Doppler-shift is required for its detection in comets from the ground. Mixing ratios C$_2$H$_6$/H$_2$O range from $\sim$ 0.1 – 2\%, with typical abundances  $\sim$ 0.6\% \citep[][and references therein]{DelloRusso2016a}, so mixing ratios of C$_2$H$_6$/CH$_4$ $\sim$1 in comets are typical.
     
     \subsubsection{C$_2$H$_2$ (acetylene)}
     Ro-vibrational lines from the $\nu_3$ and $\nu_2$ + $\nu_4$ + $\nu_5$ vibrational bands of C$_2$H$_2$ near 3 $\mu$m have been detected in many comets.  However, detection of C$_2$H$_2$ is not routine owing to its low abundances, presence of lines in regions of low atmospheric transmittance, and blending with stronger lines of other species. Mixing ratios C$_2$H$_2$/H$_2$O range from $<$ 0.03 – 0.5\%, with typical abundances  $\sim$ 0.1\% \citep[][and references therein]{DelloRusso2016a}.  The typically low abundances of C$_2$H$_2$ relative to C$_2$H$_6$ in comets may indicate the importance of H-atom addition reactions on pre-cometary ices in the nebula.
     %Because conversion of C$_2$H$_6$ $\rightarrow$ C$_2$ is very inefficient, C$_2$H$_2$, despite its relatively low abundance, is likely the major volatile source of C$_2$ detected in a large number of comets at optical wavelengths (e.g., \cite{AHearn1995, Helbert2005, Fink2009, Weiler2012}). C$_2$H$_2$ abundances are insufficient to explain measured C$_2$ abundances in some comets suggesting the presence of additional unidentified sources of C$_2$ (e.g., \cite{DelloRusso2016a}).
     
     \subsubsection{Other hydrocarbons}
     Although C$_2$H$_4$ (ethylene) emissions are numerous in $\sim$ 3.2 – 3.35 $\mu$m region, they are relatively weak, located in many areas of generally low atmospheric transmittance, and are subject to blends from stronger and more abundant species. For these reasons C$_2$H$_4$ has been detected in only two comets: 67P from \textit{Rosetta} ROSINA measurements \citep{Luspay2015, Rubin2015b, Altwegg2017b} and C/2014 Q2 (Lovejoy) from ground-based IR measurements with C$_2$H$_4$/H$_2$O~$\sim$~0.2\% \citep{DelloRusso2021}. Other more complex hydrocarbons also have strong IR bands but they have not been detected from the ground because of various factors including low abundances, spectral confusion with more abundant species (mostly CH$_3$OH), and lack of adequate fluorescence models. Many long carbon chain molecules were detected for the first time in 67P with ROSINA \citep[e.g., ][]{Altwegg2017b}, showing the complexity of hydrocarbons stored in comets.

     \subsubsection{Aromatic hydrocarbons}
     The aromatic hydrocarbons benzene, C$_6$H$_6$, and toluene, C$_7$H$_8$ were detected by ROSINA in 67P \citep{Schuhmann2019} and the latter also tentatively by the Ptolemy mass spectrometer on the Philae lander \citep{Altwegg2017b}. Furthermore, tentative detections of xylene, C$_8$H$_{10}$, and naphtalene, C$_{10}$H$_8$,  were reported by \cite{Altwegg2019}. The latter most likely associated with dust grains and released at elevated temperatures \citep{Lamy1988}.

%--------------------------------- Nicolas
  \subsection{CHO-species and complex organics}
  \label{sec:CHO-species}
  Organic molecules containing C, H and O are significant constituents of cometary comae comprising about 3--5\% relative to water. Since the first detections of methanol and formaldehyde in 1986-1990 via in-situ mass spectrometry and radio and infrared remote observations, the number and complexity of organic molecules discovered in cometary comae has steadily increased \citep{Bockelee2004}. They generally follow their detection in the Interstellar Medium, from which they may have inherited their diversity.
  Complex organic molecules (COMs) are often defined by astrophysicists as CHO-bearing molecules with six atoms or more, but there is no strict definition.
     \subsubsection{CH$_3$OH (methanol)}
  Methanol is the most abundant COM in cometary comae. It was first suggested and subsequently shown to be mainly responsible for the 3.52~$\mu$m feature in the low resolution IR spectrum of comet 1P/Halley in 1986 and in spectra of subsequently observed comets such as C/1989~X1 (Austin) and C/1990~V1 (Levy), in which its detection in the mm range via rotational lines was unambiguous \citep[][and references therein]{Bockelee2004}.
  Since then methanol has been routinely detected in comets, both in the radio via its rotational lines and with high-resolution IR spectrometers via its ro-vibrational P, Q, R branches of the $\nu_2$, $\nu_3$, $\nu_9$ bands in the 3.3-3.5$\mu$m range.
  At radio wavelengths, multiple lines (up to 70 in bright comets) are often observed and used to probe the gas temperature \citep{Biver2015,Biver2021a}. As of 2021, methanol has been detected in about 60 comets (55 in the radio and 37 in the IR, including two with in-situ mass spectrometry) with an abundance relative to water ranging from 0.5 to 6\%, within 2~au from the Sun. IR and radio observation often yield similar production rates, with methanol rich comets exhibiting 3--4\% methanol relative to water, and methanol poor comets around 1\%, but the distribution is not clearly bimodal (Fig.~\ref{fig-histo1}). %The lowest abundance measured was likely in comet C/1999~S4 (LINEAR) which disintegrated in July 2000, in which upper limits derived from both radio \citep[$<0.9$\%,][]{Bockelee2001} and IR \citep[$<0.2$\%,][]{Mumma2001a} suggest significant depletion. Comet 67P seems also relatively depleted with an integrated (over two years) production of 1.5\%  based on submillimeter data \citep{Biver2019b} to 0.5\% based on mass spectrometry \citep{Lauter2020}.
  %The apparent disagreement is mostly due to the reference water production since the two methods yield a similar peak production around $Q_{CH_3OH}=1.1\times10^{26}$ molec.s$^{-1}$. Seasonal effects have been hypothesized as the cause of the variable mixing ratio of CH$_3$OH/H$_2$O in 67P. Due to its different volatility compared to water, heliocentric variation may also be present (depletion close to the Sun \citep{Biver2011} with a higher CH$_3$OH/H$_2$O ratio beyond $\sim2.5$~au).
 
     \subsubsection{Formaldehyde (H$_2$CO)}
  Formaldehyde was first unambiguously identified in a comet on 1P/Halley with the NMS instrument on the \textit{Giotto} spacecraft. Subsequently H$_2$CO was detected in the radio in C/1989 X1 (Austin) and C/1990 K1 (Levy) \citep{Colom1992} and in the IR via its $\nu_1$ band at 3.59 $\mu$m in 153P/Ikeya-Zhang \citep{DiSanti2002}.  
  Measured abundances range from 0.12 to 1.4\% relative to water for the total production of H$_2$CO including distributed sources. 
  
     \subsubsection{Complex organics}
  Including HCOOH, complex CHO-species have been observed in several comets in the radio, for example in C/1996~B2 (Hyakutake), C/1995~O1 (Hale-Bopp), and in C/2014 Q2 (Lovejoy). In-situ mass spectrometry in comet 67P has identified COMs, some not yet detected in the interstellar medium.
  \begin{itemize}
    \item {\sl Formic acid (HCOOH)} has been detected in 10 comets with an estimated abundance between 0.01 and 0.2\% relative to water. But abundances can be uncertain by a factor of two or more owing to poor constraints on its photodissociation rate. \citep{Biver2021a};
    \item {\sl Acetaldehyde (CH$_3$CHO)} exhibits multiple lines of similar intensities which helps its detection in wide band spectroscopic surveys. It has been detected in 7 comets with a very similar abundances of 0.05-0.08\% relative to water. 
    \item {\sl Ethylene-glycol} in its lowest energy conformer {\sl aGg'-(CH$_2$OH)$_2$} has been identified in the radio spectra of 7 comets with an abundance of 0.07 to 0.35\% relative to water, which is a substantial percentage for such a large molecule. It is notably higher than measurements by \textit{Rosetta} \citep[][but it was not measured at the time of peak activity]{Rubin2019a} and measured values in star forming regions \citep{Biver2019a}.
    \item {\sl Methyl formate (HCOOCH$_3$)} has only been detected in two comets in the radio ($\sim$0.07\% relative to water). It is the most abundant isomer of mass 60, with {\sl glycoladelhyde (CH$_2$OHCHO)}, four times less abundant, and {\sl acetic acid (CH$_3$COOH)} at least two times less abundant in comet C/2014 Q2 (Lovejoy) \citep{Biver2015}. These molecules appear depleted in 67P compared to previous ground-based measurements (Table~\ref{tab:abundances}).
    \item {\sl Alcohols: Ethanol (C$_2$H$_5$OH)} has been identified in four comets with an abundance around 0.1\%, one order of magnitude lower than methanol. More complex alcohols have been identified in 67P by \textit{Rosetta}/ROSINA \citep{Altwegg2019}.
  \end{itemize}
 The more complex CHO-molecules, especially those with 3 carbons have not been detected with remote observation, due to their complex spectra and decreasing abundance with molecular complexity. Only \textit{Rosetta}/ROSINA had the sensitivity to detect species such as C$_3$H$_6$O, C$_3$H$_8$O, C$_4$H$_8$O, C$_4$H$_{10}$O, and C$_5$H$_{12}$O \citep{Altwegg2019} but other techniques are required to distinguish isomers.
     
%--------------------------------- Neil
  \subsection{Nitrogen bearing species}
  \label{sec:N-species}
  N-bearing volatiles are generally detectable both at IR and radio wavelengths. Evidence from both remote sensing and in-situ missions have established that comets contain many of the necessary ingredients for life from simple HCN and NH$_3$ to the amino acid glycine. Some N-bearing species show spatial distributions consistent with extended sources in the coma, providing further evidence for complex nitrogen molecules in comets.
  
     \subsubsection{N$_2$ (molecular nitrogen)}\label{sec:N2}
     The presence of N$_2$ in comets has until recently been a matter of debate as its direct detection from radio or IR techniques is not possible. However, N$_2^+$ can be observed at near-UV wavelengths to trace N$_2$ abundances in comets. Detections of the $\mathrm{B^2\Sigma_u^+-X^2\Sigma_g^+}$ First Negative (0,0) band of N$_2^+$ at 391.4 nm have been claimed in the tails of comets from low-resolution spectroscopic observations, first in photographic spectra \citep{Swings1956,Cochran2000} and subsequently in the coma of 1P/Halley \citep{Wyckoff1989b} and C/1987 P1 (Bradfield). However, these early detections have been questioned because at low spectral resolution N$_2^+$ emissions from the comet and the terrestrial atmosphere can be difficult to differentiate. More reliable detections of N$_2^+$ have only been reported in the last decade:  in the coma of C/2002 VQ94 (LINEAR) and C/2016 R2 (PANSTARRS) using high spectral resolution \citep{Korsun2014,Cochran2018,Opitom2019b} and in the coma of the centaur 29P/Schwassmann-Wachmann 1 from low-resolution spectroscopic observations \citep{Ivanova2016}.  In addition to the (0,0) band, the (1,1) and (0,1) bands were also detected in comet C/2016 R2 \citep{Opitom2019b}. 
     %In that same comet, a record number of 15 $\mathrm{A^2\Pi-X^2\Sigma}$ CO$^+$ bands were also detected. 
          %The presence of N$_2$ in comets has until recently been a matter of debate as its direct detection form radio or IR techniques is not possible.
    The presence of N$_2$ in comets was confirmed by its detection in 67P by \textit{Rosetta}/ROSINA \citep{Rubin2015b}, showing this highly volatile molecule is retained in the nucleus of an evolved Jupiter-family comet. 
          %The presence of N$_2$ can be inferred indirectly through the detection of N$_2^+$ in the near-UV; however, there are few firm detections, with the most definitive occurring more recently (\cite{Korsun2014, Ivanova2016, Cochran2018, Opitom2019b}).  

     While N$_2$ is surely the dominant N-bearing volatile in some comets, its extremely high volatility makes it susceptible to depletion by evolutionary effects so its abundance is probably highly variable in comets, e.g., in 67P NH$_3$ contributed more to the nitrogen reservoir than N$_2$ \citep{Rubin2019a}.

     %Another small molecule identified at 67P is N$_2$ \citep{Rubin2015b}. N$_2$ was expected at comets due to earlier remote observation of the corresponding ion at comet C/2002 VQ94 (LINEAR) by \cite{Korsun2014}.
     
     The N$_2$/CO ratio in cometary ices can be estimated from ground-based detection of CO$^+$ and N$_2^+$ emission band at optical wavelengths, as was done for three comets \citep{Korsun2014,Cochran2018,Ivanova2016}. Inferring the relative abundance of neutrals from ion emission intensity ratio is not always straightforward but \cite{Raghuram2021} showed that this can be done when the ion production is controlled by photon and photoelectron impact ionization of the neutrals (rather than ion-neutral chemistry). \textit{Rosetta} provided a direct measure of the relative abundance \citep[N$_2$/CO = 0.029~$\pm$~0.012,][and  Table~\ref{tab:abundances}]{Rubin2019a}, about five times higher than the value measured earlier in the mission at 3.1~au \citep{Rubin2015b}. The evolutionary difference between the northern and southern hemispheres \citep{Keller2015a} coupled with the different illumination conditions during the two observation phases led to differences in the measured N$_2$/CO ratio, similar to the observations of the ratios of other species \citep{Lauter2020,Combi2020}. The presence of both species puts constraints on the maximum temperature of the material inside 67P since its formation. %Both CO and N$_2$ have very low but somewhat different sublimation temperatures in their pure ice form. 
     When comparing the measured N$_2$/CO ratio to the estimated primordial solar system ratio \citep{Lodders2010} a formation temperature well below 30~K for amorphous ices \citep[cf.][]{Rubin2015a,Rubin2020} and $<$50~K for crystalline ices was derived \citep{Mousis2016b}. These results may be further modified when taking into account how these species may not be present in their pure ice forms but trapped in either CO$_2$ or H$_2$O ices \citep{Kouchi1995}. 
     
    % At the time of the \textit{Giotto} flyby of 1P/Halley, N$_2$ could not be resolved (cf. section~\ref{sec:mass-spectro}) and only an upper limit was obtained \citep{Eberhardt1988}. \cite{Geiss1988} reported a lack of nitrogen in comets and this finding remained even after the firm detection of N$_2$ in 67P with an abundance relative to water of N$_2$/H$_2$O~=~0.00089~$\pm$~0.00024 (cf. \cite{Rubin2019a} and later discussion in section~\ref{sec:N-deficiency}).

     \subsubsection{NH$_3$ (ammonia)}
     Ammonia is an important biogenic molecule as it was likely an intermediary for synthesis of amino acids on the early Earth \citep{Bernstein2002,MunozCaro2002}. NH$_3$ can be targeted at both radio and IR wavelengths in comets but it has been measured in fewer comets than HCN owing to the relative difficulty of its detection. Mixing ratios NH$_3$/H$_2$O range from $<$ 0.1 – 5\%, with typical abundances $\sim$ 0.5 - 1\% \citep[][and references therein]{DelloRusso2016a}.
     %Results from \textit{Rosetta} suggest that the breakdown of ammonium salts could be a possible additional source of NH$_3$ in cometary comae  \citep[e.g.,][]{Altwegg2020a}.
     
     \subsubsection{HCN (hydrogen cyanide)}
     HCN is the simplest nitrile parent volatile as well as an important intermediary for synthesis of biochemical compounds, and can be routinely detected in comets at both millimeter and IR wavelengths.  HCN/H$_2$O mixing ratios in comets are typically $\sim$ 0.1 – 0.2\%, with no obvious differences between Jupiter-family and long-period comets. Notably, production rates of HCN in comets derived at millimeter wavelengths are typically about a factor of two smaller than those derived at IR wavelengths \citep{MageeSauer2002, Biver2002b}. The reason for this difference has not been discovered. Modern capabilities at IR and radio wavelengths can provide sensitive spatial maps of HCN in the coma of comets. Results generally suggest that HCN in the coma comes predominately from direct sublimation from nucleus ices. 
     %HCN is likely a major volatile source of CN detected in comets at optical wavelengths \citep[e.g.,][]{AHearn1995, Schleicher2007, Fink2009}; however, derived production rates suggest that other sources in addition to HCN are needed to explain CN abundances in many comets.
     
     \subsubsection{HNC (hydrogen isocyanide)}
     Hydrogen isocyanide has been detected in several comets at millimeter wavelengths. HNC also has strong vibrational bands near 5 $\mu$m, but these transitions have not been detected in comets to date. The HNC/HCN mixing ratio has been shown to increase with decreasing heliocentric distance, with a mean value around 7\% at 1 au \citep{Lis2008}.
     
     \subsubsection{other N-bearing molecules}
     Additional N-bearing molecules (e.g., NH$_2$CHO, HNCO, CH$_3$CN, and HC$_3$N) have been detected in a few comets at radio wavelengths with mixing ratios on the order of 10\% with respect to HCN \citep[Table~\ref{tab:abundances},][]{Bockelee2017}. More complex N-bearing molecules have also been searched for in the radio and some detected by ROSINA, but in all cases they are minor contributors to the nitrogen budget in cometary atmospheres.
     \begin{itemize}
     \item \textit{CH$_3$CN and HC$_3$N} were first identified in comets Hyakutake and Hale-Bopp and have been detected in over 10 comets during the last 20 years. CH$_3$CN is routinely detected in cometary comae, typically representing 10 to 30\% of the HCN abundance. On the other hand, the abundance of HC$_3$N can be more variable: generally between only 2 and 13\% relative to HCN, its abundance in comet C/2002~X5 at 0.2 au from the Sun was found to be up to 40\% relative to HCN \citep{Biver2011}. A variation with heliocentric distance of the HC$_3$N abundance is not excluded. %These two molecular species can produce CN via photodissociation \citep{Bockelee1985}, but their measured abundances definitely show that HCN remains the major nitrile contributor.
     \item \textit{isocyanic acid, HNCO} has also been detected in over 10 comets since 1996. Its abundance (Table~\ref{tab:abundances}) can vary between about 10 and 60\% relative to HCN. This acid could also be associated with ammonium salts and its spatial distribution and the heliocentric variation of its abundance needs to be investigated.
     \item \textit{formamide, NH$_2$CHO} has only been detected in a handful of comets (Fig.~\ref{fig-histomolec}), with some uncertainties on its abundance due to an unknown lifetime \citep{Biver2021a}. Nevertheless it does not exceed $2\times10^{-4}$ relative to water in comets and formamide could also be produced in the coma \citep{Cordiner2021}.
     \end{itemize}
     More complex N-bearing molecules have also been identified in comets (Sect.\ref{sec:glycine}).
     
     \subsubsection{The apparent nitrogen depletion in comets}
     \cite{Geiss1988} reported an apparent lack of nitrogen in the coma of comet 1P/Halley when compared to the Sun, meteorites and terrestrial samples. These results were based on the abundances of the major N-bearing molecules NH$_3$ and HCN (including HNC) obtained during the \textit{Giotto} fly-by. Similarly, abundances of NH$_3$, HCN, and other minor N-bearing species in most comets measured with remote sensing techniques are consistent with this observed nitrogen deficiency (Table~\ref{tab:abundances}). Gases in the coma of 67P were analyzed \citep{Rubin2019a}, this time also including molecular nitrogen, N$_2$ \citep{Rubin2015b}, which revealed a similar under-abundance of nitrogen.
     
     The heliocentric distance dependence of NH$_3$/H$_2$O in comets shows evidence for increasing ratios for smaller heliocentric distances, up to $>$3\% for comet D/2012 S1 (ISON) \citep{Altwegg2020a,DelloRusso2016a}. This is consistent with the additional release of NH$_3$ from grains when a thermal threshold is reached. It has been suggested that ammonium salts may be present on the surface of comet 67P, similar to the surface of some observed asteroids, based on a broad spectral reflectance feature \citep{Quirico2016, Poch2020}. Ammonium salts are made up of the NH$_3$ base and an acid providing an H$^+$ to the base. Mass spectrometric measurements in the coma, most likely due to dust grains entering the ion source of ROSINA, revealed all possible sublimation products of the five different ammonium salts NH$_4^+$Cl$^-$, NH$_4^+$CN$^-$, NH$_4^+$OCN$^-$, NH$_4^+$HCOO$^-$, and NH$_4^+$CH$_3$COO$^-$ \citep{Altwegg2020a}. Direct identification of ammonium salts is difficult as they mostly dissociate into the base and corresponding acid upon sublimation of the salt. Ammonium salts may hence be released from distributed grain sources that subsequently release simpler volatiles into the coma. Assuming that the NH$_3$/H$_2$O ratio of comets close to the Sun is more representative of the actual content of nitrogen, the apparent lack of cometary nitrogen could be an observational bias related to the difficulty in directly measuring nitrogen contained in less volatile sources.
     
%--------------------------------- Nicolas
  \subsection{Sulfur bearing species}
  \label{sec:S-species}
  Eight sulfureted molecular species (or radicals) have been identified in multiple comets from remote observations primarily at radio wavelengths: H$_2$S, CS, SO, SO$_2$, OCS, H$_2$CS, NS and S$_2$, (Table~\ref{tab:abundances}, Fig.~\ref{fig-histomolec}). The OCS $\nu_3$ vibrational band has also been detected in the infrared at $4.85\mu$m \citep[e.g.,][]{DelloRusso1998, Saki2020}. S$_2$ has only been detected in the UV via electronic transitions around 295~nm \citep[e.g.][]{Reyle2003}. %H$_2$S is the major sulfur bearing molecule, but the sum of other molecules (CS, SO, SO$_2$ and OCS) can contribute equally to the sulfur budget in cometary comae. 
  Atomic S has also been observed in the UV \citep[e.g.,][]{Meier1997} but is interpreted as mostly a secondary photo-dissociation product of the other sulfureted species. The total S/O ratio observed in the volatile phase ranges from 0.5 to 2\%, in agreement with in-situ measurements in the coma of comet 67P by \textit{Rosetta}/ROSINA \citep{Rubin2019a,Calmonte2016}.

    \subsubsection{Hydrogen sulfide (H$_2$S)}
    H$_2$S, the dominant sulfur-bearing molecule in comets, has been detected in 34 comets since 1990 \citep{Bockelee1991}, almost exclusively by its rotational lines at 168.8 and 216.7~GHz, or by in-situ mass spectrometry in comet 67P. Its abundance relative to water can vary by a factor 16 from 0.09\% in the carbon-chain depleted JFC comet 21P to 1.5\% in comet Hale-Bopp. However, no systematic differences in H$_2$S abundances are seen between JFCs and OCCs. \textit{Rosetta} measured an average abundance of 1.1\% relative to water \citep{Calmonte2016} typical of a relatively H$_2$S-rich comet. In H$_2$S depleted comets, the total abundance of other sulfur-bearing species can be larger than H$_2$S itself.

    \subsubsection{Sulfur monoxide and dioxide (SO and SO$_2$)}
    The abundance of SO$_2$ has been measured at radio-wavelengths in two comets and abundances relative to water are in the range $<0.009$ to 0.23\%. SO$_2$ was also detected by \textit{Rosetta} in comet 67P (0.13\%). The linear molecule SO has stronger millimeter lines and has been detected in 8 comets. Although it is often assumed that SO is produced by the photolysis of SO$_2$ ($\beta_{0,SO_2}=2.5\times10^{-4}$s$^{-1}$ at 1~au), there is evidence of a secondary source of SO, as suggested from interferometric maps \citep{Boissier2007}. \textit{Rosetta}/ROSINA find 0.07\% of SO relative to water in 67P, with evidence of direct release from the sublimation of SO ice and not as a secondary product of SO$_2$ \citep{Calmonte2016}.

    \subsubsection{CS and CS$_2$}
    CS has been detected in comets remotely via UV and radio spectroscopy for decades \citep{Meier1997,Biver1999} with abundances between 0.03 and 0.2\%. CS is a radical for which there is little information on its photolysis properties. A photo-dissociation rate around $2.5\times10^{-5}$s$^{-1}$ at 1~au from the Sun was estimated from radio observations \citep{Boissier2007,Biver2011}. It is usually assumed that CS is a product species coming from the photo-dissociation of CS$_2$ \citep[photo-dissociation rate $\beta_{0,CS_2}=1.7\times10^{-3}$s$^{-1}$ at 1~au,][]{Jackson1986}. However some observations suggest a longer lifetime for the parent of CS (section~\ref{sec:fragment-parent}). 
 %Most observations show evidence of production in the coma at least compatible with its production from CS$_2$, but a few constraints on the parent scale-length yield $2-4\times$ longer lifetimes for the parent of CS than expected if the parent is CS$_2$. 
    \textit{Rosetta}/ROSINA detected CS$_2$ at very low abundances in 67P \citep{Calmonte2016,Rubin2019a}, but unfortunately CS could not be detected. 
 %The low CS$_2$ abundances measured at 67P compared to CS abundances measured in other comets (Table~\ref{tab:abundances}) suggests that it is not the main parent of CS. 
  
 %OCS mostly dissociates into S + CO, so it cannot be a significant source of CS in the atmosphere of the comets. H$_2$CS is not expected to be a major source of CS owing to its low abundance. The photodissociative lifetime of H$_2$CS is not known, but if it is 20 times shorter than assumed (i.e. $\beta_0=2.2\times10^{-3}$s$^{-1}$ as suggested by \cite{Rodgers2006} versus $\beta_0=1.0\times10^{-4}$s$^{-1}$ at 1~au), its abundance derived from radio observations could be 3-6 times higher; however, this is not what is suggested by the in-situ ROSINA measurements. 

%Even assuming that CS is a daughter species, all observations show evidence of a strong dependency of the CS abundance with heliocentric distance \citep{Meier1997,Biver1999,Biver2002,Biver2006,Biver2011}, roughly as $Q_{\rm CS}/Q_{\rm H_2O} \propto r_h^{-1}$, down-to heliocentric distances as low as 0.1~au. This evidence suggests that the major parent of CS is still to be found, possibly stored in the dust phase that could not be sampled by the ROSINA instrument.

    \subsubsection{OCS, Thioformaldehyde (H$_2$CS)}
    Including IR and radio techniques {\sl OCS}, has been detected in 10 comets with an abundance relative to water ranging from 0.04 to 0.4\%. An average value of 0.07\% \citep{Lauter2020} was measured in comet 67P by \textit{Rosetta}.

    {\sl Thioformaldehyde (H$_2$CS)}, because of its likely short lifetime and low abundance (0.009-0.09\% relative to water) has been detected remotely in only two comets. ROSINA measurements in the coma of 67P yield an even lower abundance of 0.003\% \citep{Rubin2019a}.

    \subsubsection{S$_2$, S$_3$, S$_4$}
    Sulfur has also been detected in the form of {\sl disulfur (S$_2$}) in cometary comae via UV spectroscopy \citep[][and references therein]{Reyle2003}, and even in the form of S$_3$ and S$_4$ \citep{Calmonte2016} in the coma of 67P by \textit{Rosetta}. %Those species having relatively short lifetimes are difficult to observe remotely. 
    UV observations yield abundances of S$_2$ ranging from 0.001 to 0.02\%, comparable to ROSINA findings (0.002\%). But some S$_2$ and most of the S$_3$ and S$_4$ detected by ROSINA are estimated to be secondary products not coming from nucleus ices but from a likely warmer dust component due to their low volatility \citep{Calmonte2016}.
    
    \subsubsection{NS, CH$_3$SH, and other sulfur-bearing molecules}
    Several other sulfur-bearing species have been identified in the coma of 67P such as {\sl methyl mercaptan or methanetiol} (CH$_3$SH) and C$_2$H$_6$S \citep{Calmonte2016} and more complex species such as CH$_4$S$_2$, CH$_4$OS, C$_2$H$_6$OS, and C$_3$H$_8$OS according to \cite{Altwegg2019}. CH$_3$SH has not yet been detected in a comet in the radio domain, but the abundance found in 67P (0.04\% relative to water) is comparable to upper limits found in other comets (0.02-0.06\%). NS was not detected in the coma of 67P but its abundance was estimated to be 0.006-0.012\% in comets C/2014~Q2 and C/1995~O1. The origin of this radical is still unknown \citep{Irvine2000}.

 %Neil: Nicolas, I removed this short paragraph on atomic sulfur as it is discussed above.
 
%--------------------------------- Martin
  \subsection{\label{sec:other-species}Other minor species}
    \subsubsection{Halogens}
    \cite{Dhooghe2017} reported the detection of three main halogen-bearing molecules HF, HCl, and HBr in the coma of 67P; the measured fragmentation pattern (cf. section~\ref{sec:mass-spectro}) indicates at least one additional Cl-bearing compound aside from the already identified HCl, CH$_3$Cl, and NH$_4$Cl \citep[cf. section~\ref{sec:fragment-small-rh}, and][] {Fayolle2017,Altwegg2020a}, and an increasing relative proportion of chlorine with distance indicates the likely presence of a distributed source \citep{DeKeyser2017}.

    \subsubsection{\label{sec:phosphorous}Phosphorus-bearing molecules}
    Phosophorous has been identified in the gas phase around 67P \citep{Altwegg2016} with PO being the main carrier while minor contributions of PN and PH$_3$ cannot be ruled out \citep{Rivilla2020}. Phosphorous is an important element for life and its presence supports the theory that comets may have contributed key molecules related to the emergence of life on the early Earth \citep{Oro1961,Schwartz2006}.

    \subsubsection{\label{sec:glycine}Glycine, amines,…}
    The presence of the most simple amino acid glycine has been established in the dust samples returned by the \textit{Stardust} mission to comet 81P/Wild~2 \citep{Elsila2009}. Glycine (C$_2$H$_5$NO$_2$) and two of its precursor molecules, methylamine (CH$_3$NH$_2$) and ethylamine (C$_2$H$_5$NH$_2$), have also been detected in the coma of 67P \citep{Altwegg2016}. Together with the suite of C, H, O, N and S-bearing organic compounds \citep[cf. Tab.~\ref{tab:abundances} and][]{Altwegg2017b} as well as phosophorous (cf section~\ref{sec:phosphorous}), comets contain elements key in prebiotic chemistry and molecules identified as possible biomarkers in the search for life elsewhere \citep{Seager2016}. More complex amino acids may form through aqueous alteration \citep{Burton2012} but were not identified in either 81P or 67P, which limits how their material is reprocessed in larger parent bodies.

%--------------------------------- Martin
    \subsubsection{\label{sec:noble-gases} Noble gases Ar, Kr, Xe}
    The noble gases argon, krypton, and xenon have been identified in the coma of 67P \citep{Balsiger2015,Marty2017,Rubin2018} with Ar abundances much lower than reported in C/1995 O1 (Hale-Bopp) by \cite{Stern2000}. Neon, on the other hand, was below the detection limit \citep{Rubin2018}. 
     
    The measured relative abundances of noble gases (Tab.~\ref{tab:abundances}) may indicate trapping at elevated temperatures if the original abundances were solar or subsequent loss of the more volatile Ar and Ne during the comet's transition from the scattered disk, through the Centaur stage, and to its current orbit \citep{Maquet2015,Guilbert2016}. 
     %However, this is at odds with the measured  N$_2$/CO ratio discussed in section~\ref{sec:N2}. 
    Another possibility, suggested by \cite{Marty2017}, is the addition of an exotic component of xenon required to explain its isotopic composition.
     %See later discussion in section~\ref{sec:noble-gas-isotopes}). 
     
  \subsection{\label{sec:O2} O$_2$ (molecular oxygen)}
  The measured O$_2$ abundance in the coma of comet 67P (3.1$\pm$1.1)\% relative to H$_2$O measured at $\sim$1.5~au pre-perihelion \citep{Bieler2015a,Rubin2019a}, made it the dominant coma species after H$_2$O, CO$_2$, and CO. Furthermore, \cite{Bieler2015a} reported a strong correlation with water despite the very different sublimation temperatures of the corresponding pure ices \citep{Fray2009}. Based on these results the presence of O$_2$ has also been inferred in the coma of 1P/Halley from measurements by \textit{Giotto} NMS \citep{Rubin2015a}. At 67P, during suitable periods when CO$_2$ was at times even the dominant molecule in the coma \citep{Lauter2020}, O$_2$ was found to be correlated to CO$_2$ \citep{Luspay2022}.
    
  There are several possible mechanisms for the formation of O$_2$ in comets \citep[cf. review by][]{Luspay2018}. \cite{Dulieu2017} proposed the dismutation of H$_2$O$_2$ during the sublimation process, where H$_2$O$_2$ would be co-produced with the water ice explaining the correlation of H$_2$O and O$_2$. On the other hand, a high conversion of 2~H$_2$O$_2\rightarrow$2~H$_2$O+O$_2$ would be required. \cite{Mousis2016a} is consistent with radiolysis of water ice as the origin of molecular oxygen with subsequent trapping in the ice, which is also consistent with the correlation between the two species. However, it is not clear these different scenarios can reproduce the observed amounts of O$_2$. 
  Yet another mechanism was proposed by \cite{Yao2017} based on Eley-Rideal reactions forming O$_2$ involving energetic water ions. However, this mechanism is inconsistent with the low fluxes of energetic H$_2$O$^+$ ions and their poor correlation with neutral water in the coma \citep{Heritier2018}. Additionally, all these production scenarios are at odds with the different oxygen isotope ratios between the two species (cf. section~\ref{sec:SOC-isotopes}).
     
  To date, the most promising scenario remains a primordial origin of O$_2$ as proposed by \cite{Taquet2016} where the O$_2$ is formed through ice grain surface chemistry and/or in the gas phase \citep{Rawlings2019}. A primordial origin is also favored by \cite{Luspay2022}.
% \add{\sl FOR MARTIN: Reference to Luspay-Kuty (2020)? and different isotopic ratios for O2 and H2O?}
     
%--------------------------------- 
%  \subsection{Elemental abundances of the gas coma (table)}
%  \label{sec:elemental}     

%-----------------------------------------------------------------------------------------------------------
%--------------------------------- Cyrielle
\section{\textbf{FRAGMENT SPECIES}}
\label{sec:fragment-species}
The first molecule identified in the coma of comets, C$_2$, is a secondary product. Secondary products are not present in cometary ices but are produced in the coma by the photo-dissociation of more complex molecules. For example, the photodissociation of water can lead to the production of OH, H, H$_2$, O, OH$^+$, O$^+$, or H$^+$ in the coma. Most secondary products can be observed at optical wavelengths and are among the easiest species to detect in the coma of comets. They are commonly used to infer the composition of cometary ices when observations of parent species are lacking. However, inferring the composition of cometary ices from the abundance of secondary products in the coma is not straightforward as the parents of some radicals have not been identified. Some secondary products may also have multiple parents or be released from organic-rich grains. Species parentage in the coma of comets is a complex problem requiring further investigation and coordination of observations with modern techniques in multiple spectral regions.

%\com{Cyrielle}{Add a table with fragment species observed, their wavelengths, and abundances}
%---------------------------------  Cyrielle
  \subsection{Radicals}
  \label{sec:radicals}
  The main radicals observed in the coma of comets are OH, CN, C$_2$, C$_3$, CH, CS, NH, and NH$_2$. The NS radical was also detected in the coma of a handful of comets. %Some ions such as CO$^+$, CO$_2^+$, N$_2^+$,  H$_2$O$^+$, CH$^+$, and OH$^+$ are also observed in the coma of some comets. 
  CO is a special case as it can be both a parent or secondary product. \cite{Feldman2004} explore in detail the main emissions of these radicals. They are typically observed through electronic transitions at optical wavelengths but some like OH also have transitions in the UV, radio, and IR. Most of these radicals are easy to observe at optical wavelengths and their abundances have been measured in the coma of hundreds of comets. The first taxonomic classifications of comets based on their composition relied on the systematic observations of these radicals.

  The most abundant of these radicals is OH, which is produced by the photo-dissociation of water. Because the OH (0-0) band around 308 nm is one of the brightest emissions in the optical spectrum of comets, OH is often used as a water tracer and abundances of other radicals observed at optical wavelengths are often measured relative to OH. Just like for parent species, the relative abundances of radicals with respect to water (or OH) varies among comets\remove{,} by more than one order of magnitude \citep{AHearn1995,Schleicher2008,Cochran2012,Langland2011,Fink2009}. CN abundances relative to OH vary between less than 0.1\% to almost 1\% while C$_2$ exhibits a wider range of abundances: in carbon-chain depleted comets (see section \ref{sec:taxonomy}), C$_2$ abundance can be lower than 0.01\%, but it can reach maximum values similar to CN for typical comets. C$_3$ is generally less abundant than C$_2$ and CN, (C$_3$/OH = 0.005 -- 0.2\%). NH and NH$_2$ have relatively similar abundances as they are both thought to originate from the photo-dissociation of NH$_3$ ranging between less than 0.1\% to a few \%. Abundances of CH are the least well constrained of the radicals observed at optical wavelengths, partly because it isn't isolated by any of the narrow-band filters commonly used and it is usually faint in most comets. Estimates of its abundance range from about 0.3 to typically a few \%. 

  High resolution spectroscopy has allowed the detailed characterization of the molecular band structure of several of these radicals.  It has also paved the way for new modelling and laboratory measurements.  For example, new models or laboratory measurements of the OH and NH$_2$ emission bands at optical wavelengths have been used to derive isotopic ratios of H, C, N, and O \citep{Rousselot2014,Hutsemekers2008}.
  Significant progress has also been made in the observations and modelling of CN, and especially the (A-X) red system, whose (0-0) band has a bandhead around 1100 nm \citep{Paganini2016,Shinnaka2017}.

  \subsection{Atoms}
  \label{sec:atoms}	

  In addition to radicals, emissions from atoms are also detected at UV and visible wavelengths. HI, OI, CI, NI, SI and the ions CII, OII, have all been observed in the coma of comets, in the 80-200 nm range via solar fluorescence \citep{Feldman2004} and electron impact dissociative excitation of their parent in the inner coma of 67P \citep{Bodewits2022,Feldman2015}. HI Lyman-$\alpha$ emission at 121.6 nm, which totals almost 98\% of far-UV emission in comets, has been used to estimate the water abundance in the coma of several dozen comets, in particular using the SWAN instrument onboard SOHO  \citep{Combi2019}. At optical wavelengths, the most prominent atomic features are the three forbidden OI emission lines at 557.73 nm, 630.03 nm, and 636.38 nm. These lines have been used for a number of years to estimate the water production rate and more recently as an indirect way to estimate the ratio between H$_2$O and CO$_2$ abundances in the coma of comets \citep{Festou1981,Schultz1992,Morgenthaler2001,McKay2013,Decock2013}.  Forbidden [CI] lines at 872.7, 982.4, and 985.0 nm, coming from the photo-dissociation of neutral C-bearing species, have also been detected in the coma of C/1995 O1 (Hale-Bopp) and C/2016 R2 (PANSTARRS) \citep{Oliversen2002,Opitom2019b}. Recently, lines at 519.79 and 520.03 nm in the coma of C/2016 R2 (PANSTARRS) were identified as forbidden nitrogen lines \citep{Opitom2019b}. To this day, [NI] lines have only been detected in the coma of a single comet. 

  The sodium D-line doublet was identified in comets over a century ago \citep{Newall1910}. This emission is difficult to observe from ground-based observatories due to telluric sodium lines. It has been mostly observed in the coma of comets well within 1~au from the Sun, manifesting as a neutral tail. Sodium detection in several comets is discussed by \citep{Feldman2004}. There is no present consensus on the origin of neutral sodium in comets as both nucleus ice and dust grain sources have been proposed \citep{Cremonese2002}. However, the detection of sodium in dust grains and in gas phase in the coma of 67P was reported from \textit{Rosetta} data \citep{Schulz2015,Wurz2015,Rubin2022}. 

  In exceptional cases when a comet passes very close to the Sun, other heavier elements are detected in the coma.  Emission lines from Ca, K, V, Cr, Mn, Fe, Co, Ni, and Cu were detected in the coma of sungrazing comet C/1965 S1 Ikeya-Seki \citep{Preston1967,Slaughter1969}. These elements are typically only detected in the coma of sungrazing comets because of the extremely high temperatures needed to release them through the sublimation of dust grains. Si was detected in the coma of comet C/2011 W3 (Lovejoy) with the UVS onboard SOHO \citep{Raymond2018}.  \cite{Fulle2007} also reported the discovery of an atomic iron tail for comet C/2006 P1 (McNaught) at a heliocentric distance of 0.3~au from observations with the STEREO spacecraft. More recently \cite{Manfroid2021} showed that FeI and NiI emission lines are present in the coma of comets at a variety of heliocentric distances, from 0.7 to $>$ 3~au. This discovery was surprising as neither iron or nickel were detected by \textit{Rosetta}. Their detection at large distances from the Sun rules out a production from the sublimation of dust grains as is the case for sungrazing comets. \cite{Manfroid2021} suggest that iron and nickel could be contained in cometary ices within organometallic complexes (e.g. [Fe(PAH)]$^+$) or carbonyls such as Fe(CO)$_5$ and Ni(CO)$_4$. 

  \subsection{Species parentage and release mechanisms}

  While some species like OH have well identified production mechanisms, it is not the case of all radicals. Some radicals have several potential parents, while for others no parent has been identified to date. Understanding the parentage of secondary products in the coma of comets provides a window into the composition of the nucleus.

    \subsubsection{Identifying parent molecules}
    \label{sec:fragment-parent}
% I have added the radical name in italics at the beginning of each paragraph - remove it if you think it is not usefull but I think it would be better to avoid lengthy sections.
% It looks better indeed. I would keep it there (Cyrielle)

   \textit{NH and NH$_2$.} It is generally believed that both NH and NH$_2$ result from the photo-dissociation of ammonia. NH would be a third-generation product, with NH$_3$ being first dissociated into NH$_2$, and subsequently into NH. Comparison of the abundances of NH, NH$_2$, and NH$_3$ to confirm this parentage is limited by a lack of simultaneous measurments of these species, the difficulty in detecting NH$_3$, and the uncertainty of NH scalelengths \citep{Fink1991,Feldman2004}.

   \textit{CN.} It has long been postulated that photo-dissociation of HCN is the dominant source of CN in the comae of comets. If this is true, then CN and HCN production rates measured simultaneously should be consistent with each other. However, for some comets, derived HCN abundances are significantly lower than those reported for CN \citep{Fray2005,DelloRusso2009,DelloRusso2016b}. Inconsistencies between HCN and CN production rates were also observed in coma of 67P, from measurements with the ROSINA mass spectrometer \citep{Hanni2020}. \cite{Bockelee1985} argued that the spatial profile of CN in the coma did not match a Haser model assuming release by the photodissociation of HCN. This is strong evidence that another source other than HCN plays a role in the production of CN in the coma of at least some comets. Other parent molecules have been investigated by \cite{Fray2005} who find that C$_2$N$_2$, HC$_3$N, and  CH$_3$CN, have lifetimes consistent with the parent of CN, even though their abundances in comets are likely insufficient to explain CN abundances. %C$_2$N$_2$ was only recently identified by \textit{Rosetta} in the coma of 67P but its abundance is also insufficient to account for the observed CN \citep{Hanni2021}. 
   Sublimation of carbon-rich dust grains, CHON particles, or other macro-molecules in the coma have also been suggested as potential extended sources of CN. 
   %A CN source from dust grains was first suggested by \cite{AHearn1986} to explain the existence of CN jets in the coma of 1P/Halley, but the exact production mechanism and potential specific sources need further investigation.

   \textit{C$_2$ and C$_3$.} The formation pathways of C$_2$ and C$_3$ are even less clear. 
 %This lack of understanding of the chemical process producing C$_2$ and C$_3$ prevents us from interpreting the composition differences between comets observed at optical wavelengths in terms of abundance of nucleus ices, and hinders the ability to link comets to their places of formation. 
   Molecules like C$_2$H$_2$ and C$_2$H$_6$ were suspected to be potential parents of C$_2$ \citep[e.g.,][]{Jackson1976} even before they were first detected at IR wavelengths in the 1990's \citep{Brooke1996,Mumma1996}. The photodissociation of C$_3$ itself is also thought to contribute to the formation of C$_2$. C$_3$ parents are less obvious but C$_4$H$_2$ and C$_3$H$_4$ were proposed in the 60s-70s by \citet{Swings1965} and \citet{Stief1972}. Other potential parents for the two radicals include HC$_3$N and C$_2$H$_4$. Because the spatial distribution of C$_2$ in the inner coma of comets is often flatter than what would be expected from the simple photo-dissociation of a parent molecule, \cite{Combi1997} suggested that C$_2$ might be released from a distributed source like CHON grains or by a two-step process. To this day, attempts to link C$_2$ and C$_3$ to potential parents through photochemical modelling have had mixed results: \cite{Helbert2005} concluded that C$_3$ could have C$_3$H$_4$ as main parent in C/1995 O1 (Hale-Bopp) while \cite{Weiler2012} found that the breakdown of C$_2$H$_6$ does not produce C$_2$ efficiently enough to be a major parent in most comets. As pointed out by \cite{Holscher2015}, some of the reactions in the chemical models used to estimate the origin of C$_2$ and C$_3$ have poorly known rate coefficients. 
 %More recently, \cite{Helbert2005} used a chemical network including C$_2$H$_2$, C$_2$H$_6$, and C$_3$H$_4$ as the main parent molecules of C$_2$ and C$_3$ to model their column densities at large heliocentric distances ($>$2.8~au) in the coma of in C/1995 O1 (Hale-Bopp) and concluded that C$_3$ could have C$_3$H$_4$ as main parent, while Using an updated chemical network applied to observations of comets closer to the Sun, \cite{Weiler2012} found that the breakdown of C$_2$H$_6$ does not produce C$_2$ efficiently enough to be a major parent in most comets. As pointed out by \cite{Holscher2015}, some of the reactions in the chemical models used to estimate the origin of C$_2$ and C$_3$ have poorly known rate coefficients. %They need to be improved in the future with new laboratory measurements and ab initio computations to allow for more realistic modelling of the chemical pathways leading to the production of these radicals. 
   Additionally, only some of the parents considered in the chemical networks mentioned above have been detected from ground-based observations of comets: HC$_3$N, C$_2$H$_6$, C$_2$H$_2$ \citep[see, e.g.,][]{Bockelee2017}. %This makes the comparison between parent and daughter molecules abundances very difficult. 
   \textit{Rosetta} detected many saturated molecules and even more unsaturated ones containing at least 2 carbon atoms that could be potential C$_2$ and C$_3$ parents \citep{Schuhmann2019}. Among all potential parents, C$_3$H$_4$ is not the most abundant, so it is likely not the only (or even primary) C$_3$ source in the coma of 67P. \cite{Altwegg2019} conclude that there are probably many parents leading to the production of these two radicals, with their exact contributions likely variable from comet to comet. 

   \textit{CS.} Another radical with uncertain parent is CS. It has been assumed to generally come from the photo-dissociation of CS$_2$. Most observations show evidence of production in the coma at least compatible with its production from CS$_2$, but a few constraints on the parent scale-length yield $2-4\times$ longer lifetimes \citep[$\tau_0 = 1/\beta_0 = 1000-2500$s,][respectively]{Biver1999,Roth2021} for the parent of CS than expected if the parent is CS$_2$. The low CS$_2$ abundances measured at 67P by \textit{Rosetta}/ROSINA compared to CS abundances measured in other comets (Table~\ref{tab:abundances}) suggests that it is not the main parent of CS.

   \subsubsection{Distributed sources and production versus r$_h$}
   \label{sec:fragment-small-rh}

   Few comets are observed spectroscopically at small heliocentric distances ($r_h$) and information on long term variability of composition from small heliocentric distances to greater than 1 au is lacking. However, there is emerging evidence from global molecular associations within the comet population that production rates of some molecules increase at smaller heliocentric distances relative to others \citep[e.g.,][]{DelloRusso2016a}. These species, including C$_2$H$_2$, H$_2$CO, and NH$_3$, have been traditionally considered as parents released from nucleus ices. From infrared observations of D/2012 S1 (ISON), \cite{DiSanti2016} report an increase of the NH$_3$, NH$_2$, HCN, H$_2$CO, and C$_2$H$_2$ to H$_2$O ratio within 0.5~au while from optical observations \citep{McKay2014} report increase of CN, C$_2$, CH, and C$_3$ abundances relative to H$_2$O. From Rosetta measurements, \cite{DeKeyser2017} found changes of the relative abundance of the hydrogen halides HF and HCl with decreasing heliocentric distance in the coma of 67P. Observations also show evidence of a strong dependency of the CS abundance with heliocentric distance in the coma of comets \citep[][and references therein]{Biver2006,Biver2011}, roughly as $Q_{\rm CS}/Q_{\rm H_2O} \propto r_h^{-1}$, down-to heliocentric distances as low as 0.1~au. A possible explanation for this behavior is significant release of these species from a distributed source of less volatile grains once a thermal threshold is reached. Because of the relative difficulty of detecting species like C$_2$H$_2$ and NH$_3$ in comets, confirming extended release through spatial distributions is challenging; however, there is evidence of extended release of these from spatial studies in bright comets \citep[e.g.,][]{DelloRusso2016b, DelloRusso2021,DiSanti2016}.  The presence of ammonium salts (NH$_4^+$X$^-$) in cometary ices was inferred based on observations with \textit{Rosetta}/ROSINA \citep{Altwegg2020a} and laboratory experiments \citep{Poch2020}. Sublimation of ammonium salts, which happens at higher temperature than for species like HCN or even water, could account for the  additional source of HCN and NH$_3$ at small heliocentric distances. %Ammonium salts are discussed in more details in Section \ref{sec:N-deficiency}.

   Other evidence also seem to indicate that some volatiles usually considered as parent volatiles could be released from extended sources even at typical heliocentric distances beyond 1 au. OCS showed a broad spatial distribution that is evidence of a possible extended source in C/1995 O1 Hale-Bopp \citep{DelloRusso1998}, but high signal-to-noise measurements of OCS are rare so it is unclear if this behavior is typical. Chemical models also suggest that much of the HC$_3$N and NH$_2$CHO seen in comets at radio wavelengths could be produced from reactions in the coma \citep[e.g.,][]{Cordiner2021}. Some comets have an inferred upper limit for the SO$_2$ abundance that is lower than the measured abundance for SO \citep[e.g., C/2014~Q2,][]{Biver2015}, which also suggests that there is another source of SO, or that SO is a parent molecule. Further interferometric observations are needed to test the sources of SO in a larger subset of comets.

   Radio observations have also shown that the abundance of H$_2$CO relative to water \citep[e.g., ][]{Biver2002} and the HNC/HCN ratio \citep{Lis2008} increase with decreasing heliocentric distance. While infrared observations indicate significant nucleus sources of formaldehyde (H$_2$CO) \citep{DelloRusso2016a}, measurements from the \textit{Giotto} flyby \citep{Meier1993} and radio and interferometric data point to an extended emission of formaldehyde for several comets \citep{Colom1992,Biver1999,Roth2021}. Millimeter studies and ALMA observations suggest that the unknown formaldehyde parent would have a scalelength between 1000 and 6800 km \citep{Bockelee2000,Cordiner2014,Cordiner2017}, which is shorter than the dissociation scalelength of CH$_3$OH, which mostly dissociates into H$_2$CO according to \citep{Huebner2015}. Thermal degradation of polyoxymethylene \citep[POM,][]{Fray2006} has been proposed as parent source of formaldehyde. Similarly, radio interferometric data indicate that HNC is mostly produced in the coma within a few thousands of km \citep{Cordiner2014,Roth2021} and could be released from the degradation of nitrogen-rich grains \citep{Cordiner2017} or come from the photodissociation of a more complex nitrile parent \citep{Rodgers2001}.
   More studies are necessary to better quantify the importance of extended sources for the production of species like OCS, H$_2$CO, HNC, or NH$_3$ and their variation among comets.

%-----------------------------------------------------------------------------------------------------------
%--------------------------------- Neil
\section{\textbf{VARIATIONS OF COMA COMPOSITION}}
\label{sec:var-composition}
Cometary composition is often deduced by measurements obtained at few (or one) points in time. These snapshots are frequently the only information that is available on the chemistry of a particular comet. Recently, longer timescale measurements of comets have had been increasingly obtained. It is clear in many cases from ground-based observations obtained over long time periods and \textit{Rosetta} measurements of 67P over much of its orbit that measured relative abundances of volatile species often vary with time.  This variability within a comet must be accounted for when comparing measured compositions between comets. A key question is whether these temporal effects seen in individual comets due to heterogeneous active areas are signatures preserved from comet formation (natal) or have evolved through various processes over time.
%--------------------------------- Nicolas
  \subsection{Differences in composition between comets}
  \label{sec:difference-composition}
   
 \begin{figure*}[ht!]
 \begin{center}
 \includegraphics[angle=0,width=\textwidth]{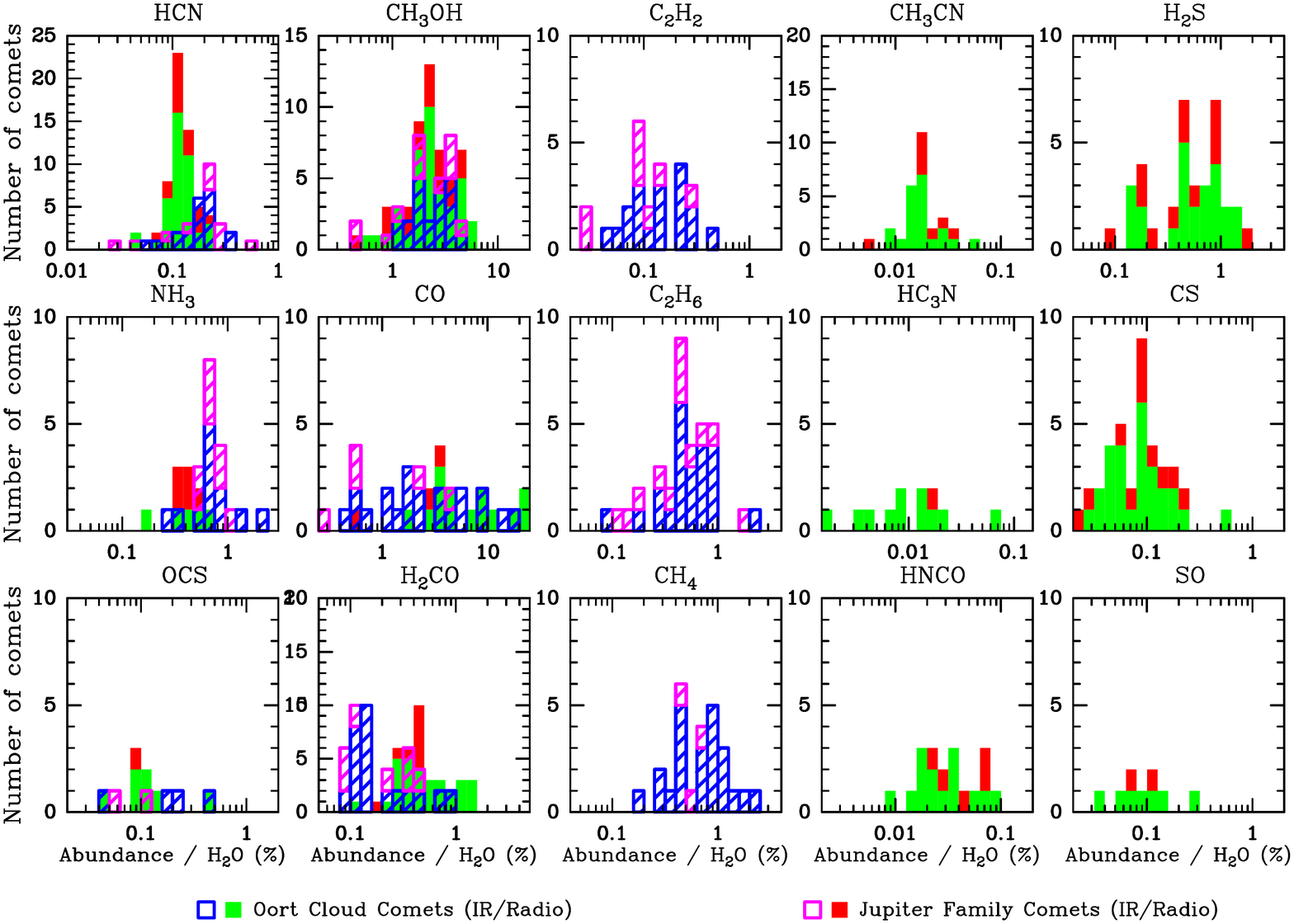}%\vspace{-1cm}
 \caption{Histograms (number of comets in each bin of abundance relative to water) of molecules observed in several comets both in the radio and in the infrared. Radio data correspond to filled bars and infrared are hatched. Blue-green colors are for comets originating from the Oort cloud (OCC), on periodic to parabolic orbits of any inclination. Red-pink colors are for JFCs in low inclination orbit with short period. HCN shows more systematic scatter from infrared data while in the other cases similar behavior is observed from the two techniques, infrared being able to sample comets with lower CO abundance. Excepted for CO which is less abundant on average in JFC \citep[by a factor 4][]{DelloRusso2016a}, there is no significant difference between JFCs and OCCs, with a scatter in abundance that can reach one order of magnitude.}
 \label{fig-histo1}
 \end{center}
 \end{figure*}

 %\begin{figure*}[ht!]
 %\begin{center}
 %\includegraphics[angle=0,width=0.8\textwidth]{fighistoh2scssoch3cnhc3nhnco.pdf}%\vspace{-1cm}
 %\caption{Histograms for 6 molecules observed in the radio in at least 10 comets. scales as in Fig.~\ref{fig-histo1}.}
% \label{fig-histo2}
% \end{center}
% \end{figure*}
 
 %\begin{figure*}[ht!]
 %\begin{center}
 %\includegraphics[angle=0,width=0.8\textwidth]{fighistoircxhy.pdf}%\vspace{-6cm}
 %\caption{Histograms for hydrocarbons observed in the infrared range. scales as in Fig.~\ref{fig-histo1}.}
 %\label{fig-histo3}
 %\end{center}
 %\end{figure*}
 
% \begin{figure*}
% \begin{center}
% \includegraphics[angle=0,width=0.8\textwidth]{fig4-aj285163-schleicher2008.pdf}%\vspace{-3cm}
% \caption{Histograms of abundances relative to OH or CN based on %observations of radicals in the visible range from %\cite{Schleicher2008}, based on a updated restricted dataset since %\cite{AHearn1995}. Comets highlighted in black are those of %``classical'' composition in contrary to carbon-chain depleted ones %(empty bars). Comet 96P is the one shown with diagonal striping.}
% \label{fig-histo4}
% \end{center}
% \end{figure*}

 \begin{figure*}
 \begin{center}
 \includegraphics[width=\textwidth]{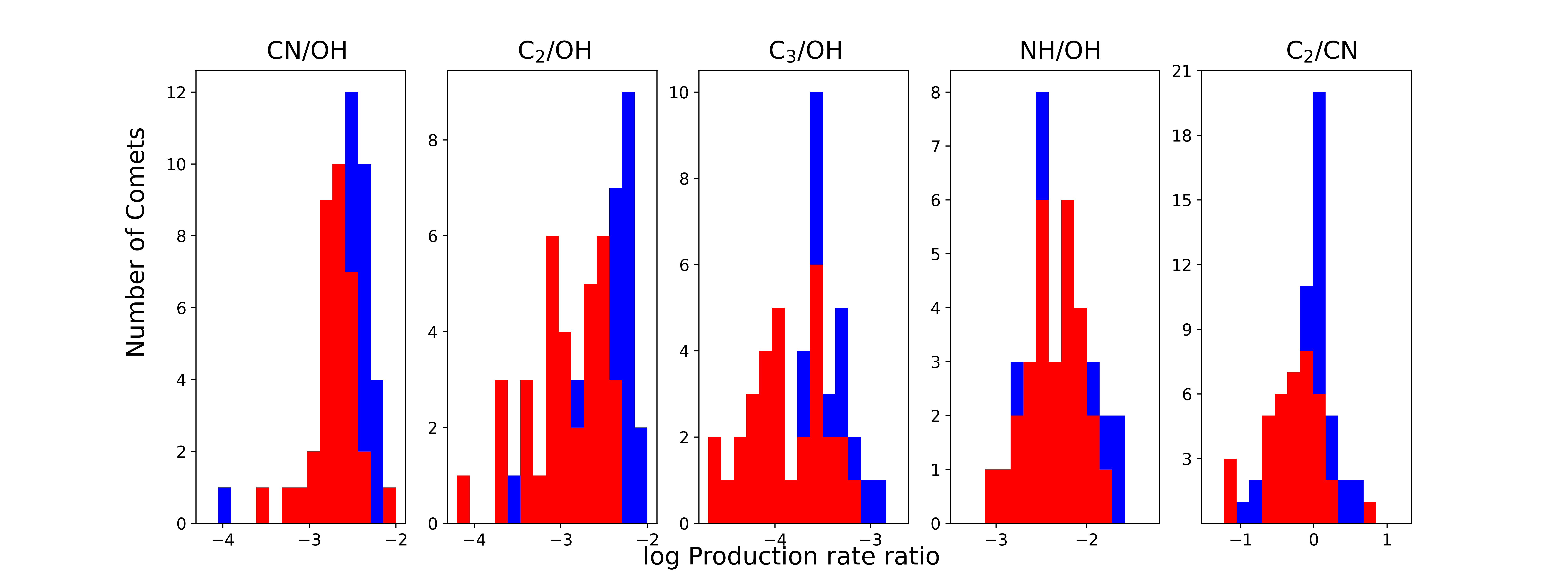}%\vspace{-3cm}
 \caption{Histograms of abundances relative to OH or CN based on observations of radicals in the visible range from \cite{Osip2010}, based on the data set of \cite{AHearn1995}. Blue color is for comet originating from the Oort cloud (OCC), on periodic to parabolic orbits of any inclination. Red color is for JFC in low inclination orbit with short period.}
 \label{fig-histo4}
 \end{center}
 \end{figure*}
 
    \subsubsection{Taxonomic classes}
    \label{sec:taxonomy}
    Following the work done by \cite{AHearn1995} based on narrowband visible photometry of product species in 85 comets and a subsequent update on over 153 comets \citep{Schleicher2008}, IR and radio spectroscopy based datasets also enable classification of comets according to the composition of parent volatiles.
    \cite{DelloRusso2016a} and \cite{Lippi2021} have established a taxonomy based on the composition of 20 to 30 comets observed at IR wavelengths, using the abundances of HCN, CH$_4$, C$_2$H$_6$, C$_2$H$_2$ CH$_3$OH, H$_2$CO, CO and NH$_3$. Earlier attempts based on radio spectroscopy haven't established significant sub-grouping based on, e.g., principal component analysis of the abundances of four to eight molecules. Difficulties are often linked to the consistency of the dataset, dealing with the largest number of comets versus dealing with a more limited number of objects with more precise abundance determinations for a larger number of molecules.
    Histograms showing the distribution of abundances relative to water (or OH) based on visible, IR and radio data show that for most molecules the distribution of abundances is typically Gaussian with a total width smaller than a factor $\sim$10 (Figs.~\ref{fig-histo1}--\ref{fig-histo4}). Evidence from the optical taxonomy from A'Hearn et al. (1995) shows that a group of comets is depleted in carbon-chain (C$_2$ and C$_3$) molecules, by typically one order of magnitude (Fig.~\ref{fig-histo4}). \cite{DelloRusso2016a} has established that arbitrary groupings of comets can be made based on their hydrocarbon, CH$_3$OH, HCN, NH$_3$, CO and H$_2$CO content and, as established by \cite{AHearn1995} for product species, also showed that the heliocentric dependence of parent abundances are also variable. %CS and HNC observed in the radio are also likely secondary products with strong heliocentric variation of their abundances. C$_2$, CN and NH might also come from the sublimation of grains in the coma \citep{DelloRusso2016a,AHearn1995}. This may explain why depleted C$_2$H$_2$ abundances in the IR are not necessarily correlated with the carbon-chain depleted group of comets from optical measurements. 
    Comets depleted in C$_2$H$_2$ in the IR are not necessarily correlated with the carbon-chain depleted group of comets from optical measurements. There is however some evidence that carbon-chain depleted comets have a rather low abundance of methanol in radio investigations \citep{Biver2021b}. However, for some molecules the distribution shown in Figs.~\ref{fig-histo1}--\ref{fig-histo4} might be better fitted by two Gaussians which suggests that additional data are needed to characterize molecular trends in comets.
 
 %    \subsubsection{Links with dynamical families}
     %CO has a wide range of abundances in comets (Fig.~\ref{fig-histo1}), which is also related to the dynamical class, as short-period JFCs are on average less abundant in CO. This is likely due to their dynamical history which brought them from the Kuiper belt into orbits with multiple close solar passages that can potentially deplete hypervolatile CO over time.
     %Also, according to \cite{AHearn1995} most carbon depleted comets are JFCs, i.e. originating from the Kuiper Belt. But it seems that links with dynamical origin -- JFCs with low inclination orbits versus OCCs with any inclination, comprising Halley-family periodic comets, long period and dynamical new comets from the Oort Cloud -- stops here.
     %Fig.~\ref{fig-histo1}--\ref{fig-histo3} shows distribution for both JFCs and OCCs, and besides the case for CO, the distribution of abundances for other species looks more similar to within the sample size limitations of the data to date. This might reflect the latest model of early solar system dynamical evolution, showing that Kuiper Belt and Oort Cloud objects were accreted in overlapping regions of the solar system\citep{Brasser2013}, although on a sufficiently large domain to get the dispersion in abundances that we observe.
     
     \subsubsection{Outliers}\label{Sec:outliers}
     Even considering the dispersion in molecular abundances seen within the comets observed over the last decades, a few comets show more extreme differences of one or two orders of magnitude in some molecular abundances.
     Comet C/2016~R2 (PanSTARRS) seems to be a member of a class of ``blue'' comets comprising also C/2002 VQ$_{94}$ (LINEAR), C/1961 R1 (Humason), C/1908 R1 (Morehouse) and possibly 29P/Schwassmann-Wachmann 1. A few other cases based on visible spectroscopy are given in \citet{Cochran2000}. As observed at 2.7~au from the Sun, the coma of comet C/2016~R2 shows a very unusual composition, strongly enriched in CO, N$_2$ and CO$_2$, compared to other comets even at the same heliocentric distance \citep[Table~\ref{tab-cometco} and ][]{Biver2018,McKay2019,Korsun2014}. The CO:N$_2$:CH$_3$OH:HCN ratio are typically orders of magnitude different from other comets. Explanations for such differences could be that either these comets are fragments of a differentiated TNO that were formed mostly from the volatile part of its surface, or that they formed further away than most of the comets, beyond the CO/N$_2$ ice lines \citep{Mousis2021}.
     The interstellar comet 2I/Borisov is also relatively enriched in CO compared to HCN \citep{Cordiner2020}, but still much closer to the tail of the distribution observed in solar system comets.

     Optical observations have also revealed potential compositional outliers such as comets 96P/Machholz and C/1988 Y1 (Yanaka): \cite{Schleicher2008} found them very depleted in CN and C$_2$ with abundances relative to OH typically two orders of magnitude lower than the average value measured in comets while \citep{Fink2009} found them typical regarding their NH$_2$ content.

%--------------------------------- Martin
  \subsection{Heterogeneity, evolution of the coma} \label{sec:heterogeneity}
%--------------------------------- Nicolas (13 july)
    \subsubsection{Temporal variability, outbursting and split comets}
     %In order to investigate the heterogeneity of comets, we look for variation in coma composition over both short- and long-term time intervals.
     %Monitoring of the composition as a function of time and heliocentric distances has been done for a few comets: C/1995~O1 (Hale-Bopp) was followed over a wide range of heliocentric distances \citep{Biver2002}, revealing that some molecules were more abundant in the coma closer to the Sun, and comet 67P was intensively monitored in-situ during two years by \textit{Rosetta} \citep{Biver2019a,Lauter2020}. In other comets observed at IR to radio-wavelength \citep[e.g.,][]{Feaga2014,DiSanti2016}, some molecules can be more abundant closer to the Sun or after perihelion due to a seasonal evolution. Similar changes of relative abundances of radicals have also been observed for some comets at optical wavelengths \citep{Opitom2015,McKay2014}. 
     Secular evolution in composition with time can also be investigated either on a short or longer time scales for periodic comets. For example, 2P/Encke \citep{Radeva2013, Roth2018} showed large compositional changes between two apparitions (2003 and 2017), showing a substantial decrease of C$_2$H$_6$ and CH$_3$OH by factors of 8 and 4 respectively, whereas H$_2$CO abundances relative to water increased by a factor of 3. However, these observations were not obtained during the same orbital phase and might be related to seasonal changes on the nucleus of 2P. On the other hand, comet 21P/Giacobini-Zinner was observed at three apparitions close to its perihelion time, in 1998, 2005, and 2018, and showed only minor short-term daily variations in abundances, with about the same average coma composition during each apparition \citep{Weaver1999, Mumma2000, DiSanti2013, Roth2020, Biver2021b,Moulane2020}. Generally the carbon-chain depleted comets within the visible taxonomy (previous Section) show this characteristic behavior at each perihelion which is indicative that a specific coma composition is generally characterizing each comet.
    
     Comet compositional heterogeneity can often be rigorously tested when a comet breaks up or fragments. First, fragmenting comets release volatiles from the previously protected interior of the nucleus that is likely more indicative of primitive material. Second, when a comet completely disintegrates or loses a significant amount of its mass, subsequent measurements of coma material should be representative of the bulk nucleus. Third, when comets fragment into pieces that are large enough to separately investigate, the chemistry of different parts of the original nucleus can be sampled and compared. 73P/Schwassmann-Wachmann 3 split in several pieces in 1995, and subsequently came within 0.08 au of the Earth in 2006 allowing observations with high sensitivity. Fragments C, B, and G showed a very similar composition from radio \citep{Biver2008}, IR \citep{DelloRusso2007}, and visible investigations \citep{Schleicher2011}.
     Measurements so far suggest that the degree of heterogeneity within comet nuclei is variable, and the likely cause of time variation (selective sublimation due to temperature threshold, seasonal variation due heat wave propagation or modification of near surface composition after multi-perihelion passages) might not exclude a more homogeneous composition seen in many comets. 
     
%--------------------------------- Martin
     
     \subsubsection{Local differences in the coma}
     Seasonal and diurnal variations in the (relative) abundances of volatile parent species have been observed in H$_2$O and CO$_2$ (cf. section~\ref{sec:CO2-driver}) but also in many other species \citep{Feaga2007,Hassig2015}. Non-homogeneity has also been reported for the radical species OH, [O I], CN, NH, and NH$_2$, albeit more diffuse given their origin from dissociation of cometary parent molecules \citep{Bodewits2016}. \cite{LeRoy2015} reported relative abundances for a suite of species measured at comet 67P at 3.1~au inbound. In this comparative study strong differences in the coma composition above the northern summer versus the southern winter hemisphere were obtained. CO varied from 2.7\% to 20\% and CO$_2$ from 2.5 to 80\% with respect to H$_2$O north to south. \cite{Bockelee2015b} reported a range of 2 to 30\% for CO$_2$ using VIRTIS-H but this may be explained by the difference in the measurement technique, taking into account that local measurements can lead to much higher variations compared to remote sensing observations representing global integrated densities along a column (cf. section~\ref{sec:intro}). When the southern winter hemisphere was poorly illuminated, increased ratios of CO, CO$_2$, CH$_4$, C$_2$H$_2$, C$_2$H$_6$, HCOOH, HCOOCH$_3$, HCN, OCS and CS$_2$ with respect to H$_2$O were measured, whereas for some other organics the differences were less striking \citep{LeRoy2015}. A difference was also noted from ground-based measurements of the global CN abundance \citep{Opitom2017}. This was expected due to the low volatility of water relative to these species. However, CO, CO$_2$, CH$_4$, C$_2$H$_2$, C$_2$H$_6$, HCN, OCS, and CS$_2$ were also more abundant in absolute numbers above the southern winter hemisphere at 3.1~au inbound. This hints at a heterogeneous nucleus, most likely caused by evolutionary processes as the southern hemisphere is subject to significantly higher rates of erosion per orbit \citep{Keller2015a}. A plausible explanation is that erosion exposes fresh material from the comet's interior to sublimation. However, once the southern hemisphere became more active during the short and intense summer months, the measured ratios with respect to water above the south at $\sim$1.5~au \citep{Rubin2019a} were similar to within a factor of a few to the ones measured above the northern hemisphere at $\sim$3.1~au \citep{LeRoy2015}. This indicates that the measured heterogeneity is also illumination dependent and hence temperature driven and occurs on seasonal as well as diurnal time-scales \citep{Hassig2015}, further modified by recondensation of H$_2$O \citep{DeSanctis2015}. It may also provide clues as to how different species are embedded inside the ices of the nucleus. %cf. next section.

     \subsubsection{Groups of Molecules sharing similar behaviors}
     \label{sec:MoleculeCorrelations}
     Cometary volatile species exhibit different outgassing patterns when it comes to seasonal \citep[see, e.g.,][]{Biver2002} and diurnal patterns \citep[cf.][]{Hassig2015}. It is no surprise that the volatility of the different ices can play a crucial role in the outgassing behavior of the nucleus. Hence, the relative abundances of the major volatiles CO and CO$_2$ with respect to H$_2$O tend to be lower near the comet's perihelion compared to larger heliocentric distances \citep[cf.][]{Ootsubo2012}.
    
     \textit{Rosetta} allowed an investigation of the relative abundances of species on diurnal and seasonal time-scales. Very early in the mission, substantial amounts of O$_2$ were detected \citep[cf. section~\ref{sec:O2} and][]{Bieler2015a}. Interestingly, there was also high correlation between the production of O$_2$ and H$_2$O, despite the vastly different sublimation temperatures of the corresponding pure ices \citep{Fray2009}. On the other hand O$_2$ showed much lower correlation to CO and N$_2$. A similar picture was also obtained for other volatiles, e.g. NH$_3$ and H$_2$O \citep{Lauter2020,Biver2019a} and among the noble gases Ar, Kr, Xe together with N$_2$ \citep{Balsiger2015,Rubin2018}.
    
     \cite{Gasc2017} investigated the change in outgassing of H$_2$O, CO$_2$, CO, H$_2$S, CH$_4$, HCN, O$_2$, and NH$_3$ as a function of heliocentric distance during the outbound path past the second equinox. As mentioned above, H$_2$O, O$_2$, and NH$_3$ dropped rapidly as the comet moved away from the Sun.  CO$_2$, CO, H$_2$S, CH$_4$, and HCN, on the other hand, dropped much more gradually and furthermore exhibited strong heterogeneity where the region of peak outgassing did not follow the sub-solar latitude. Little correlation to the pure-ice sublimation temperatures were observed, i.e., O$_2$ ($\sim$30~K) behaved very similar to H$_2$O ($\sim$144~K) but very different from CH$_4$ ($\sim$36~K). The strongest decrease in outgassing was observed for water while the CO$_2$ activity dropped by the smallest amount of the species listed above. 
    
     Volatile production rates measured over large timescales suggested two distinct ice phases in 67P, associated with either H$_2$O or CO$_2$ release \citep{Hassig2015, Fink2016, Gasc2017}. Two types of ices, a polar phase dominated by H$_2$O and an apolar phase dominated by CO and CO$_2$, have also been observed in the interstellar medium and young stellar objects \citep{Boogert2015,Mumma2011}. At 67P, CO$_2$ appeared correlated with C$_2$H$_6$, CO, H$_2$S, and CH$_4$ \citep{Luspay2015, Hassig2015, Gasc2017}, whereas H$_2$O appeared correlated with CH$_3$OH, NH$_3$, and O$_2$ \citep{Luspay2015, Gasc2017}. Some relationships varied with time; for example, HCN was sometimes correlated with CO$_2$ \citep{Gasc2017} and at other times with H$_2$O \citep{Luspay2015}. The correlation of O$_2$ with H$_2$O furthermore suggests that polarity was not the sole reason for this behavior. On the other hand, the different formation and processing mechanisms of some of these species, such as grain surface, gas phase chemistry, radiolysis, and thermal processing are likely key. Because the drop in outgassing of all these cometary parent molecules is bracketed by the two major species \citep{Rubin2019a,Lauter2020,Combi2020}, H$_2$O and CO$_2$, \cite{Gasc2017} suggested that all minor species are trapped in different proportions within H$_2$O and CO$_2$ and then released during phase transitions or co-desorbed during sublimation \citep{Kouchi1995, Mousis2016a}. It is possible that at one time minor species were present in their pure ices but that they are not anymore (or only to a very limited degree), with cometary activity governed by water and carbon dioxide. In this scenario, outgassing may be further modified by recondesation and resublimation processes \citep[][]{DeSanctis2015}. 
    
     For both comets and the interstellar medium, laboratory measurements are key to understanding this behavior \citep{Kouchi1995, Collings2004,Laufer2017}. However, mixed ice experiments in the laboratory are quite challenging for species present in trace amounts, i.e. $\ll$1\% with respect to H$_2$O and/or CO$_2$. But this corresponds to exactly the situation encountered at 67P. Furthermore, as the volatile coma structure in comets is explored by spacecraft it becomes important to connect these high-resolution spatial results with the global coma results obtained for a large number of comets from remote sensing observations.
    
     \subsubsection{Present day composition: Natal or evolved?}
    
     Ices within comet nuclei have been protected since formation by an overburden of material; thus, cometary volatiles observed today likely retain signatures from their birth. Yet comets are not perfectly preserved early solar system relics, as each nucleus has experienced a unique processing history over its long life, especially (but not only) during close passages to the Sun. Determining the degree to which comets retain their natal character is a fundamental question, but one that is difficult to answer. First, comets from the Jupiter family and the Oort cloud have quite different dynamical histories; however, comet forming regions were vast which leads to significant expected and observed compositional diversity within each dynamical class. Second, while some systematic compositional differences have been seen between comets from different dynamical classes on average, this could plausibly reflect either different evolutionary histories or distinct formative regions.
    
     As volatile abundances have been determined in an increasing number comets, there are several avenues of investigation that allow natal versus evolutionary effects in comets to be tested. (1) Determining abundances of the most volatile parents (e.g., N$_2$, CO, CH$_4$) shows the proficiency of comets in preserving these low sublimation temperature (hypervolatile) species. (2) Observations of Jupiter-family comets over multiple apparitions with modern instrumentation that allow detailed compositional information to be obtained is now feasible. (3) Spatial distributions of parent molecules in the coma provide evidence for how ices are associated or sequestered in the nucleus. Determining spatial distributions in many comets can test which spatial properties and possible ice associations in the nucleus are comet-specific and which are global among the population of comets.

%----------------------------------------------------------------------------------------------------------- 
%--------------------------------- Nicolas
\section{\textbf{ISOTOPIC AND ORTHO-TO-PARA RATIOS}}
\label{sec:isotopes}
 Isotopic ratios provide key information on the origin of cometary material. The abundances of the heavier elements in a molecule were determined by the reservoir of heavy isotopes such as HD, atomic D or H$_2$D$^+$, present during formation of these ices and the chemical reactions that happened during this time \citep[e.g.,][]{Taquet2013}. Reactions in the gas phase or in grain-gas interactions lead to various fractionation mechanisms in the ISM. Then when the planets formed in a warmer environment in the protoplanetary disk, further fractionation occurred due to sublimation, photodissociation screening, mixing, and recondensation at various distances and temperatures.  The results of these temperature and processing conditions, including isotopologues abundances, may have been frozen in cometary ices. So, isotopic ratios measured in the sublimating gas phase today may reveal their original value 4.6 Gy ago. However, the impact of isotopic exchange in the ice phase and during sublimation is a subject of debate.

 The ortho-to-para ratio (OPR) or spin temperature may be an estimate of the temperature of the molecules when they condensed into the icy grains that formed cometary nuclei. The change of spin state of a molecule is in principle strictly forbidden during collisions or radiative transitions and can only occur through chemical reactions, so present-day spin temperatures may be unchanged since cometary ices formed.

%--------------------------------- Nicolas
  \subsection{D/H ratios}
  \label{sec:D-to-H}
  D/H is the isotopic ratio of greatest interest as hydrogen is the most abundant element in the universe and the difference in the atomic mass (a factor of two) is the largest, causing significant differences in D/H ratios in various molecules. The main molecular reservoir of deuterium in the cold universe and Solar System is HD, but in comets it is HDO. Strong enrichment in deuterium in water (D/H=10$^{-4}$ to 10$^{-3}$) and other molecules is observed in star forming regions \citep[e.g.,][]{Drozdovskaya2021,Jensen2021} and in the Solar System. The reference D/H on Earth, the Vienna Standard Mean Ocean Water (VSMOW) value, is $1.558\times10^{-4}$, while in the local interstellar medium (ISM) the D/H value in molecular hydrogen is $2\times10^{-5}$. The HDO abundance in cometary ices is of key interest, especially when considering that comets may have contributed a significant amount of water to the young Earth. The D/H in water is half the HDO/H$_2$O, since there are two H atoms that can be substituted by D: D/H = 1/2$\times$(HDO/H$_2$O or DHO/H$_2$O).
  
\begin{figure*}[ht!]
 %\begin{center}
 \includegraphics[angle=0,width=0.9\textwidth,trim=0 190 0 5,clip]{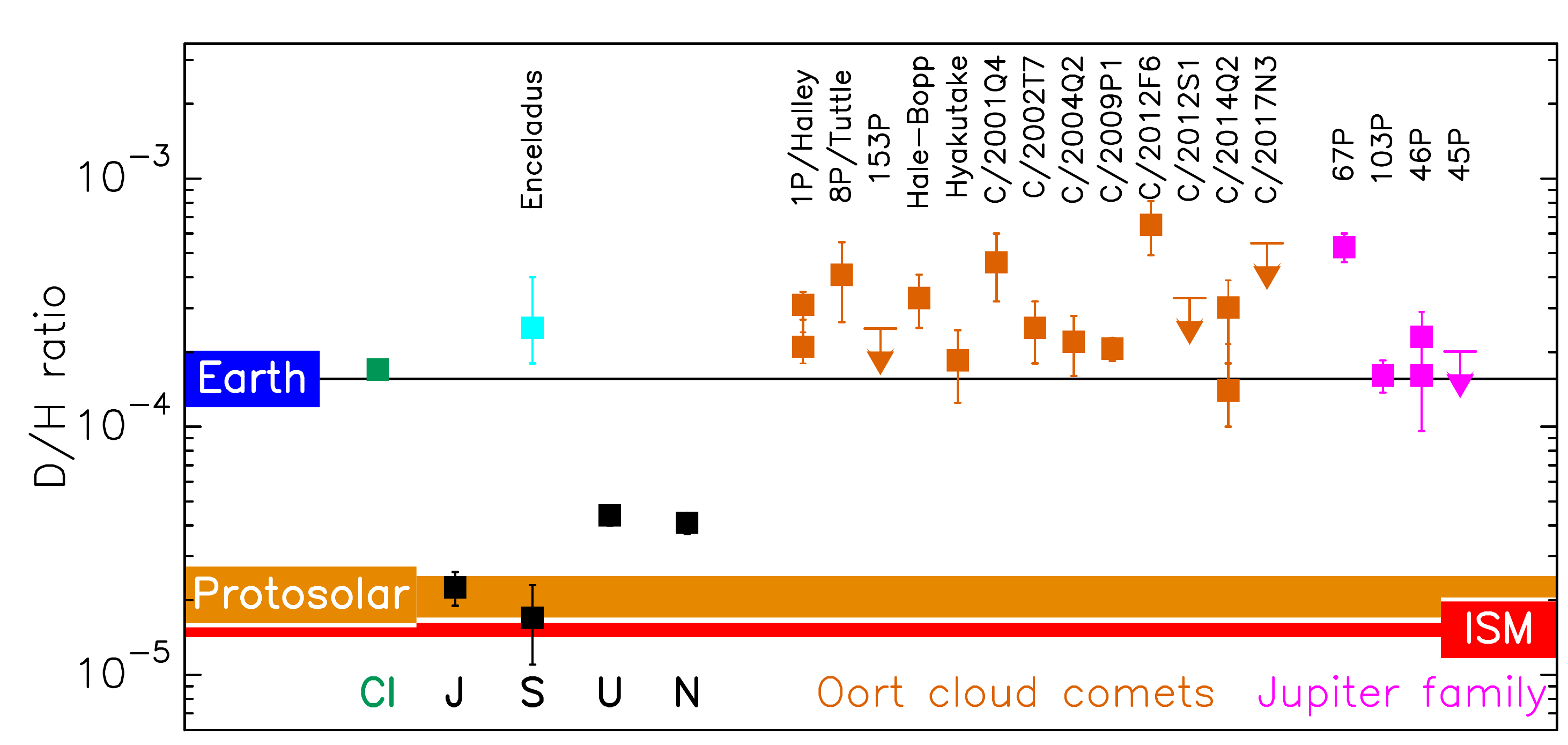}%\vspace{-2cm}
 \caption{Values of the D/H in solar system objects, in H$_2$ for giant planets and the protosolar and ISM reservoirs or in water for Earth, CI meteorites, Enceladus and comets. Adapted from \citet{Lis2019}, with data from \citet[][ and references therein]{Waite2009,Hartogh2011, Altwegg2019,Biver2016,Paganini2017,Biver2022,Lis2019}}
 \label{fig-dhratio}
 %\end{center}
 \end{figure*}

  D/H in cometary water has been measured in a dozen comets (Fig.~\ref{fig-dhratio}) with sensitive upper limits obtained in a few others. Cometary D/H varies between one and four times the VSMOW value as measured by different techniques. Radio spectroscopy samples HDO in the coma generally over thousands of km around the nucleus via the $J_{K_aK_c}=1_{01}-0_{00}$ line at 464.9~GHz, the $1_{10}-1_{01}$ line at 509.3~GHz, or the $2_{11}-2_{12}$ line at 241.6~GHz. HDO production rates determined at radio wavelengths are subject to uncertainty in the excitation model. IR spectroscopy (as well as UV spectroscopy of OD) is in theory less sensitive to coma temperature since it integrates several rovibrational lines within a vibrational band (electronic band in the UV). In both cases precise determination of the HDO/H$_2$O, requires the detection of several HDO lines as well as the simultaneous detection of the water. However, H$_2$O and HDO are typically sampled through observations of lines in different spectral regions that require additional instrument settings. %In all cases the large differences between the number of H$_2$O and HDO molecules (three orders of magnitude) present in the coma makes these measurements sensitive to non-linearity and optical thickness of H$_2$O lines. 
  Values of the D/H in water measured in comets have no apparent correlation to the dynamical origin (Jupiter-Family vs Oort cloud long period). It has been suggested recently by \citet{Lis2019} that the D/H could be anti-correlated with hyperactivity of the comet. For example, comets with an equivalent active surface larger than 100\% would have terrestrial D/H while those with small active surface like comet 67P have higher D/H. More measurements of D/H in comets are needed to test if this effect is real and whether this could be dictated by differences in formation conditions between hyperactive and low activity comets.

  D/H has also been measured in a handful of other molecules in the volatile phase: DCN was remotely detected in C/1995 O1 Hale-Bopp \citep{Meier1998b} with a D/H eight times the VSMOW value. Measurements with ROSINA on-board the \textit{Rosetta} spacecraft have found even higher enrichment in deuterium in other molecules such as D$_2$O (D$_2$O/HDO =$1.8\pm0.9$\%), HDS, CH$_3$OD/CH$_2$DOH. These are often expected as fractionation processes in the ISM and star forming regions yield similar enrichments \citep{Drozdovskaya2021}. D/H measurements in cometary comae are presented in Table~\ref{tab-dhratio}.
  
\begin{table}[ht!]
\caption[]{D/H in cometary molecules}\label{tab-dhratio}
\vspace{0.1cm}
%\begin{center}
\begin{tabular}{lll}
\hline
Deuterated molecule & D/H value & comet \\  
\hline
HDO      & $1.4-6.5\times10^{-4}$  & Several   \\ 
%D$_2$O/HDO     & $180\pm90\times10^{-4}$  &  67P$^a$   \\
HDCO     & $<70\times10^{-4}$        & C/2014~Q2$^b$ \\
HDS      & $12\pm3\times10^{-4}$  &  67P$^{a*}$   \\
         & $<170\times10^{-4}$        & C/2014~Q2$^b$ \\
         & $<80\times10^{-4}$        & 17P/Holmes$^x$ \\
DCN      & $23\pm4\times10^{-4}$  &  Hale-Bopp$^{c,d}$   \\
NH$_2$D  & $11\pm2\times10^{-4}$  &  67P$^a$   \\
         & $<400\times10^{-4}$  & Hale-Bopp$^f$   \\
CH$_3$D  & $24\pm3\times10^{-4}$  &  67P$^{a}$ \\
         & $<64\times10^{-4}$  &  C/2004~Q2$^e$   \\
         & $<50\times10^{-4}$  &  C/2004~Q2$^f$   \\
C$_2$H$_5$D  & $24\pm3\times10^{-4}$  &  67P$^{a}$ \\
CH$_3$OD  & \vline~$140\pm30\times10^{-4}$ & 67P$^g$ \\ \smallskip
CH$_2$DOH & \multicolumn{2}{l}{\vline~(for same deuteration on each H)} \\
CH$_3$OD  & $<300\times10^{-4}$ & Hale-Bopp$^d$ \\ 
          & $<100\times10^{-4}$ & C/2014~Q2$^x$ \\ 
CH$_2$DOH & $<80\times10^{-4}$ & Hale-Bopp$^d$ \\ 
          & $<35\times10^{-4}$ & C/2014~Q2$^x$ \\ 
\hline
\multicolumn{3}{c}{Doubly deuterated species (see text for D$_2$O)} \\
\hline
%\remove{\mbox{D$_2$O     & $44\pm14\times10^{-4}$  &  67P$^a$   \\} }
CHD$_2$OH & \vline~$110\pm10\times10^{-4}$ & 67P$^g$ \\ \smallskip
CH$_2$DOD & \multicolumn{2}{l}{\vline~(for same deuteration on each H)} \\
\hline
\end{tabular}
Refs.: $^a$\cite{Muller2022}, $^b$\cite{Biver2016}, $^c$\cite{Meier1998b},
$^d$\cite{Crovisier2004b}, $^e$\cite{Kawakita2009}, $^f$\cite{Bonev2009}, $^g$\cite{Drozdovskaya2021}, $^x$unpublished, $^*$ Note that the value in \citet{Altwegg2019} and in the abstract of \citet{Altwegg2017a} is wrong by a factor 2.\\
%\end{center}
\end{table}

%--------------------------------- Cyrielle
  \subsection{Abundance of $^{15}$N in comets}
  \label{sec:15N}
  
  The ratio of $^{14}$N/$^{15}$N across the solar system shows significant diversity, ranging from the proto-solar value of 441 \citep{Marty2011}, to the terrestrial value of 272 \citep{Anders1989}, to material very enriched in $^{15}$N ($^{14}$N/$^{15}$N $\sim$ 50) in insoluble organic matter in carbonaceous chondrites \citep{Bonal2010}. The origin of the diversity in $^{15}$N across the solar system, and in comets, is still poorly understood. 

  The $^{14}$N/$^{15}$N isotopic ratio has been measured in several comets from ground-based observations using HCN, CN, and NH$_2$ \citep{Bockelee2015a}. Initial measurements made for comet C/1995 O1 Hale-Bopp using sub-millimeter detection of HCN \citep{Jewitt1997,Ziurys1999} derived a value of 323$\pm$46, consistent with the terrestrial value while the value measured for CN was over two times lower: 140$\pm$ 35 \citep{Arpigny2003}. Subsequent measurements for a dozen comets made from high resolution spectroscopy of CN \citep{Jehin2009}, together with almost simultaneous measurements in comet 17P/Holmes from both HCN and CN, and a re-analysis of the Hale-Bopp data \citep{Bockelee2008}, found an isotopic ratio consistent with the original CN measurement. For a sample of 20 comets of different origins and observed at different distances from the Sun \cite{Manfroid2009} measured an average $^{14}$N/$^{15}$N=147.8$\pm$5.7. They also point out that the isotopic ratios are remarkably constant (within the uncertainties) for all comets, irrespective of their origin.  

\begin{figure}[ht!]
 \begin{center}
 \includegraphics[trim=30 70 110 60,clip,width=\columnwidth]{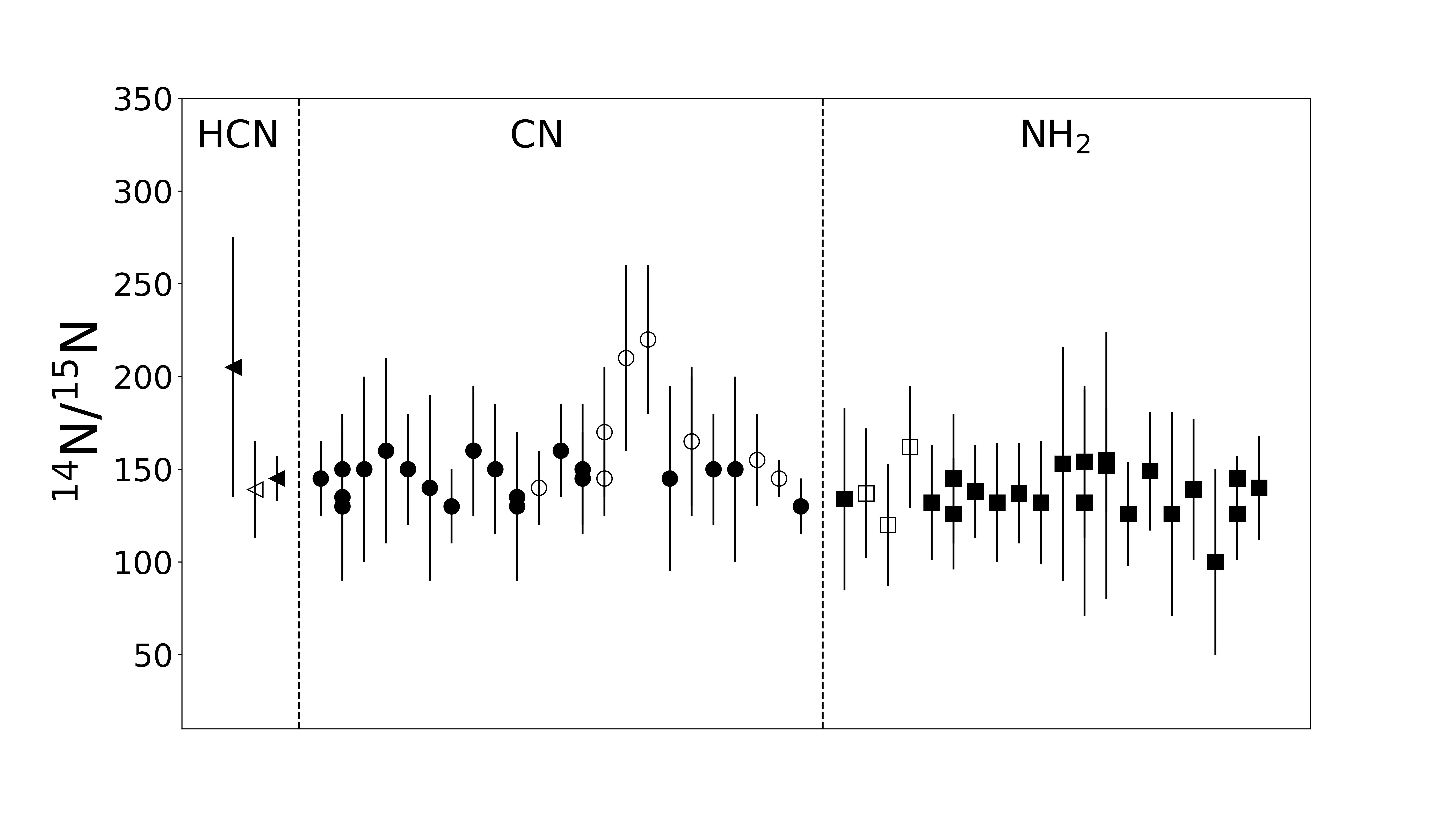}%\vspace{-3cm}
 \caption{$^{14}$N/$^{15}$N isotopic ratios in comets measured using HCN, CN, and NH$_2$. Full symbols represent Oort Cloud comets while open symbols represent Jupiter Family comets from the Kuiper Belt. Data from \cite{Bockelee2008,Manfroid2009,Jehin2011,Moulane2020,Yang2018,Shinnaka2016,Biver2016,Shinnaka2014,Rousselot2015,Shinnaka2016b}.}
 \label{fig-Nisotope}
 \end{center}
 \end{figure}

  Ammonia is another major nitrogen reservoir in cometary ices but the lack of sensitivity of current instrumentation has prevented the measurement of $^{15}$NH$_3$ so far. The $^{14}$N/$^{15}$N was only recently indirectly measured in NH$_2$, assumed to be produced by the photodissociation of NH$_3$. The measurement, performed both on co-added spectra of several comets and on individual comets is consistent within the uncertainties with that measured for C$^{14}$N/C$^{15}$N \citep{Rousselot2014,Shinnaka2016}, as can be seen in Fig. \ref{fig-Nisotope}. The first detection of molecular nitrogen in the coma of a comet by \textit{Rosetta}/ROSINA and the subsequent measurement of a $^{14}$N/$^{15}$N revealed a value consistent with the ratio measured for CN and NH$_3$ \citep{Altwegg2019}.

  The origin of the large variation of N fraction across the solar system and the enrichment of $^{15}$N in comets is still debated. The existence of two distinct nitrogen reservoirs, one atomic (likely leading to the formation of HCN and NH$_3$) and one molecular (N$_2$) has been suggested \citep{HilyBlant2017}. We refer the reader to \citet{Marty2022} for a more in depth discussion of nitrogen fractionation in the early solar system.
 
%--------------------------------- Martin
  \subsection{Other isotopic ratios: S, O, C}\label{sec:SOC-isotopes}
  Isotopic ratios of carbon, oxygen, and sulfur are available for a suite of comets \citep[cf review by][]{Bockelee2015a}. Table~\ref{tab:CarbonIsotopes} provides $^{12}$C/$^{13}$C ratios measured in comets and corresponding references. Carbon isotope ratios are often consistent within uncertainties with the the terrestrial V-PDB (Vienna Pee Dee Belemnite) reference. As discussed in section~\ref{sec:radicals}, a subset of the ratios provided in Table ~\ref{tab:CarbonIsotopes} are measured in radicals (C$_2$, CN) and may hence reflect a mixture of different parent species. This further complicates the identification of any species-dependent differences based on the chemical origin. Compared to the case of nitrogen (cf. section~\ref{sec:15N}) only limited isotopic anomalies have been identified so far in comets. One notable exception is the case of H$_2$CO which shall be discussed later.
  
     \begin{table}[ht!]
     \renewcommand{\tabcolsep}{0.32 cm}
     \caption[]{\label{tab:CarbonIsotopes} Carbon isotope ratios in comets}
     %\begin{center}
     \vspace{0.3cm}
     \begin{tabular}{llc}
     \hline
     Species & Comet  & $^{12}$C/$^{13}$C \\
           & reference: & (89$^a$/89$^b$) \\
     \hline 
     C$_2$ & C/1975 V1 (West) & 60$\pm$15$^c$ \\
        %\cline{2-3} 
        & 4 comets & 93$\pm$10$^d$ \\
        %\cline{2-3}
        & C/2002 T7 & 85$\pm$20$^e$ \\
        %\cline{2-3}
        & C/2001 Q4 & 80$\pm$20$^e$ \\
        %\cline{2-3}
        & C/2012 S1 & 94$\pm$33$^f$ \\
     \hline
     CN & 1P/Halley & 65$\pm$9$^g$ \\
        &  & 89$\pm$17$^h$ \\
        &  & 95$\pm$12$^i$ \\
        %\cline{2-3}
        & C/1995 O1 (Hale-Bopp) & 90$\pm$15$^j$ \\
        &  & 165$\pm$40$^k$ \\
        %\cline{2-3}
        & C/1990 K1 Levy & 90$\pm$10$^d$ \\
        %\cline{2-3}
        & C/1989 X1 Austin & 85$\pm$20$^d$ \\
        %\cline{2-3}
        & C/1989 Q1 O-L-R & 93$\pm$20$^d$ \\
        %\cline{2-3}
        & 122P/de Vico & 90$\pm$10$^l$ \\
        %\cline{2-3}
        & 88P/Howell & 90$\pm$10$^m$  \\
        %\cline{2-3}
        & 9P/Tempel 1 & 95$\pm$15$^n$ \\
        %\cline{2-3}
        & 17P/Holmes & 90$\pm$20$^o$ \\
        %\cline{2-3}
        & 21 comets & 91$\pm$4$^p$ \\
        %\cline{2-3}
        & 103P/Hartley 2 & 95$\pm$15$^q$ \\
        %\cline{2-3}
        & C/2012 F6 & 95$\pm$25$^r$ \\
        %\cline{2-3}
        & C/2015 ER$_{61}$ & 100$\pm$15$^s$ \\
        %\cline{2-3}
        & 21P/G.-Z. & 100$\pm$10$^t$ \\
    \hline
    HCN & C/1995 O1 & 111$\pm$12$^u$ \\
        &  & 94$\pm$8$^o$ \\
        %\cline{2-3}
        & 17P/Holmes & 114$\pm$26$^o$ \\
        %\cline{2-3}
        & C/2014 Q2 & 109$\pm$14$^v$ \\
        %\cline{2-3}
        & C/2012 F6 & 124$\pm$64$^v$ \\
        %\cline{2-3}
        & C/2012 S1 &  88$\pm$18$^w$ \\
        \hline
    CO & 67P & 86$\pm$9$^x$ \\
        \hline
    CO$_2$ & 67P & 84$\pm$4$^y$ \\
        \hline
    CH$_4$ & 67P & 88$\pm$10$^z$ \\
        \hline
    C$_2$H$_6$ & 67P & 93$\pm$10$^z$ \\
        \hline
    H$_2$CO & 67P & 40$\pm$14$^{aa}$ \\
        \hline
    CH$_3$OH & 67P & 91$\pm$10$^{aa}$ \\
     \hline
     \end{tabular}
     %\end{center}
      $^a$V-PDB terrestrial \citep{Meija2016} and $^b$solar \citep{Lodders2010}, in comets: $^c$\cite{Lambert1983}, $^d$\cite{Wyckoff2000}, $^e$\cite{Rousselot2012}, $^f$\cite{Shinnaka2014}, $^g$\cite{Wyckoff1989}, $^h$\cite{Jaworski1991} , $^i$\cite{Kleine1995}, $^j$\cite{Lis1997}, $^k$\cite{Arpigny2003}, $^l$\cite{Jehin2004} , $^m$\cite{Hutsemekers2005}, $^n$\cite{Jehin2006}, $^o$\cite{Bockelee2008}, $^p$\cite{Manfroid2009}, $^q$\cite{Jehin2011}, $^r$\cite{Bockelee2017}, $^s$\cite{Yang2018}, $^t$\cite{Moulane2020} , $^u$\cite{Jewitt1997}, $^v$\cite{Biver2016}, $^w$\cite{Cordiner2019}, $^x$\cite{Rubin2017}, $^y$\cite{Hassig2017}, $^z$\cite{Muller2022}, and $^{aa}$\cite{Altwegg2020b}.
     \end{table}

  Table~\ref{tab:OxygenIsotopes} shows measured $^{16}$O/$^{18}$O isotopic ratios including $^{16}$O/$^{17}$O in H$_2$O and O$_2$ and corresponding references. Most $^{16}$O/$^{18}$O ratios have been measured in cometary water and are in most cases consistent within 1-$\sigma$ with the terrestrial VSMOW reference. 
  %(Vienna Standard Mean Ocean Water). 
  \textit{Rosetta} revealed oxygen isotope ratios in a few additional species \citep{Altwegg2020b}. These initial results reveal a remarkable heterogeneity among the different chemical groups. Both the $^{16}$O/$^{17}$O and $^{16}$O/$^{18}$O ratios in H$_2$O and O$_2$ differ, on the other hand, the ratios $^{18}$O/$^{17}$O~=~4.5$\pm$1.0 (O$_2$) and 4.9$\pm$0.6 (H$_2$O) are both consistent with each other and within 1-$\sigma$ of the VSMOW reference (5.3). CO$_2$ and CH$_3$OH have $^{16}$O/$^{18}$O ratios comparable to H$_2$O, but the ratios obtained for H$_2$CO, SO, SO$_2$, and OCS are about half while that for O$_2$ remains somewhere in between. Furthermore, \cite{Schroeder2019a} analyzed oxygen isotopes over the course of the full \textit{Rosetta} mission and revealed no variation in the $^{16}$O/$^{18}$O ratio as a function of cometary activity.

     \begin{table}[ht!]
     \renewcommand{\tabcolsep}{0.2 cm}
     %\begin{center}
     \caption[]{\label{tab:OxygenIsotopes} Oxygen isotope ratios in comets}
     \vspace{0.3cm}
     \begin{tabular}{llcc}
     \hline
     Species & Comet  & $^{16}$O/$^{17}$O & $^{16}$O/$^{18}$O \\
        & reference: & (2632$^a$/2798$^b$) & (498.7$^a$/530$^b$) \\
    \hline
    H$_2$O & 1P/Halley & & 518$\pm$45$^c$ \\
        & & & 470$\pm$40$^d$ \\
        %\cline{2-4} 
        & 153P & & 530$\pm$60$^{e,f}$ \\
        %\cline{2-4}
        & C/2001 Q4 & & $\sim$530$\pm$60$^f$ \\
        %\cline{2-4}
        & C/2002 T7 & & $\sim$550$\pm$75$^f$ \\
        &           & & 425$\pm$55$^g$ \\
        %\cline{2-4}
        & C/2004 Q2 & & 508$\pm$33$^f$ \\
        %\cline{2-4}
        & C/2012 F6 & & 300$\pm$150$^h$ \\
        %\cline{2-4}
        & C/2009 P1 & & 523$\pm$32$^i$ \\
        %\cline{2-4}
        & 67P & 2347$\pm$191$^m$ & 445$\pm$45$^j$ \\
        \hline
    O$_2$ & 67P & 1544$\pm$308 & 345$\pm$40$^k$ \\
        \hline
    CO$_2$ & 67P &  & 494$\pm$8$^l$ \\
        \hline
    CH$_3$OH & 67P &  & 495$\pm$40$^k$ \\
        \hline
    H$_2$CO & 67P &  & 256$\pm$100$^k$  \\
        \hline
    SO & 67P &  & 239$\pm$52$^k$ \\
        \hline
    SO$_2$ & 67P &  & 248$\pm$88$^k$ \\
        \hline
    OCS & 67P &  & 277$\pm$70$^k$ \\
        \hline
     \end{tabular}
     %\end{center} 
     $^a$VSMOW terrestrial \citep{Clayton2003} and $^b$solar \citep{McKeegan2011}, in comets: $^c$\cite{Balsiger1995}, $^d$\cite{Eberhardt1995}, $^e$\cite{Lecacheux2003}, $^f$\cite{Biver2007}, $^g$\cite{Hutsemekers2008}, $^h$\cite{Bockelee2017}, $^i$\cite{Bockelee2012}, $^j$\cite{Schroeder2019a}, $^k$\cite{Altwegg2020b}, $^l$\cite{Hassig2017}, $^m$\cite{Muller2022}.
     \end{table}

  Table~\ref{tab:SulfurIsotopes} lists the measured $^{32}$S/$^{33}$S and $^{32}$S/$^{34}$S ratios and corresponding references. The number of observed comets is limited and the observed species are radicals such as CS or the S$^+$ ion originating from fragmentation of various S-bearing species inside the \textit{Giotto} NMS (cf. section~\ref{sec:mass-spectro}). Nevertheless, these results are consistent with the $^{32}$S/$^{34}$S measurements of cometary parent species obtained at 67P: the $^{32}$S/$^{34}$S ratios are consistent with the terrestrial reference samples within uncertainties and also the ratio $^{32}$S/$^{34}$S~=~21.6$\pm$2.7 \citep{Paquette2017} measured in cometary dust by \textit{Rosetta}/COSIMA \citep[COmetary Secondary Ion Mass Anaylzer;][]{Kissel2007}. However, deviations from the terrestrial and solar system references have been observed in the $^{32}$S/$^{33}$S ratios, namely the $^{33}$S isotope was found to be depleted in H$_2$S, OCS, and CS$_2$. Similar to the oxygen isotopes, the deviations are notable and in excess of what is typically found in the material in the inner solar system aside from the much less abundant presolar grains \citep{Hoppe2018}.

    \begin{table}[ht!]
    \renewcommand{\tabcolsep}{0.15 cm}
    %\begin{center}
    \caption[]{\label{tab:SulfurIsotopes} Sulfur isotope ratios in comets}
     \vspace{0.3cm}
    \begin{tabular}{llcc}
    \hline
        Species & Comet  & $^{32}$S/$^{33}$S & $^{32}$S/$^{34}$S \\
        & reference: & (126.9$^a$/126.7$^b$) & (22.6$^a$/22.5$^b$) \\
     \hline
     S$^+$ & 1P/Halley & & 23$\pm$6$^c$ \\
        \hline
     CS & C/1995 O1 & & 27$\pm$3$^d$ \\
        %\cline{2-4}
        & C/2014 Q2 & & 24.7$\pm$3.5$^e$ \\
        %\cline{2-4}
        & C/2012 F6 & & 20$\pm$5$^e$ \\
        \hline 
      H$_2$S & C/1995 O1 & & 16.5$\pm$3.5$^f$ \\
        %\cline{2-4}
        & 46P/Wirtanen & & 20.6$\pm$2.9$^g$ \\
        %\cline{2-4}
        & 67P & & \\
        & ~~(10/2014) & 187$\pm$9$^h$ & 23.6$\pm$0.4$^h$ \\
        & ~~(05/2016) & 132$\pm$3$^h$ & 22.4$\pm$1.6$^h$ \\
        \hline
        OCS & 67P & & \\
        & ~~(10/2014) & & 25.1$\pm$1.3$^h$ \\
        & ~~(03/2016) & & 21.7$\pm$4.0$^i$ \\
        & ~~(05/2016) & 165$\pm$12$^h$ & 22.8$\pm$0.3$^h$ \\
        \hline
        CS$_2$ & 67P & & \\
        & ~~(10/2014) & 157$\pm$7$^h$ & 24.3$\pm$0.7$^h$ \\
        & ~~(05/2016) & 151$\pm$8$^h$ & 25.3$\pm$0.6$^h$ \\
        \hline
        SO & 67P & & 23.5$\pm$2.5$^i$ \\
        \hline
        SO$_2$ & 67P & & 21.3$\pm$2.1$^i$ \\
        \hline
     \end{tabular}
     %\end{center}
     $^a$V-CDT terrestrial \citep{Ding2001} and $^b$solar \citep{Lodders2010} and in comets: $^c$\cite{Altwegg1995}, $^d$\cite{Jewitt1997}, $^e$\cite{Biver2016}, $^f$\cite{Crovisier2004b}, $^g$\cite{Biver2021a}, $^h$\cite{Calmonte2017}, $^i$\cite{Altwegg2020b}. Note that the 1P/Halley value represents a mixture of species fragmenting into S (cf. section \ref{sec:mass-spectro}). Also note that \cite{Calmonte2017} analyzed 2 periods for H$_2$S, OCS, and CS$_2$, one early (10/2014) and the other late (05/2016) in the Rosetta mission.
     \end{table}

  \cite{Altwegg2020b} reported significant $^{18}$O and $^{13}$C enrichment in H$_2$CO compared to CH$_3$OH and terrestrial/solar system standards, while the $^{18}$O/$^{13}$C ratios are in agreement. This is remarkable as both molecules were thought to originate from the same pathway, i.e., the hydrogenation of CO \citep{Watanabe2002}. There are observations in the interstellar medium that reveal a similar picture, comparable carbon isotope ratios in CO and CH$_3$OH but significant deviations in H$_2$CO \citep{Wirstrom2011a,Wirstrom2011b}. Large isotopic fractionation has also been modeled and found in other O-bearing species in the gas phase of cold clouds (10~K) in the interstellar medium \citep{Loison2019}. Hence, despite the resemblance to the findings at 67P, the question remains to what extent these results can be related to each other.
  
  Furthermore the difference in both the $^{16}$O/$^{17}$O and $^{16}$O/$^{18}$O ratios in H$_2$O and O$_2$ is important and suggests that O$_2$ does not directly originate from H$_2$O (cf. section~\ref{sec:O2}). This favors a primordial origin of O$_2$ based chemistry occurring on grain surfaces \citep{Taquet2016}. The $^{16}$O/$^{18}$O isotope ratio varies by up to a factor of 2 in the observed species. For the S-bearing subset, SO, SO$_2$, and OCS, the $^{32}$S/$^{34}$S isotope ratio, however, is consistent with terrestrial and solar reference material.
  
  %A more in-depth discussion on the implications of the measured isotope ratios can be found in a following chapter. 
  %\com{Martin}{I think it's Marty et al.? Depends of course on the topics that will be discussed in that paper.}, the review by \cite{Hoppe2018,Rubin2020} and others.

%--------------------------------- Martin
  \subsection{\label{sec:noble-gas-isotopes}Noble gas isotopes}
  %While comet's have often been suggested to have contributed water to the Earth's oceans \citep{Hartogh2011}, they likely also made significant contributions to the terrestrial organic inventory \citep{Oro1961}, \citep{Anders1977} and atmosphere in the case of noble gases \citep{Marty2016}. 
  The relative abundances of a suite of noble gas isotopes reported in section~\ref{sec:noble-gases} were obtained by \textit{Rosetta}. Argon and krypton isotope ratios seem to be in agreement with the solar and terrestrial reference material listed in table~\ref{tab:NobleGasIsotopes} within measurement uncertainties. For xenon, however, deviations from these standards were measured in 67P. In particular $^{129}Xe$ was enriched while both $^{134}$Xe and $^{136}$Xe were considerably depleted. Taking cometary Xe, mixing it with with solar wind Xe, and taking mass-dependent isotopic fractionation into account, \citet{Marty2017} and \citet{Marty2022} estimated a contribution of 22.5$\pm$5\% of cometary xenon to the terrestrial atmosphere.
  
  %The depleted heavy Xe isotopes in 67P resemble a previously postulated, but not yet identified primordial component labelled U-Xe \citep{Pepin2000}. Mixing U-Xe with solar wind Xe and taking mass-dependent isotopic fractionation into account, \cite{Marty2017} estimated a contribution of 22.5$\pm$5\% of cometary xenon to the terrestrial atmosphere. Based on the relative abundances of the noble gases, water, organics, and other species in 67P, it was estimated that the contribution of cometary water to Earth and carbon to its crust was minor ($<$1\% and a few \% at max for nitrogen) while it may have been substantial for organic compounds \citep{Marty2016,Rubin2019b}.

  \begin{table*}
  \renewcommand{\tabcolsep}{0.113 cm}
  %\begin{center}
  \caption[]{\label{tab:NobleGasIsotopes} Noble gas isotope ratios in comet 67P and reference material.}
  \vspace{0.3cm}
  \begin{tabular}{lccccccl}
  \hline
        Species & Isotopes & & & & & & Reference \\ 
  \hline
        Argon & $^{36}$Ar/$^{38}$Ar & & & & & & \\
        \hline
        67P & 5.4$\pm$1.4 & & & & & & \cite{Balsiger2015} \\
        solar wind & 5.470$\pm$0.003 & & & & & & \cite{Heber2012} \\
        terr atm. & 5.319$\pm$0.008 & & & & & & \cite{Lee2006} \\ \hline

        Krypton & $^{84}$Kr/$^{80}$Kr & $^{84}$Kr/$^{82}$Kr & $^{84}$Kr/$^{83}$Kr & $^{84}$Kr/$^{86}$Kr & & \\
        \hline
        67P & 23$\pm$14 & 4.9$\pm$0.4 & 5.3$\pm$0.4 & 3.4$\pm$0.2 & & & \cite{Rubin2018} \\
        solar & 24.4 & 4.88 & 4.93 & 3.31 & & & \cite{Lodders2010} \\ 
        terr atm. & 24.9 & 4.92 & 4.96 & 3.30 & & & \cite{Aregbe2001} \\ \hline

        Xenon & $^{132}$Xe/$^{128}$Xe & $^{132}$Xe/$^{129}$Xe & $^{132}$Xe/$^{130}$Xe & $^{132}$Xe/$^{131}$Xe & $^{132}$Xe/$^{134}$Xe & $^{132}$Xe/$^{136}$Xe & \\
        \hline
        67P & 13$\pm$4 & 0.7$\pm$0.1 & 5.4$\pm$0.9 & 1.2$\pm$0.1 & 4.2$\pm$0.9 & 8.6$\pm$3.0 & \cite{Marty2017} \\
        %U-Xe & 11.9 & 0.96 & 6.05 & 1.21 & 2.84 & 3.64 & \cite{Pepin2000} \\ 
        solar & 11.8 & 0.96 & 6.02 & 1.21 & 2.73 & 3.35 & \cite{Lodders2010} \\ 
        terr atm. & 14.1 & 1.02 & 6.62 & 1.27 & 2.58 & 3.04 & \cite{Valkiers1998} \\
  \hline
  \end{tabular}
  %\end{center}
  \end{table*}
  
  \subsection{\label{sec:halogen-isotopes}Halogen isotopes}
  The measured isotopic ratios of $^{35}$Cl/$^{37}$Cl and $^{79}$Br/$^{81}$Br in 67P were consistent with solar system values when taking the uncertainties into account which points to molecular cloud chemistry. The isotope ratio $^{35}$Cl/$^{37}$Cl did not change throughout the mission \citep{Dhooghe2021}.
  
%--------------------------------- Neil
  \subsection{Ortho-to-para and spin state ratios}
  \label{sec:opr}
  Molecules that have identical nuclei having non-zero nuclear spin, especially hydrogen atoms having a spin of 1/2, can exist in different energy levels due to their total spin value $I$. Molecules such as water, formaldehyde or ammonia can be in two different spin states, ($I=1$ {\sl ortho}, or 3/2) or ($I=0$ {\sl para}, or 1/2), or A and E states and can be characterized by their {\sl ortho-to-para} (OPR), or A/E abundance ratio. CH$_4$ can be in three different spin states ($I=$ 0, 1 or 2). H$_2$O, the most common species where OPRs are measured, achieves a statistical equilibrium value of 3/1 for temperatures above $\sim$ 50 K, whereas the para species is increasingly favored as the temperature decreases below 50 K. OPRs have shown variability in measured values in comets \citep[e.g.,][]{Faure2019}, with values for $T_{spin}$ consistent with statistical equilibrium above 50 K in some cases, while others are sometimes clustered close to 30~K \citep[][and references therein]{Bockelee2004,Kawakita2009}. To what extent differences in OPRs among comets relate to differences in the formation temperatures of ices in the solar nebula or sublimation and coma processes remains an open question \citep[e.g.,][] {Bonev2007, Bonev2013, Faure2019, Aikawa2022, Bodewits2022}. There is some evidence that once OPRs are locked into H$_2$O ice upon its formation, conversion is difficult \citep[e.g.,][]{Miani2004}, so it is possible that OPRs have remained stable since ices were incorporated into the nucleus and represent the chemical formation temperature of cometary ices \citep[e.g.,][]{Kawakita2005}.  However, other laboratory results suggest OPRs are not ancient but instead are reset to their high-temperature equilibrium value after sublimation, independent of formation processes \citep{Hama2018}. 

\section{\textbf{PERSPECTIVES}}
\label{sec:perspectives}
 Over the last fifteen years since Comets II was published, increasing remote sensing capabilities have greatly increased the number of comets where the volatile composition and spatial distributions in the coma have been determined. Additionally, in-situ analysis through comet missions have allowed the study of volatile release and coma spatial distributions with unprecedented detail. The \textit{EPOXI} flyby provided the highest spatial resolution measurements of how volatiles were released from the nucleus of 103P/Hartley 2 \citep{AHearn2011} near perihelion, and supporting ground-based studies showed that spatial distributions in the global coma were consistent with spacecraft observations \citep[e.g.,][]{DelloRusso2011, Mumma2011b, Bonev2013, Kawakita2013}. The two-year \textit{Rosetta} rendezvous with 67P allowed the chemical composition of a comet to be determined with unprecedented detail and many new and more complex volatiles were detected through observations with the ROSINA mass spectrometer \citep[e.g.,][]{Rubin2019a, Altwegg2017b}. \textit{Rosetta} and \textit{EPOXI} observations also revealed distinct outgassing behavior for 67P and 103P/Hartley 2.

 %Based on the ground-breaking \textit{Rosetta}/ROSINA results, it is likely that a future spacecraft encounter with a comet will include a mass spectrometer, allowing a detailed comparison to results from the \textit{Rosetta} mission. A high priority in the coming decades is the collection and return of a comet nucleus sample, and eventually a cryogenic sample that includes volatiles. This will allow analysis with the most advances laboratory instrumentation on Earth, providing the next leap in our knowledge of the volatile content and structure of ices in a comet. Despite these likely future advances, mission studies will remain rare and confined to a very few objects within a compositionally diverse population. Therefore, remote sensing techniques will remain important in studying the composition and diversity of comets as a population. Technological advances in ground-based facilities and instrumentation will continue to push towards improved sensitivity, efficiency, spectral coverage, and spatial resolution which increases the number and variety of comets available for both compositional and coma spatial studies.

%--------------------------------- Nicolas
  \subsection{The state of molecular line databases}
  \label{sec:databases}
  Knowledge of the synthetic spectrum of a molecule or radical is essential to identify the species and compute its abundance in cometary comae. Knowing the fragmentation patterns in mass spectrometers is also necessary to evaluate the amount of the considered molecular species observed. There are still many unidentified lines in cometary spectra, especially in the high spectral resolution optical and IR spectra where the number of lines can be large.
  Laboratory measurements and ab-initio calculations are regularly providing new catalogs of lines, available in databases such as the JPL one \citep[\url{https://spec.jpl.nasa.gov},][]{Pickett1998} or CDMS \citep[\url{https://cdms.astro.uni-koeln.de/},][]{Muller2005} for rotational transition\add{s}, and HITRAN \citep[\url{https://lweb.cfa.harvard.edu/HITRAN/},][]{Gordon2017} or GEISA \citep[\url{https://cds-espri.ipsl.upmc.fr/ etherTypo/index.php?id=950\&L=1}, ][]{Jacquinet2016} for ro-vibrational lines. Other databases are available but there is always a need for new data to be able to search for, detect, and quantify new molecular species or radicals in cometary atmospheres. For example, acetaldehyde was present in spectra of C/1995~O1 (Hale-Bopp) but could only be identified later when accurate frequencies where available \citep{Crovisier2004b}. %The knowledge of its full spectrum with precise line frequencies (in the radio an accuracy of 100 to 10~kHz can be necessary to study the line shapes and determine if lines overlap) has enabled detection of acetaldehyde in several comets by averaging multiple lines. 
  Several molecular species identified by ROSINA in the coma of 67P, such as CH$_3$OSH, CH$_3$S$_2$H, C$_3$H$_7$NH$_2$ \citep[][ Table 4]{Altwegg2019}, do not yet have published spectra. Similarly, data on line frequencies and strengths are also needed to extract new species from already existing spectra such as in the case of $^{15}$NH$_2$ \citep{Rousselot2014}. There is also an on-going need for basic chemical information such as provided by the National Institute of Standards and Technology (NIST, chemistry WebBook: \url{https://webbook.nist.gov/chemistry}), %\cite[][]{NIST}) 
  for the fragmentation patterns of molecules in mass spectrometers. %, and especially for the various ionisation energies used by the instruments.
  
  In all cases the progress in the detection and determination of abundances of more and more complex species in cometary atmospheres, enabled by increased performance of observatories, cannot come without laboratory work and the release of new molecular data \citep[see ][]{Poch2022}.
  
%--------------------------------- New subsection: Nicolas (Apr. 2022)
  \subsection{\label{sec:ISMcomparison}Comparison to ISM and pathway for new searches}
  Comets are considered as possible samples of interstellar material from which the solar system formed. Molecular complexity found in comets has allowed comparison to the composition of nearby star forming regions, especially the low mass protostar IRAS 16293-2422(B) \citep{Biver2015,Biver2019a,Drozdovskaya2019} in which abundances of COMs relative to CH$_3$OH measured in several comets (C/2014 Q2, 46P, 67P) are of the same order of magnitude. Inner regions of solar-type protostars \citep[e.g.,][]{Taquet2015} and those with strong polar jets like L1157-B1 \citep{Lefloch2017} reveal a complex chemistry that likely results from a process similar to the sublimation of comet ices.
  The heritage of cometary ices from the ISM is also reviewed in previous chapters \citep{Bergin2022,Aikawa2022}. Such similarities in composition drives our search for new molecules in cometary comae, since many more complex species have been already detected in the ISM \citep{McGuire2018}. This requires more and more sensitivity as abundances generally decrease with molecular complexity (e.g. number of C atoms in the molecule) \citep{Crovisier2004b}.
  On the other hand, addressing the bulk abundances in cometary comae and ices is probably easier and less model dependent than in the ISM, due to the proximity of the source, optical thinness of the lines and access to in-situ observations via space missions. 

%--------------------------------- Martin
  \subsection{\label{sec:futuremissions}Open questions and future prospects}
  
  Despite the substantial insights gained in comet composition over the years, many fundamental questions remain unanswered \citep{Thomas2019}. For instance, what are the processes governing outbursts and activity in general and their associated time and size scales. This includes how volatiles are distributed within the nucleus and the relationship between the composition of the nucleus and measurements in the coma. Another question is how comets evolve as they are processed in the inner solar system. The stratigraphy of the outgassing layer, affected by erosion and fall-back material, and the interaction of the gas with the dust cover and how this sets the dynamics of the neutral gas in the coma are also open questions. What is the interrelation between the neutral gas, dust, and plasma (e.g. the lifting and acceleration of dust by neutral gas and the formation of a diamagnetic cavity, the innermost region around an active comet devoid of solar wind particles and fields)? 
  
  Some of these questions will be addressed by future missions \citep{Snodgrass2022} such as the Chinese ZengHe mission \citep{Zhang2021}, which will visit MBC 311P/PanSTARRS, or ESA's Comet Interceptor (CI) mission \citep{Snodgrass2019}, which proposes to fly by a long-period comet (LPC), or, if possible, a dynamically new comet (DNC) or an interstellar object (ISO). Further mission concepts have been proposed \citep{Thomas2019} and are discussed in an accompanying chapter. A high priority in the coming decades is the collection and return of a comet nucleus sample, and eventually a cryogenic sample that includes volatiles. This will allow analysis with the most advances laboratory instrumentation on Earth, providing the next leap in our knowledge of the volatile content and structure of ices in a comet.
  
  Despite these likely future advances, mission studies will remain rare and confined to a very few objects within a diverse population. Therefore, remote sensing techniques will remain important in studying the composition and diversity of comets as a population. Technological advances in ground and space-based facilities and instrumentation will continue to push towards improved sensitivity, efficiency, spectral coverage, and spatial resolution which increases the number and variety of comets available for both compositional and coma spatial studies. In addition to improvements to existing facilities, new telescopes including the James Webb Space Telescope \citep[JWST; ][]{Kelley2016}, the Vera C. Rubin Observatory \citep{Jones2009,Schwamb2018}, the Thirty Meter Telescope \citep{Skidmore2015}, and the European Extremely Large Telescope \citep[ELT;][]{Jehin2009,Brandl2008} are expected to become available in the coming years and further advance our understanding of the composition and activity of comets.
  
  Finally, we are now starting to be able to compare the composition of solar system comets to that of comets formed around other stars. Two interstellar objects have been detected crossing through our solar system. One of them was active and its composition could be compared to that of solar system comets. The existence of comets around other stars, inferred for the fist time in the late 1980s, also provides us with a way to assess the composition of comets in other planetary systems. Comparing the composition of exocomets to solar system comets remains difficult, as they are observed under different circumstances, either when they pass very close to their stars (like sungrazing comets) or directly in the debris disk. Observations of interstellar comets and the study of exocomets with upcoming facilities will give us the opportunity to understand how different formation conditions influence the composition of comets and the nature of our solar system compared to its neighbors.

  \label{sec:return-sample}
\vskip .5in
\noindent \textbf{Acknowledgments.} \\

MR was funded by the State of Bern and the Swiss National Science Foundation (200020\_182418, 200020\_207312).

%---------------------------------
\bibliographystyle{sss-three}
%\bibliographystyle{sss-full}
%\printbibliography
\bibliography{bibliography.bib}
% run >bibtex comet-atmosphere-chemistry to get latex working on the file
\end{document}